\documentclass[]{amsart}
\usepackage{amsfonts}
\usepackage{amssymb}
\usepackage{amsmath}
\usepackage{bbm}
\usepackage{graphicx}
\usepackage{color}
\usepackage{mathtools}
\usepackage{chngcntr}
\counterwithin{figure}{section}
\usepackage{caption}
\usepackage{subcaption}

\usepackage[italian, english]{babel}
\usepackage[T1]{fontenc} 

\usepackage[bookmarks=true]{hyperref}
\usepackage{bookmark}

\usepackage[toc, page]{appendix}

\DeclarePairedDelimiter{\abs}{\lvert}{\rvert}

\DeclareMathOperator{\Span}{span}

\DeclareMathOperator{\cotan}{cotan}
\DeclareMathOperator{\dd}{d}
\DeclareMathOperator{\acos}{acos}

\newtheorem{proposition}{Proposition}[section]
\newtheorem{remark}{Remark}[section]

\graphicspath{{figures/}}

\def \R {{\mathbb {R}}}
\def\S{{\mathbb{S}}}
\def \N {{\mathbb N}}

\def \PO {{\mathbb {R}^3\rtimes\mathbb{S}^2}}

\makeatletter
\newcommand\thankssymb[1]{\textsuperscript{\@fnsymbol{#1}}}
\makeatother

\title{ Good continuation in 3D: \\the neurogeometry of stereo   vision}

\author{M. V. Bolelli$^{1}$}
\author{G. Citti$^1$}
\author{A. Sarti$^2$}
\author{S. W. Zucker $^3$}

\thanks{$^1$Department of Mathematics, University of Bologna, Italy}
\thanks{$^2$CAMS, CNRS - EHESS, Paris, France.}
\thanks{$^3$Departments of Computer Science and Biomedical Engineering, Yale University, New Haven, CT, United States.}

\begin{document}
\maketitle
\begin{abstract}
Classical good continuation for image curves is based on $2D$ position and orientation. It is supported by the columnar organization of cortex, by psychophysical experiments, and by rich models of (differential) geometry. Here we extend good continuation to stereo. We introduce a neurogeometric model, in which the parametrizations involve both spatial and orientation disparities. Our model provides insight into the neurobiology, suggesting an implicit organization for neural interactions and a well-defined $3D$ association field. Our model sheds light on the computations underlying the correspondence problem, and illustrates how good continuation in the world generalizes good continuation in the plane.


\bigskip

\end{abstract}


\section{Introduction}

Binocular vision is the ability of the visual system to provide information about the three-dimensional environment starting from two-dimensional retinal images.  Disparities are among the main cues for
depth perception and stereo vision but, in order to extract them, the brain needs to determine which features in the right eye correspond to those in the left eye, and which do not. This generates a coupling problem, which is usually referred to as the \textit{stereo correspondence problem}. Our goal in this paper is to develop a perceptual organization approach to stereo, extending good continuation for planar curves to that for $3D$ spatial curves. 

Orientation good continuation in the plane (retinotopic coordinates) is one of the foundational principles of Gestalt perceptual organization. It enjoys an extensive history \cite{wagemans2012century}. It is supported by psychophysical investigations (e.g., \cite{HHF03, FHH93, elder2002ecological, geisler2001edge, lawlor2013third}), which reveal connections to contour statistics; it is supported by physiology (orientation selectivity), which reveals the role for long-range horizontal connections \cite{BZSF97}; and it is supported by computational modeling (\cite{BZ04, SCP07}), which reveals a key role for geometry.  
Historically, good continuation in depth is much less well developed than good continuation in the plane, despite having comparable origins. Quoting Koffka \cite[p. 161-162]{K63} :

\begin{quote}
    ...a perspective drawing, even when viewed monocularly, does not give the same
vivid impression of depth as the same drawing if viewed through a stereoscope
with binocular parallax ... for in the stereoscope the tri-dimensional force of the
parallax co-operates with the other tri-dimensional forces of organization; instead
of conflict between forces, stereoscopic vision introduces mutual reinforcement.
\end{quote}
\noindent
Our specific goal in this paper is to develop good continuation in depth analogously to the models of contour organization in two dimensions. Psychophysical investigations suggest this should be feasible (\cite{khuu2016perception, deas2014gestalt, deas2015perceptual, uttal2013visual}); our focus will be more mathematical. 

Specifically, in Koffka's words, we seek to develop a computational model of "mutual reinforcement." Although only one dimension higher than contours in the plane, contours extending in depth raise subtle new issues; this is why a computational model can be instructive. First among the issues is the choice of coordinates which, of course, requires a mathematical framework for specifying them. In the plane, position and orientation are natural; smoothness is captured by curvature or the relationship between nearby orientations along a contour. For stereo, there is monocular structure in the left eye and in the right.  Spatial disparity is a standard variable relating them, and it is well known that primate visual systems represent this variable directly \cite{poggio1995mechanisms}. Spatial disparity is clearly a potential coordinate. However, other physiological aspects are less clear. The columnar architecture so powerful for contour organization in the plane is monocular; it may well be a different story for contours in depth where, at least in V1, there are ocular dominance bands (see next Section). Nevertheless, orientationally-selective cells provide the input for stereo so, at a minimum, both positional disparity and orientation -- one orientation for the right eye and (possibly) another for the left -- should be involved. While it is traditional to assume only "like" orientations are matched \cite{HW62, MP79, bridge2001responses, chang2020experience, nelson1977discrimination}, our sensitivity to orientation disparity questions this, making orientation disparity another putative variable. We shall show that orientations do play a deep role in stereo, but that it is not necessarily efficent to represent them as a disparity. Furthermore, this settles a classical debate in stereo psychophysics about orientation: since its physiological realization could be confounded with disparity gradients \cite{mitchison1990mechanisms, cagenello1993anisotropies}, orientation may be redundant. This is not the case, since it is the orientation of the "gradient" that matters. Thus we can describe the main technical goal of this paper: to provide a representation of the geometry of spatial disparity and orientation in support of using good continuation in a manner that both incorporates the biological "givens" and provides a rigorous foundation for the stereo correspondence problem. As has been the case with curve organization, we further believe that our modeling will illuminate the underlying connectomics of stereo, at least at the earlier stages, even if the columnar organization is not clear. This would be important if, as is the case with orientation columns in the mouse \cite{ringach2016spatial}, the neural architectural support for stereo is laid out only implicitly in the connections (rather than in explicit columns).

Hubel and Wiesel introduced disparity-tuned neurons \cite{HW70}. They observed that single units could be driven from both eyes and that it was possible to plot separate receptive fields (RF) for each eye. We emphasize these monocular RFs are tuned to orientation; see  \cite{CD01} for more on the physiology. The architecture and the neural connections of the visual cortex underlying the establishment of stereoscopic correspondence in binocular vision have recently been studied in \cite{P16}, and a review of neural models can be found in \cite{R15}.

The classical model for expressing the left/right-eye receptive field combination is the \textit{binocular energy model (BEM)}, first introduced in \cite{AOF99S}. It encodes disparities through the receptive profiles of simple cells, raising the possibility of both position and phase disparities  \cite{JR15}.  However Read and Cumming \cite{RC07}, building upon \cite{AOF99}, showed that phase disparity neurons tend to be strongly activated by false correspondence pairs. Therefore, it is widely concluded, the most relevant disparity in the receptive fields is the position alone. This, however, neglects the orientation difference between the two eyes \cite{NKB77}, neglecting the orientation disparity. Although there are attempts to extend the energy model to incorporate binocular differences in receptive-field orientation \cite{BCP01}, they are limited. The geometrical model we will present incorporates orientation differences directly.

Many other mathematical models for stereo vision based on neural models have been developed. Many have observed (e.g., \cite{MP79}) that orientations should match between the two eyes, although small differences are allowed. This, of course, assumes the structure is frontal-parallel. Subsequently,  Jones and Malik \cite{JM92} used a set of linear filters tuned to different orientations (and scales) but their algorithm was not built on aneurophysiological basis. Subsequently, Zucker et al. \cite{AZ00, LZ03, Z14} built a more biologically-inspired model that addressed the connections between neurons. Their differential-geometry model employed position, orientations and curvatures in $2D$ retinal planes, modeling binocular neurons with orientations given by tangent vectors of Frenet geometry. Our results here are related, although the geometry is deeper. (We develop this below.)  A more recent work, based on differential geometry and precisely Riemannian geometry, is developed in \cite{N18}.
Before specifying these results, however, we introduce the specific type of geometry that we shall be using. It follows directly from the columnar organization often seen in predators and primates.

\subsection{Columnar architectures and sub-Riemannian geometry}

We propose a sub- Riemannian model for the cortical-inspired geometry underlying stereo vision based on the encoding of positional disparities and orientation differences in the information coming from the two eyes. We build on neuromathematical models, starting from the work of Hoffmann \cite{H89} and Koenderink-van Doorn \cite{KVD97}, with particular emphasis on the neurogeometry of monocular simple cells (\cite{CS06, P08, PT99,   CSS11, SC15, SCP07}). 

\begin{figure}[tbh]
\begin{subfigure}[b]{0.39\textwidth}
         \centering
         \includegraphics[width=\textwidth]{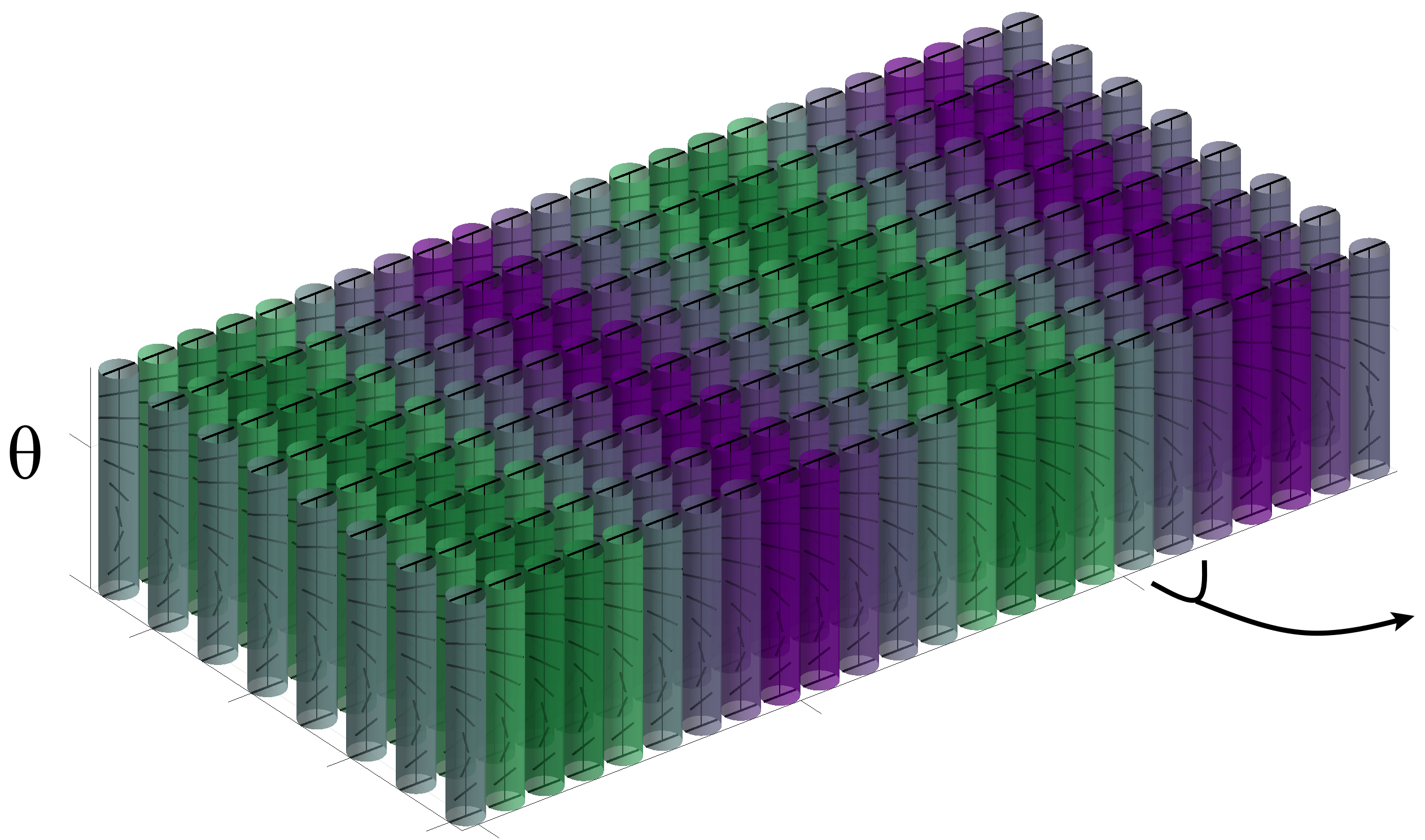} 
         \caption{}
\end{subfigure}
\begin{subfigure}[b]{0.26\textwidth}
         \centering
         \includegraphics[width=\textwidth]{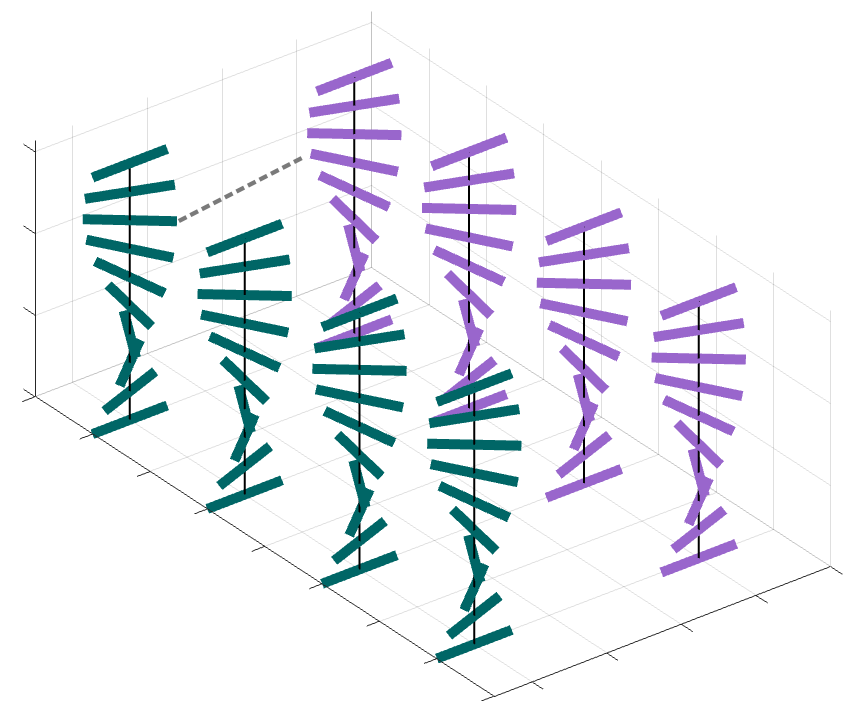}
         \caption{}
\end{subfigure}
\begin{subfigure}[b]{0.33\textwidth}
         \centering
         \includegraphics[width=\textwidth]{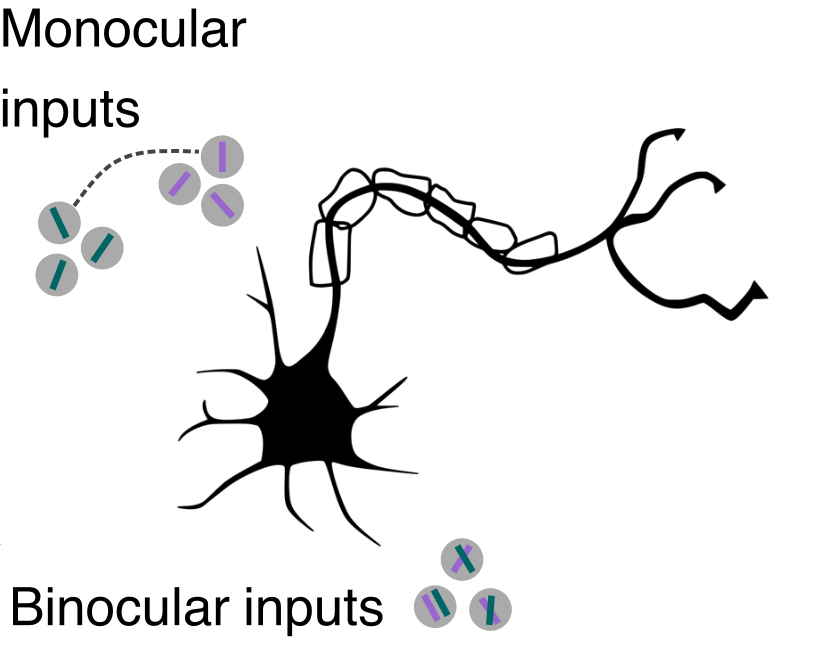}
         \caption{}
\end{subfigure}
      \caption{Cartoon of visual cortex, V1, superficial layers. (A) Macroscopic organization: A number of (abstracted) orientation hypercolumns, colored by left-eye (green)/right-eye (purple) dominant inputs. The color grading emphasizes that at the center of the ocular dominance bands the cells are strongly monocular, while at the boundaries they become binocularly-driven. (B) A zoom in to a few orientation columns showing left and right monocular cells at the border of ocular dominance bands. Cells in these nearby columns will provide the anatomical substrate for our model. (C) More recent work shows that both monocular and binocular inputs matter to these cells (redrawn from \cite{scholl2022binocular}, using data from ferret). This more advanced wiring suggests the connection structures in our model.}
      \label{fig:bands}
\end{figure}

To motivate our mathematical approach, it is instructive to build on an abstraction of visual cortex. We start with monocular information, segregated into ocular dominance bands \cite{levay1975pattern} in layer 4; these neurons have processes that extend into the superficial layers.  We cartoon this in Fig.~\ref{fig:bands}, which shows an array of orientation hypercolumns arranged over retinotopic position. It is colored by dominant eye inputs; the binocularly-driven cells tend to be closer to the ocular dominance boundaries, while the monocular cells are toward the centers.
A zoom emphasizes the orientation distribution along a few of the columns near each position; horizontal connections (not shown) effect the interactions between these units. This raises the basic question in this paper: {\em what is the nature of the interaction among groups of cells representing different orientations at nearby positions and innervated by inputs from the left and right eyes?}  The physiology suggests (Fig.~\ref{fig:bands}(right)) the answer lies in the interactions among both monocular and  binocular cells; our model specifies this interaction, starting from the monocular ones.

\subsection{Informal Setup and Overview}

Since much of the paper is technical, we here specify, informally, the main ingredients of the model and the results. We first list several of the key points, then illustrate them directly.  

\begin{itemize}
    \item Stereo geometry enjoys a mathematical structure that is a formal extension of plane curve geometry. 
    In the plane, points belonging to a curve are described by an orientation at a position, and these are naturally represented as elements (orientation, position) of columns.
    In our model, these become abstract fibres. The collection of fibres across position is a fibre bundle. Elements of the (monocular) fibre can be thought of as neurons. 
    \item For stereo, we shall need fibres that are a "product" of the left and right-eye monocular columns. The natural coordinates on the stereo fiber bundle are position, positional disparity and orientations from the left and right eyes respectively, which describe fiber over each position. 
    \item The columnar organization of the stereo system, beyond what is shown in the Figure \ref{fig:bands}, is completely unclear. While visual area MT is suggestive of columns for direction of motion (\cite{maunsell1983functional, deangelis1998cortical}) and perhaps V4 for slant (\cite{hinkle2002three}), there is no direct evidence of which we are aware in V1 for spatial or orientation disparity columns. This is the reason why models can be insightful.
    \item Curvature provides a kind of "glue" to enable transitions from points on fibres to nearby points on nearby fibres. These transitions specify "integral curves" through the stereo fibre bundle.
    \item The integral curve viewpoint provides a direction of information flow (information diffuses through the bundle) thereby suggesting underlying circuits. 
    \item The integral curves formalize {\em association field} models. Their parameters describe the spray of curves that is well in accordance with $3D$ curves as studied in psychophysical experiments in \cite{HF95, HHK97, KHK16}. 
    \item Our formal theory resolves several conjectures in the literature \cite{KGS05, KGSYM05, LZ06}.
    \item Our formal theory provides a new framework for specifying the correspondence problem, by illustrating how good continuation in the 3-D world generalizes good continuation in the 2-D plane. This is the point where consistent binocular-binocular interactions are most important. 
    \item Our formal theory has direct implications for understanding torsional eye movements. It suggests, in particular, that the rotational component is not simply a consequence of development, but that it helps to undo inappropriate orientation disparity changes induce by eye movements. This provides a novel role for Listing's Law, and is treated in a companion paper (in preparation).
\end{itemize}

\begin{figure}[tbh]
     \begin{subfigure}[b]{0.45\textwidth}
         \centering
         \includegraphics[width=\textwidth]{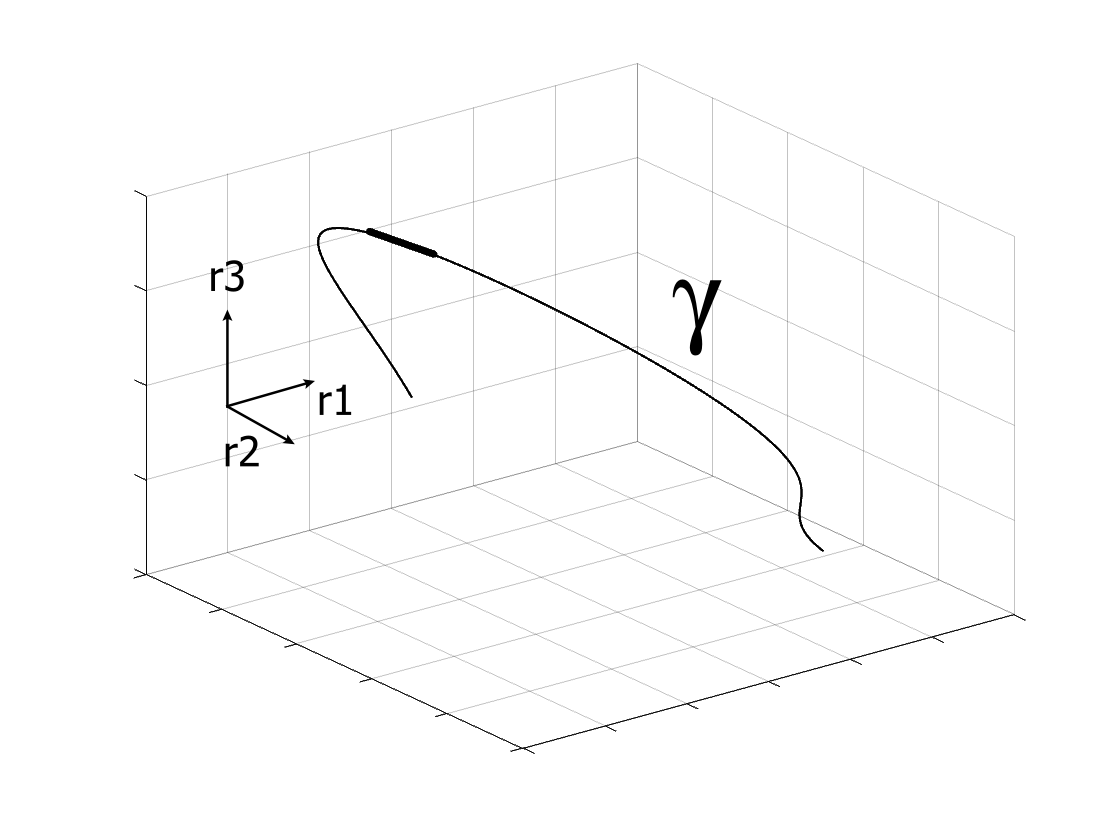}      
     \end{subfigure}
     \vline
     \begin{subfigure}[b]{0.45\textwidth}
         \centering
         \includegraphics[width=\textwidth]{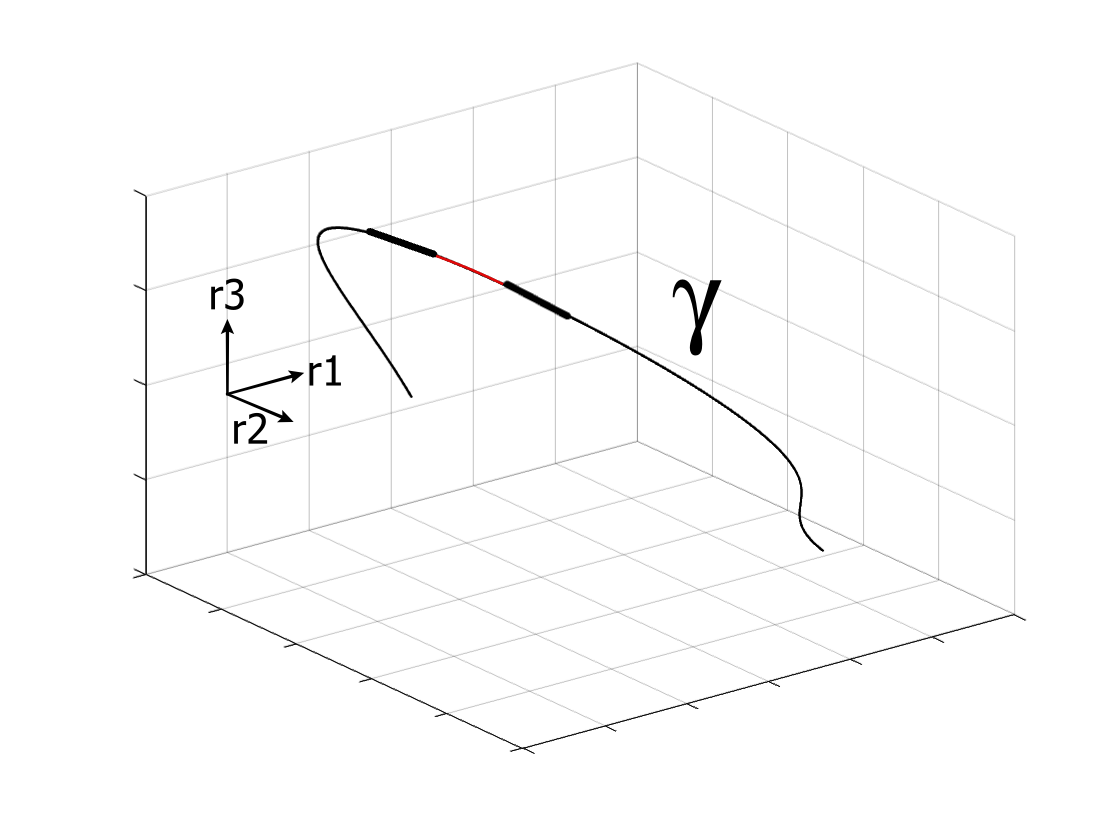}
              \end{subfigure}
              
      \begin{subfigure}[b]{0.45\textwidth}
         \centering
         \includegraphics[width=\textwidth]{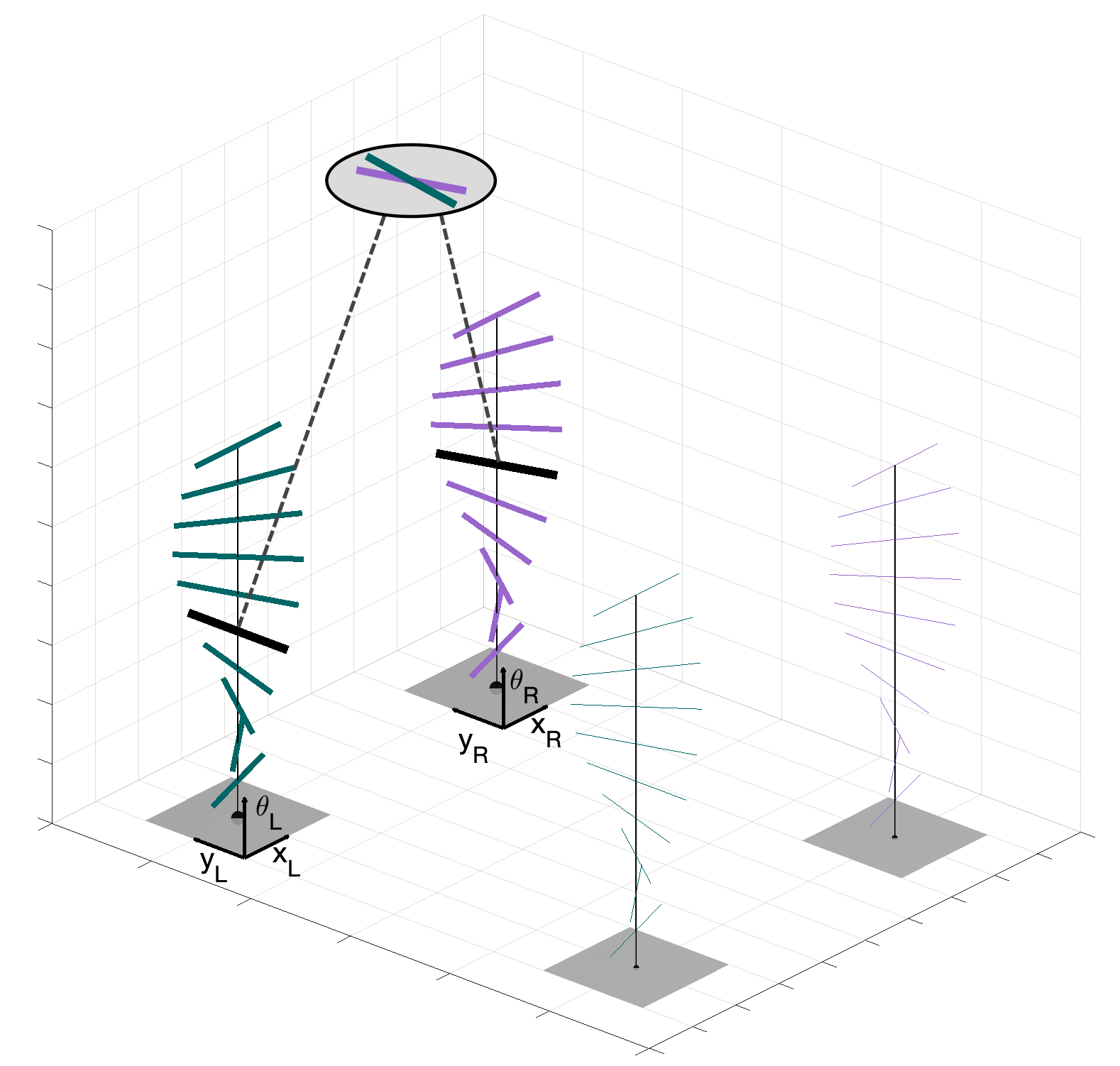}
         \caption{}
      \end{subfigure}
     \vline
          \begin{subfigure}[b]{0.45\textwidth}
         \centering
         \includegraphics[width=\textwidth]{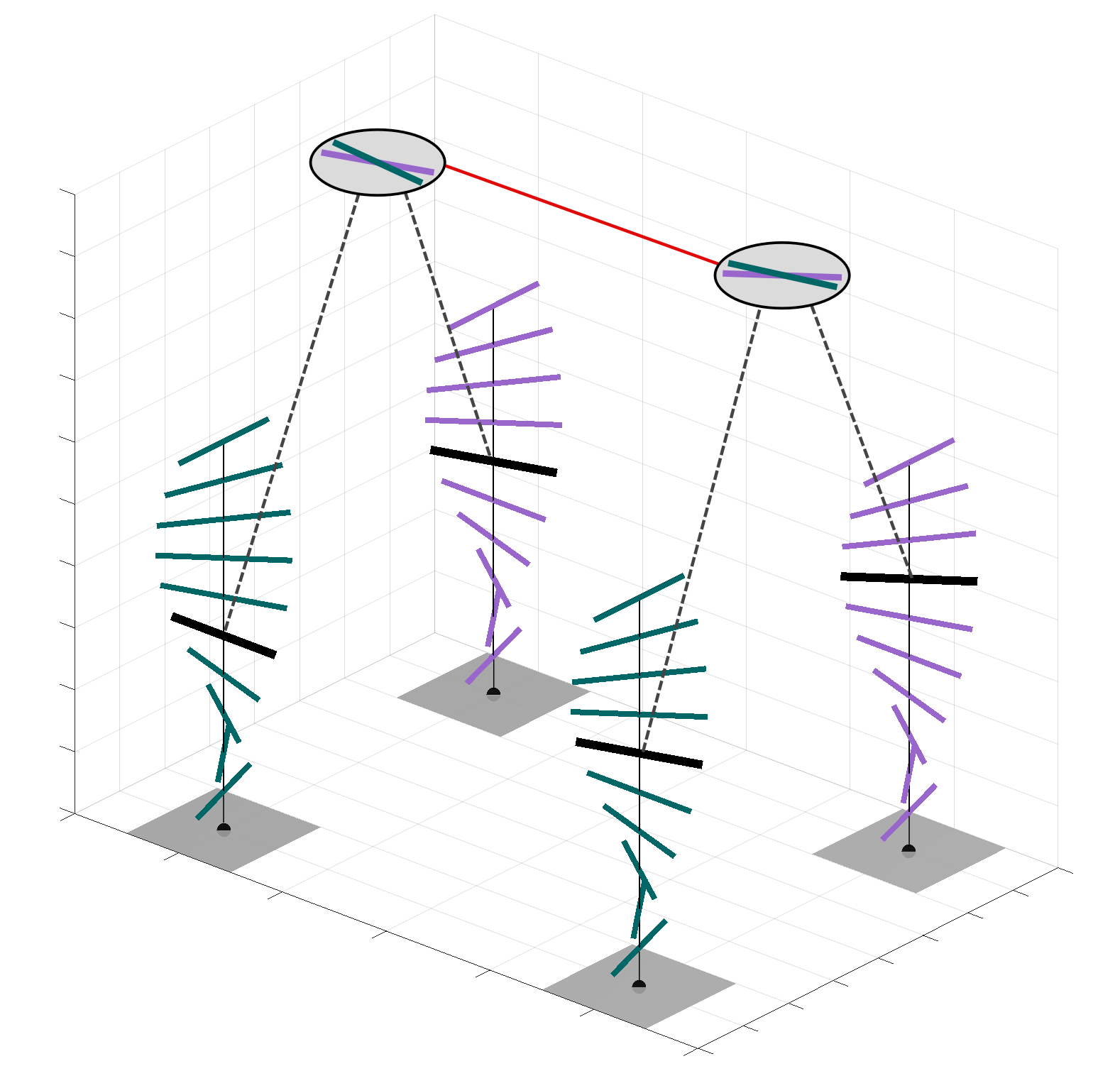} 
         \caption{}
              \end{subfigure}
      \caption{(A) Stereo projection of the highlighted tangent vector to the stimulus $\gamma \in \R^3$ in the left-eye innervated and right-eye innervated monocular orientation columns. (Each short line denotes a neuron by its orientation preference.) Joint activity across the eyes, which denotes the space tangent, is illustrated by the binocular neuron (circle). Note the two similar but distinct monocular orientations. Connections from the actively stimulated monocular neurons to the binocular neuron are shown as dashed lines.  (B) Stereo projection of a consecutive pair of tangents to the stimulus $\gamma \in \R^3$ in the left and right retinal columns. Each space tangent projects to a different pair of monocular columns because of the spatial disparity. Consistency in the responses of these four columns corresponds to consistency between the space tangents attached nearby positions along $\gamma$. This consistency is realized through the binocular neural connection (solid line).}
      \label{fig:gamma}
\end{figure}

We now illustrate these ideas (Fig.~\ref{fig:gamma}). Consider a 
three-dimensional stimulus as a space curve $\gamma : \R \longrightarrow \R^3$, with a unitary tangent at the point of fixation. Since the tangent is the derivative of a curve, the binocular cells naturally encode the unitary tangent direction $\dot \gamma $ to the spatial $3D$ stimulus $\gamma$. This space tangent projects to a tangent orientation in the left eye\footnote{We are here being loose with language. By a tangent orientation in the left eye, we mean the orientation of a left-eye innervated column in V1}, and perhaps the same or a different orientation in the right eye. A nearby space tangent projects to another pair of monocular tangents, illustrated as activity in neighboring columns. Note how connections between the binocular neurons support consistency along the space curve. It is this consistency relationship that we capture with our model of the stereo association field.

\begin{figure}[tbh]
\includegraphics[width=\textwidth]{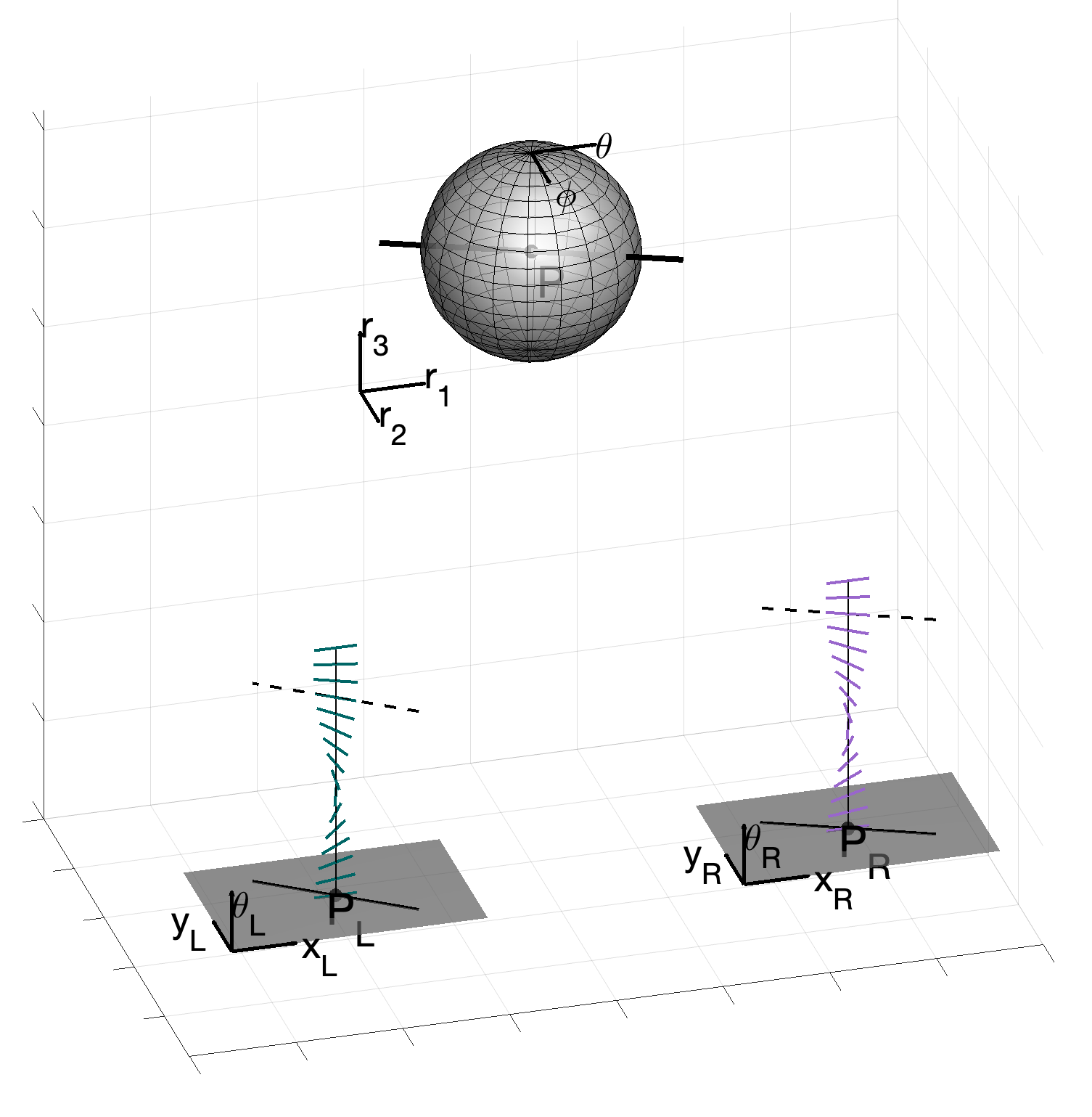}
      \caption{The full geometry of stereo. Note how the stereo correspondence problem allows to establish the relationship between the $3D$ tangent point $(P, \theta, \phi)$ and the projections $p_L$ and $p_R$, the disparity and the orientations $\theta_L$ and $\theta_R$.}
      \label{fig:sphere}
\end{figure}

Since space curves live in 3D, two angles are required to specify its space tangent at a point. In other words, monocular tangent angles span a circle in the plane; space tangent angles span a 2-sphere in 3D.
In terms of the projections into the left-eye and the right-eye, the space tangent can be described by the parameters $n = (\theta, \varphi)$ of $\S^2$ (Fig.~\ref{fig:sphere}).  Thus, we can suitably describe the space of stereo cells -- the full set of space tangents at any position in the $3D$ world -- as the manifold of positions and orientations $\PO$. Moving from one position in space to another, and changing the tangent orientation to the one at the new position, amounts to what is called a {\em group action} on the appropriate manifold. We informally introduce these notions in the next subsection; a more extensive invitation to these ideas is in Appendix \ref{app:intro_SR}.

\subsubsection{Sub-Riemannian Geometry}

We live in a $3D$ world in which distances are familiar; that is, a space of points with a Euclidean distance function defined between any pair of them. 
Apart from practical considerations we can move in any direction we would like. Cars, however, have much more restricted movement capabilities. They can move forward or backward, but not sideways. To move in a different direction, cars must turn their wheels. Here is the basic analogy: in  cortical space information can move to a new retinotopic position in a tangent direction, or it can move up or down a column (orientation fibre) to change direction. Moving in this fashion, from an orientation at a position to another orientation at a nearby position, is clearly more limited than arbitrary movements in Euclidean space. Euclidean geometry, as above, is an example of a Riemannian geometry; the limitations involved in moving through a cortical columnar space specify a sub-Riemannian geometry. Just as cars can move along a roads that are mostly smooth,  excitatory neurons mainly connect to similarly "like" (in orientation) excitatory neurons. This chain of neurons indicates a path through sub-Riemannian space; 
the fan of such paths is the cortical connectivity which can be considered the neural correlate of association fields.
Again, for more information please consult Appendix \ref{app:intro_SR}.

\begin{figure}[tbh]
     \centering
              \includegraphics[width=\textwidth]{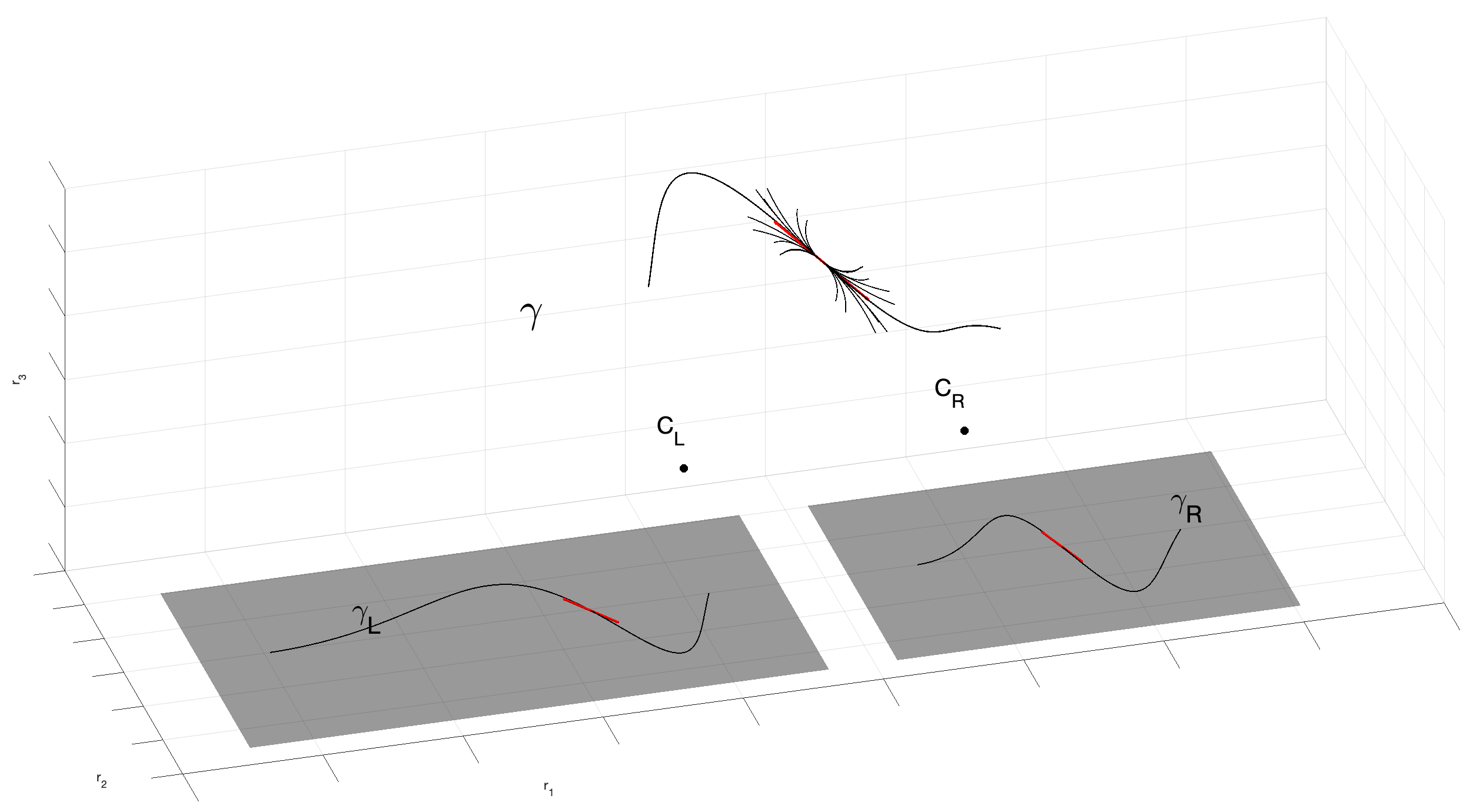}
                       \caption{Main result of the paper. The three-dimensional space curve $\gamma$ is enveloped by the $3D$ the association field centered at a point. Formally, this association field is a fan of integral curves in the sub-Riemmanian geometry computed entirely within the columnar architecture.  (It is specifically described by equation \eqref{integralCurves} with varying $c_1$ and $c_2$ in $\R$, but that will take some work to develop.) }
        \label{fan}
\end{figure}

Moving now out to the world, we must be able to move between all points. Repeating the above metaphor more technically, we equip $\PO$ with a group action of the three-dimensional Euclidean group of rigid motions $SE(3)$. Notice, importantly, that this group is now acting on the product space of positions and orientations.  A bit more is required, though, since the geometry of the stereo vision is not solved only with these punctual and directional arguments. As we showed in Fig.~\ref{fig:gamma} there is the need to take into account the relationships between nearby tangents; in geometric language this involves a suitable type of connections. It is therefore natural to look at integral curves of the sub-Riemannian structure, which encode in their coefficients the fundamental concept of $3D$ curvature and torsion. An example of this is shown in Fig~\ref{fan}. Notice how the $3D$ association field envelopes a space curve, in the same way that a $2D$ association field envelopes a planar curve. This figure illustrates, in a basic way, the fundamental result in this paper.

\subsection{Overview of Paper}

The paper is organized as follows: in \text{Section \ref{sec:preliminaries}}, we describe the geometrical and neuro-mathematical background underlying the problem of stereo vision. In particular, we review the standard stereo triangulation technique to relate the coordinate system of one retina with the other, and put them together in order to reconstruct the three-dimensional space. Then, we briefly review the classical neurogeometry of monocular simple cells selective for orientation and the underlying connections. The generalization of co-circularity for stereo is also introduced.  In \text{Section 
\ref{sec:model_for_stereo}}, starting from binocular receptive profiles, we introduce the neuro-mathematical model for binocular cells. First we present the cortical fiber bundle of binocular cells. It follows the differential interpretation of the binocular profiles in terms of the neurogeometry of the simple cells, and we show how this is well in accordance with the results of the stereo triangulation. Then, we give a mathematical definition of the manifold $\PO$ with the sub-Riemannian structure. Finally, we study the integral curves and the suitable change of variables that allow us to switch our analysis from cortical to external space.  In \text{Section \ref{sec:validation}} we proceed to the validation of our geometry with respect to psychophysical experiments. We combine information about the psychophysics of $3D$ perception and formal conjectures; it is here that we formulate a $3D$ association field analogous to the $2D$ association field. At the end,  we show an example of a lifting of a stimulus and how our integral curves properly connect corresponding points. This illustrates the use of our model as a basis for solving the correspondence problem.

\section{Stereo vision and neuro-mathematical background}
 \label{sec:preliminaries}
 
\subsection{Stereo geometry}
\label{triang}
In this subsection we briefly recall the geometrical configuration underlying $3D$ vision, to define the variables that we use in the rest of the paper, mainly referring to \cite[Ch.\ 6]{F93}. For a complete historical background see \cite{H12, HR95}.

\subsubsection{Stereo variables}

\begin{figure}[tbh]
\begin{center}
\includegraphics[width= 0.6\textwidth]{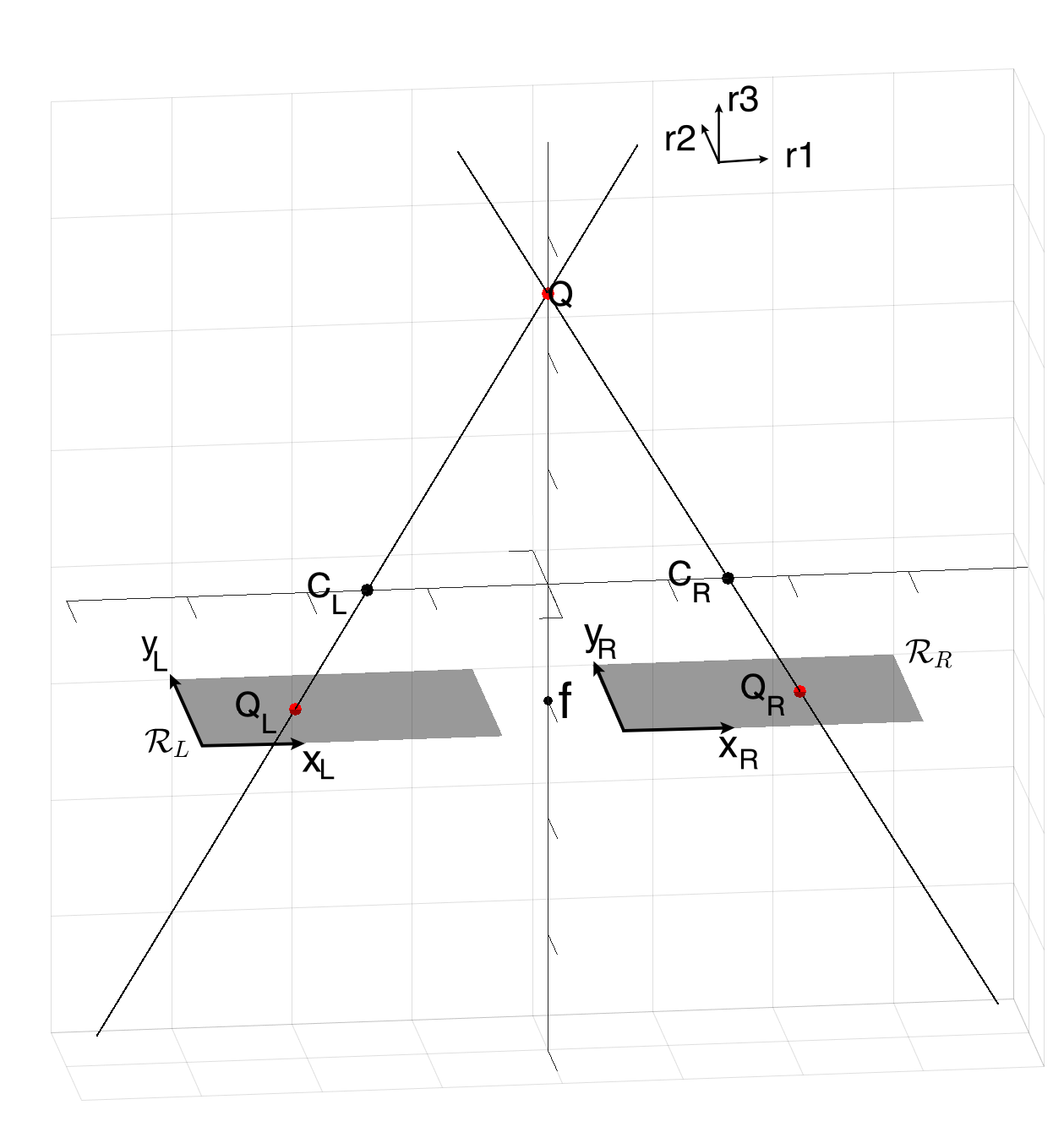}
\end{center}
\caption{Reconstruction of the $3D$ space point $Q$ through points $Q_L$ the retinal plane $\mathcal{R}_L$ and $Q_R$ in , $\mathcal{R}_R$.}
\label{3DreconstructionV}
\end{figure}

We consider the global reference system $(O, i, j, k)$ in $\R^3$,  with $O = (0,0,0)$,  and coordinates $(r_1, r_2, r_3)$. We introduce the optical centers $C_L=(-c, 0,0)$ and $C_R=(c, 0,0)$, with $c$ real positive element, and we define two reference systems: $(C_L, i_L, j_L)$,  $ (C_R, i_R, j_R)$, the reference systems of the retinal planes $\mathcal{R}_L$ and $\mathcal{R}_R$ with coordinates respectively $(x_L, y)$, $(x_R, y)$. In the global system we suppose the retinal planes to be parallel and to have equation $r_3=f$, with $f$ denoting the focal length. This geometrical set-up is shown in Figure \ref{3DreconstructionV}. 

\begin{remark} If we know the coordinate of a point $Q=(r_1, r_2, r_3)^T$ in $\R^3$, then it is easy to project it in the two planes via perspective projection, having $c$ the coordinate of the optical centers and $f$ focal length. This computation defines two projective maps $\Pi_L$ and $\Pi_R$, respectively, for the left and right retinal planes: 
\begin{equation}
\label{projEqs}
\begin{aligned}
\Pi_{L} : & \hspace{0.6cm}\R^3 &\longrightarrow & \hspace{1cm }\R^2 & \hspace{1cm} &\Pi_{R} : & \R^3 &\longrightarrow & \R^2  \\
& \begin{pmatrix} r_1\\ r_2\\ r_3 \end{pmatrix}& \mapsto & \begin{pmatrix}  \frac{f(r_1+c)}{r_3} \\  \frac{fr_2}{r_3}\end{pmatrix},  & \hspace{0.5cm} &  &\begin{pmatrix} r_1\\ r_2\\ r_3 \end{pmatrix}& \mapsto & \begin{pmatrix}  \frac{f(r_1-c)}{r_3} \\  \frac{fr_2}{r_3}\end{pmatrix}.  \\
\end{aligned}
\end{equation}
\end{remark}
\proof 
A point on the left retinal plane of local coordinates $(x_L, y)^T$ has global coordinates $Q_L=(-c + x_L, y, f )^T$, and it corresponds to a point $Q=(r_1, r_2, r_3)^T$ in the Euclidean $\R^3$ such that  $C_L$, $Q_L$ and $Q$ are aligned. This means that the  vectors $Q_L - C_L= (x_L, y, f)^T$ and $Q - C_L=(r_1 + c, r_2, r_3)^T$ are parallel,
obtaining the following relationships:  
\vspace{0.25cm}
\begin{equation}
x_L = f\frac{r_1 + c}{r_3}, \quad y= f \frac{r_2}{r_3}. 
\end{equation}
\vspace{0.25cm}
Analogously, considering $Q_R$ and $C_R$, we get:
\vspace{0.25cm}
\begin{equation}
x_R = f\frac{r_1 - c}{r_3}, \quad y= f \frac{r_2}{r_3}.
\end{equation}
\endproof

In a standard way, the \textit{horizontal disparity} is defined as the differences between retinal coordinates 
\begin{equation}
\label{dispDEF}
d := \frac{x_L-x_R}{2},
\end{equation} up to a  scalar factor. Moreover, it is also possible to define the coordinate $x$ as the average of the two retinal coordinates $x:= \frac{x_L+x_R}{2}$, leading to the following change of variables: 
\begin{equation}
\label{changeVar3DNeur}
\begin{cases}
x = \frac{fr_1}{r_3}\\
 y = \frac{fr_2}{r_3}\\
  d = \frac{fc}{r_3} \\
\end{cases}
\qquad \longleftrightarrow \qquad
\begin{cases}
r_1 =    \frac{x c}{d}\\
r_2 = \frac{y c}{d}\\
r_3 =\frac{ fc}{d}\\
\end{cases}, 
\end{equation}
where the set of coordinates $(x, y, d)$ is known as \textit{cyclopean coordinates} \cite{J71}. 

\subsubsection{Tangent estimation}
\label{tangentPrediction}
Corresponding points in the retinal planes allow to project back into $\R^3$. An analogous reasoning can be done for the tangent structure: if we have tangent vectors of corresponding curves in the retinal planes, it is possible to project back and recover an estimate of the $3D$ tangent vector. Let us recall here this result; a detailed explanation can be found in \cite{F93}.

\begin{remark}Let $\gamma_L $  and $\gamma_R$ be corresponding left and right retinal curves; i.e., perspective projections of a curve $\gamma \in \R^3$ through optical centers $C_L$ and $C_R$ with focal length $f$. Knowing the left and right retinal tangent structures, it is possible to recover the direction of the tangent vector $\dot \gamma$.
\end{remark}

\proof Starting from a curve $\gamma \in \R^3$, we project it in the two retinal planes obtaining $\gamma_L = \Pi_L (\gamma)$  and $\gamma_R = \Pi_R (\gamma)$ from eq. \eqref{projEqs}.
The retinal tangent vectors are obtained through the Jacobian matrix
\footnote{The Jacobian matrix $(J_\Pi)_p$ evaluated at point $p$ represents how to project displacement vectors (in the sense of derivatives or velocities or directions). In details, if $\dot \gamma(t)$ is the displacement vector in $\R^3$, then the matrix product $(J_{\Pi})_{\gamma(t)}\dot \gamma(t)$ is another displacement vector, but in $\R^2$. In other words, the Jacobian matrix is the differential of $\Pi$ at every point where $\Pi$ is differentiable; common notation includes $J_\Pi$ or $D\Pi$.}
of the left and right retinal projections 
$\dot \gamma_{L,R} (t)= (J_{\Pi_{L, R}})_{\gamma(t)}\dot \gamma(t)$: 
\begin{equation}
\label{projDeriv}
\dot \gamma_R(t) =  \begin{pmatrix} \frac{f(\gamma_3\dot\gamma_1+(c-\gamma_1)\dot\gamma_3)}{\gamma_3(t)^2} \\ 
\frac{f(\gamma_3\dot\gamma_2-\gamma_2\dot\gamma_3)}{\gamma_3^2} \end{pmatrix}, 
\dot \gamma_L(t) =\begin{pmatrix} \frac{f(\gamma_3\dot\gamma_1-(c+\gamma_1)\dot\gamma_3)}{\gamma_3(t)^2}\\
 \frac{f(\gamma_3\dot\gamma_2-\gamma_2\dot\gamma_3)}{\gamma_3^2} \end{pmatrix}.
\end{equation}
Extending the tangent vectors and the points into $\R^3$, we get  $\tilde t_{L} =  (\dot \gamma_{{L}1}, \dot \gamma_{{L}2} , 0)^T$, and $ \tilde m_{L} = (\gamma_{{L}1}-c, \gamma_{{L}2}, f)^T$, and 
$U_{t_{L}}= (P_{L})^{-1}\tilde m_{L}\times (P_{L}^{-1})\tilde t_{L}$, with the projection matrix $P_{L}=\begin{pmatrix} 1& 0 & -c/f \\ 0 & 1 & 0 \\ 0& 0 & 1 \end{pmatrix}$. The same reasoning holds for the right structure, with projection matrix $P_{R}=\begin{pmatrix} 1& 0 & c/f \\ 0 & 1 & 0 \\ 0& 0 & 1 \end{pmatrix}$ .

Then $U_{t_R}\times U_{t_L}$ is a vector parallel to the tangent vector $\dot\gamma$:
\begin{equation}
\label{FaugerasTang}
\begin{aligned}
U_{t_R}\times U_{t_L} &= \left( \underbrace{\frac{f^42c(\dot\gamma_2\gamma_3-\dot\gamma_3\gamma_2) }{\gamma_3^4}}_{\lambda(t)}\dot\gamma_1, \frac{f^42c(\dot\gamma_2\gamma_3-\dot\gamma_3\gamma_2)}{\gamma_3^4}\dot\gamma_2,\frac{f^42c(\dot\gamma_2\gamma_3-\dot\gamma_3\gamma_2) }{\gamma_3^4}\dot\gamma_3 \right)^T\\
& = \lambda(t) \left( \dot \gamma_1(t), \dot \gamma_2(t) , \dot \gamma_3(t)\right)^T\\
&= \lambda(t) \dot \gamma(t). \\
\end{aligned}
\end{equation}
\endproof

\subsection{Elements of neuro-mathematics}

We now provide background on the geometric modeling of the monocular system, and good continuation in the plane. Our goal is to illustrate the role of sub-Riemannian geometry in the monocular system, which will serve as the basis for generalization in the stereo system starting from the neuro-mathematical model of Citti and Sarti \cite{CS06}.

\subsubsection{Classical neurogeometry of simple cells}
\label{CSmodel}
We model the activation map of a cortical neuron's receptive field (RF) by its receptive profile (RP) $\varphi$. A classical example is the receptive profiles of simple cells in V1, centered at
position $(x, y)$ and orientation $\theta$, modeled (e.g in \cite{BCS14, D85, JP87})
as a bank of Gabor filters $\varphi_{\{x, y, \theta \}}$, which act on a visual stimulus. 

\begin{figure}[tbh]
     \centering
     \begin{subfigure}[b]{0.27\textwidth}
         \centering
         \includegraphics[width=\textwidth]{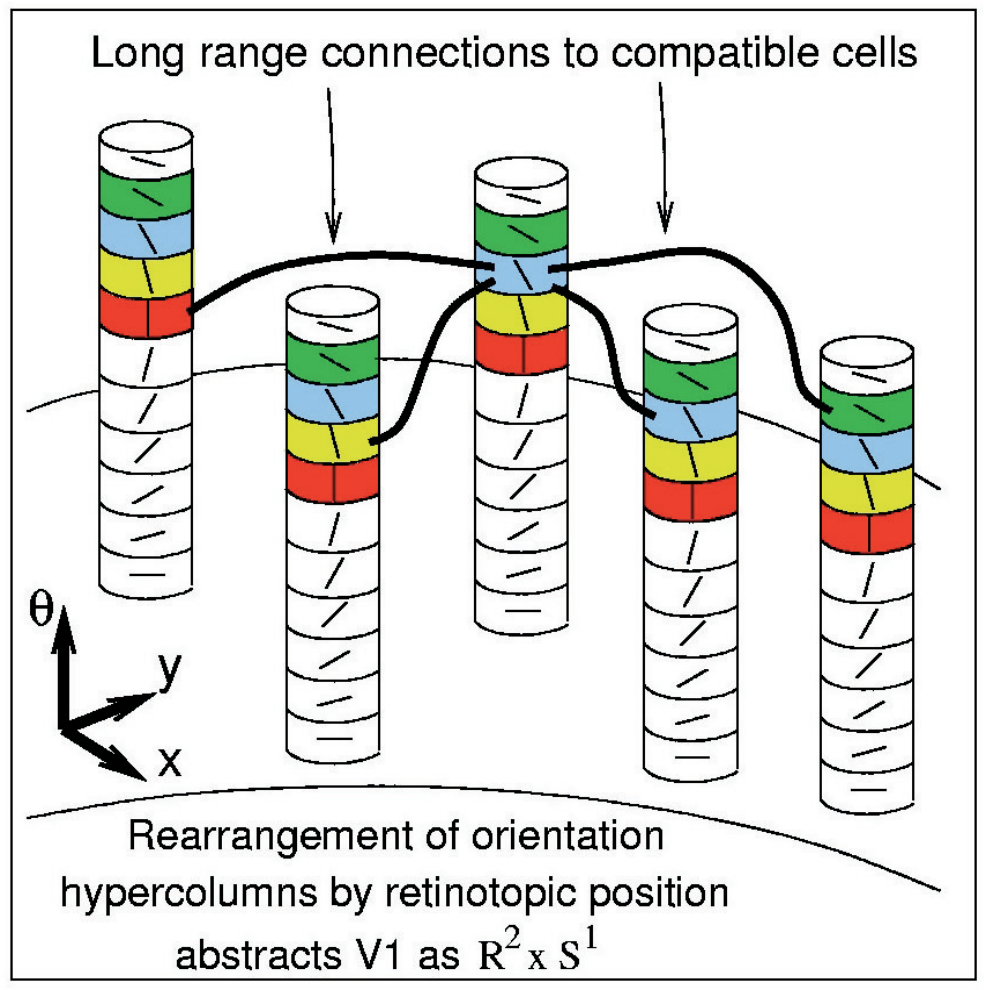}
         \caption{}
     \end{subfigure}
     \begin{subfigure}[b]{0.4\textwidth}
         \centering
         \includegraphics[width=\textwidth]{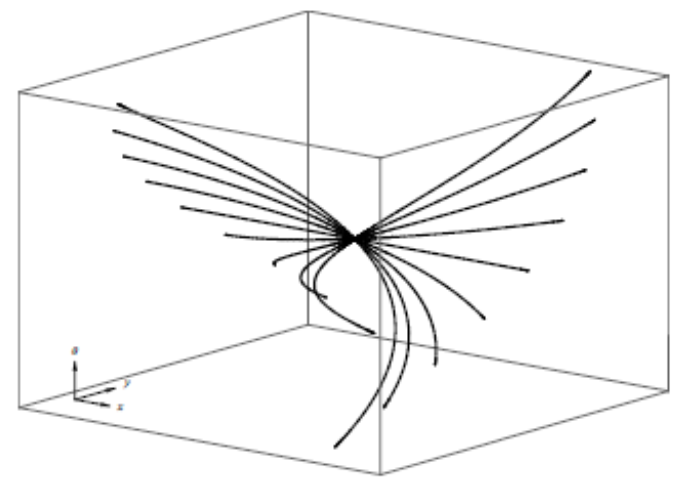}
         \caption{}
     \end{subfigure}
     \begin{subfigure}[b]{0.29\textwidth}
         \centering
         \includegraphics[width=\textwidth]{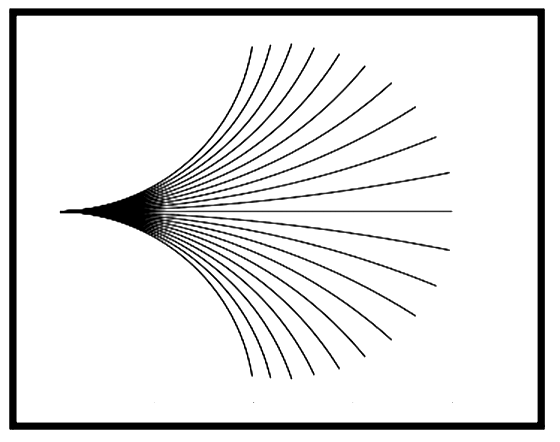}
         \caption{}
     \end{subfigure}

             \caption{ (A) Orientation columns of cells in $(x, y, \theta)$ coordinates. Long-range horizontal connections between cells relate an
orientation signal  at position $(x, y, \theta)$ to another orientation at $(x', y', \theta')$. Figure adapted from \cite{BZ04}. (B) Horizontal integral curves in $\R^2\times \S^1$ generated by the sub-Riemannian model  \cite{CS06}. (C) Projection of the fan of the integral curves in the $(x, y)$ plane. Figure adapted from \cite{CS06}.}
        \label{simpleCells}
\end{figure}

Formally, it is possible to abstract the primary visual cortex as $\R^2\times \S^1$, or position-orientation space, thereby naturally encoding the Hubel/Wiesel hypercolumnar structure \cite{HW62}. An example of this structure is displayed in image (A) of Figure \ref{simpleCells} from \cite{BZ04}. 

\begin{figure}[tbh]
     \centering
     \begin{subfigure}[b]{0.45\textwidth}
         \centering
         \includegraphics[width=\textwidth, height=4cm]{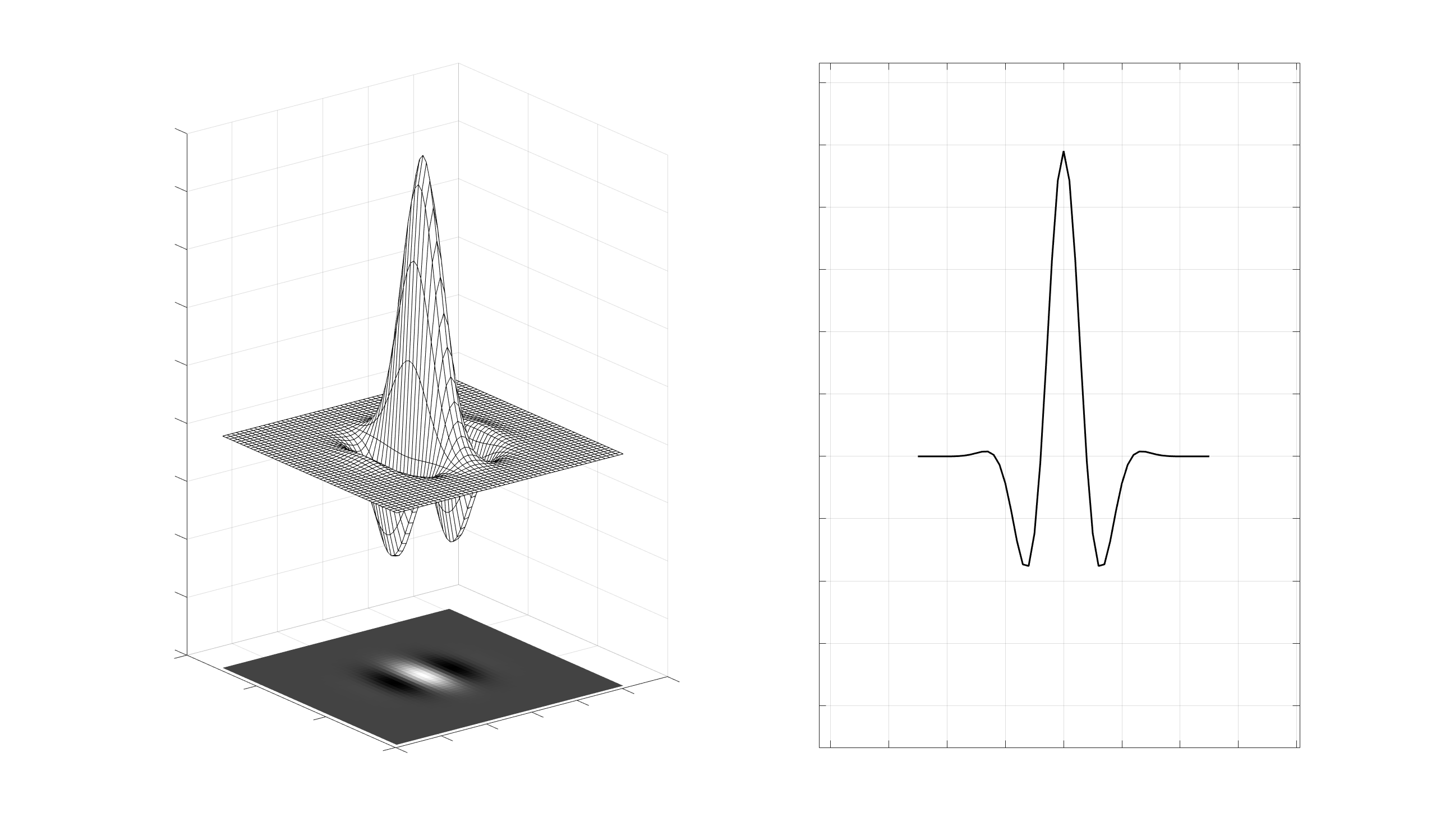}
         \caption{}
     \end{subfigure}
     \begin{subfigure}[b]{0.45\textwidth}
         \centering
         \includegraphics[width=\textwidth, height=4cm]{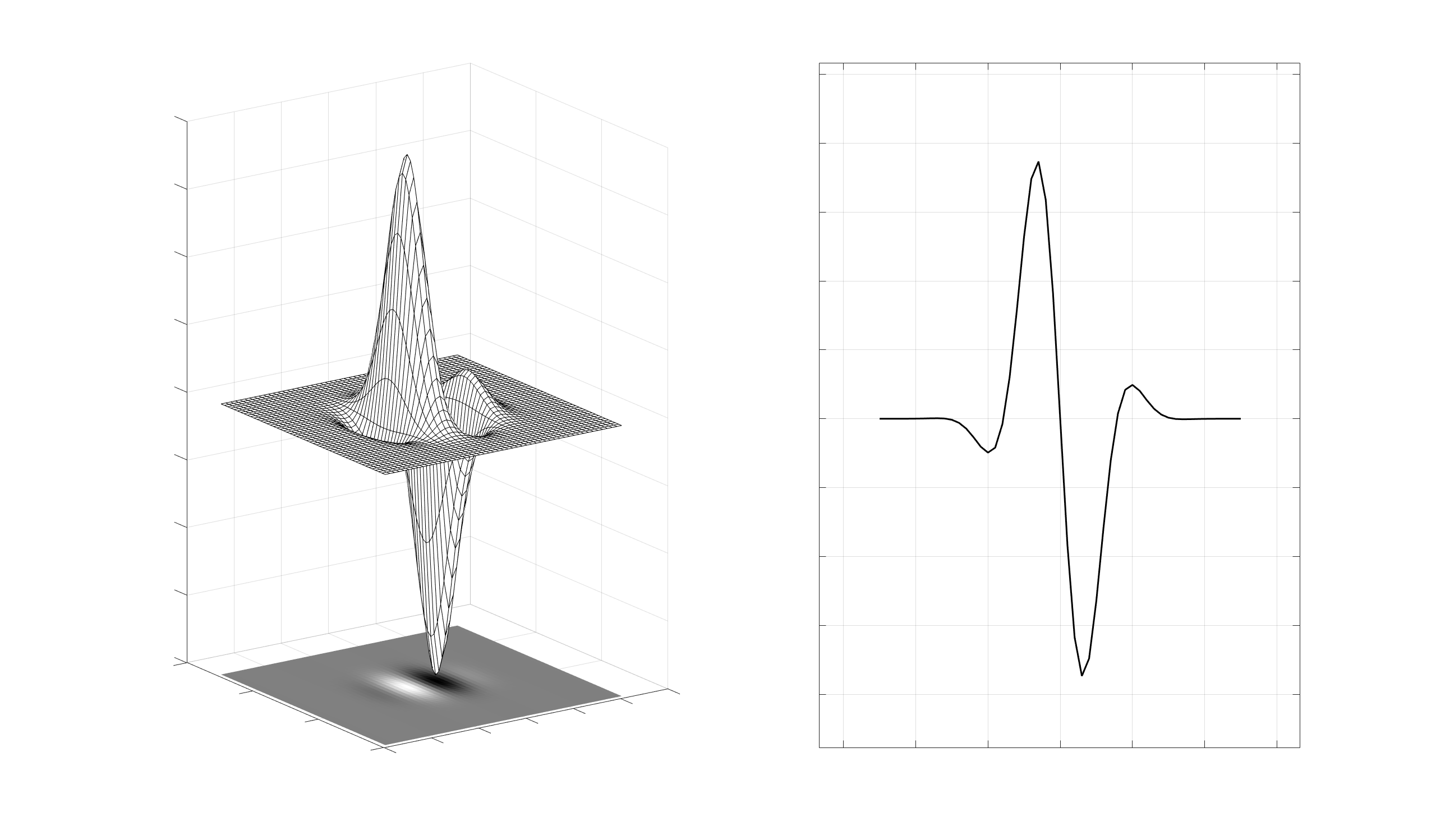}
         \caption{}
     \end{subfigure}
             \caption{Even (A) and odd (B) part of Gabor function: the surface of the two-dimensional filters, their common bi-dimensional representation and a mono-dimensional section. }
        \label{gabor}
\end{figure}

Following the neuro-mathematical model of Citti and Sarti \cite{CS06}, the set of simple cells RPs can
be obtained via translations of vector $(x, y)^T$ and
rotation of angle $\theta$ from a unique "mother" profile $\varphi_{0}(\xi, \eta)$
\begin{equation}
\varphi_0(\xi, \eta) = \exp \left( \frac{2\pi i \xi}{\lambda}\right) \exp \left( -\frac{\xi^2 + \eta^2}{2\sigma^2}\right),
\end{equation}
a Gabor function with real (even) and imaginary (odd) parts (Figure \ref{gabor}).
Translations and rotations can be expressed as: 
\begin{equation}
T _{(x, y, \theta)}(\xi, \eta) = \begin{pmatrix}\cos\theta & -\sin\theta \\ \sin \theta & \cos  \theta \\ \end{pmatrix} \begin{pmatrix} \xi \\ \eta \end{pmatrix} + \begin{pmatrix} x \\ y \end{pmatrix},
\end{equation} where $T_{(x, y, \theta)}$ denotes the action of the group of rotations and
translations $SE(2)$ on $\R^2$. This group operation associates to every point $(\xi, \eta)$ a new point $(\tilde x, \tilde y)$, according to the law $(\tilde x, \tilde y) = T_{(x, y, \theta)}(\xi, \eta)$.
Hence a general RP can be expressed as 
\begin{equation}
\varphi_{(x, y, \theta)}(\xi, \eta) = \varphi_0 (T^{-1}_{(x, y, \theta)}(\xi, \eta)),
\end{equation}
and this represents the action of the group $SE(2)$ on the set of receptive profiles. 

The retinal plane $\mathcal{R}$ is identified with the $\R^2$ plane, whose coordinates are $(x, y)$. When a visual stimulus $I: \mathcal{R} \longrightarrow \R^+$ of intensity $I(x, y)$ activates the retinal layer, the neurons centered at every point $(x, y)$ produce an output $O(x, y, \theta)$, which can be modeled as the integral of the signal $I$ with the set of Gabor filters:
\begin{equation}
\label{RP}
O(x, y, \theta)=\int_{\mathcal{R}}\varphi_{\{x, y, \theta \}}(\xi,\eta)I(\xi,\eta)d\xi d\eta,
\end{equation} where the function $I$ represents the retinal image.

For $(x, y)$ fixed, we will denote $\bar \theta$ the point of maximal response:
\begin{equation}
\max_{\theta}\abs{O(x, y, \theta)} = \abs{O(x, y, \bar \theta)}.
\end{equation}
We will then say that the point $(x, y)$ is lifted to the point $(x, y, \bar \theta)$. This is extremely important conceptually to understand our geometry: it illustrates how an image point, evaluated against an simple cell RP, is lifted to a "cortical" point by introducing the orientation explicitly. If all the points of the image are lifted in the same way, the level lines of the
$2D$ image $I$ are lifted to new curves in the $3D$ cortical space $(x, y, \bar \theta)$.

We shall now introduce a set of directions for moving on the cortical space $(x, y, \bar \theta)$, in the sense of vector fields. This is important because it will be necessary to move within this space, across both positions and orientations. Biologically, such movements would be the flow of information from one cell in a column to another cell in a nearby column.

To begin, in the right hand side of the equation \eqref{RP} the integral of
the signal with the real and imaginary part of the Gabor filter is expressed. The two families of cells have
different shapes, hence they detect (or play a role in detecting) different features. 
Since the odd-symmetry cells suggest boundary detection, we concentrate on them, but this is mainly a convenience for computation. The output of a simple cell can then be locally approximated as
$ O(x, y, \theta) = - X_{3,p}(I_\sigma)(x,y)$, where $p = (x, y, \theta) \in SE(2)$, $I_\sigma$ is a smoothed version of $I$, obtained by convolving it with a Gaussian kernel, and 
\begin{equation}
\label{X3vf}
X_{3, p} = -\sin\theta\partial_x + \cos\theta\partial_y, 
 \end{equation}
 is the directional derivative in the direction $\vec{X}_{3, p} = (-\sin\theta, \cos\theta, 0)^T$. From now on, we will denote (by a slight abuse of notation) $\omega^\star:= \vec{X}_{3,p}$ to remind the reader familiar with the language of $1$-forms the correspondence of these quantities, and the relation with the Hodge star operator.
 \footnote{The purpose of introducing this notation is also to motivate an implication of the mathematical model in \cite{CS06}; see Appendix \ref{app:proofProp_sec} for explanation.}

Now, think of vector fields as defining a coordinate system at each point in cortical space. Then, in addition to above, the vector fields orthogonal to $X_{3, p}$ are:
\begin{equation}
X_{1, p} = \cos\theta\partial_x + \sin\theta\partial_y,\quad X_{2, p} = \partial_\theta
\end{equation}
and they define a 2-dimensional admissible tangent bundle\footnote{as defined in Appendix \ref{srDef_appendix}} to $\R^2\times \S^1$. 
One can define a scalar product on this space by imposing the orthonormality of $X_{1, p}$ and $X_{2, p}$: this determines a sub-Riemannian structure on $\R^2\times \S^1$. 

\begin{figure}[tbh]
     \centering
     \begin{subfigure}[b]{0.25\textwidth}
         \centering
         \includegraphics[width=\textwidth]{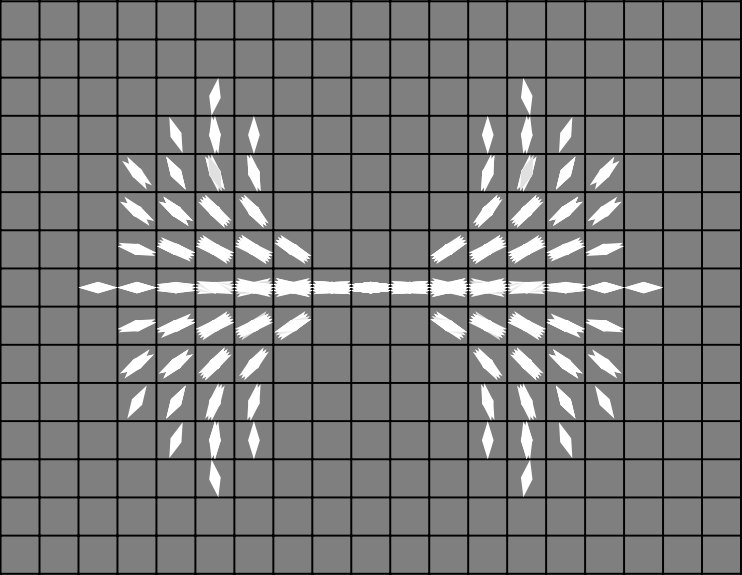}
         \caption{}
     \end{subfigure}
     \hfill
     \begin{subfigure}[b]{0.29\textwidth}
         \centering
         \includegraphics[width=\textwidth]{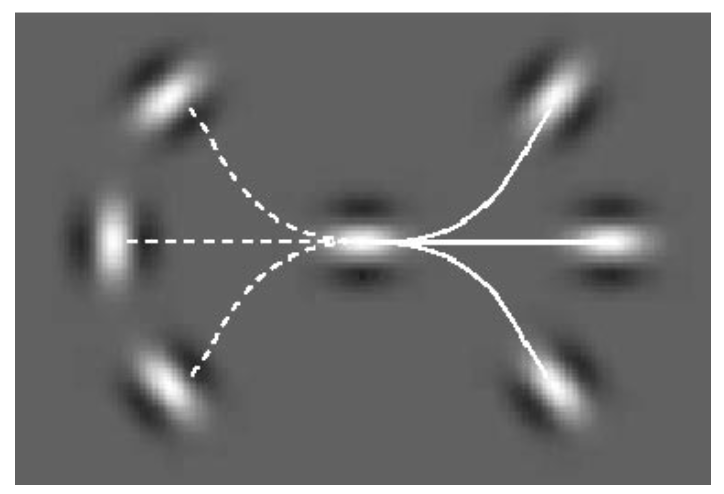}
         \caption{}
     \end{subfigure}
     \hfill
     \begin{subfigure}[b]{0.25\textwidth}
         \centering
         \includegraphics[width=\textwidth]{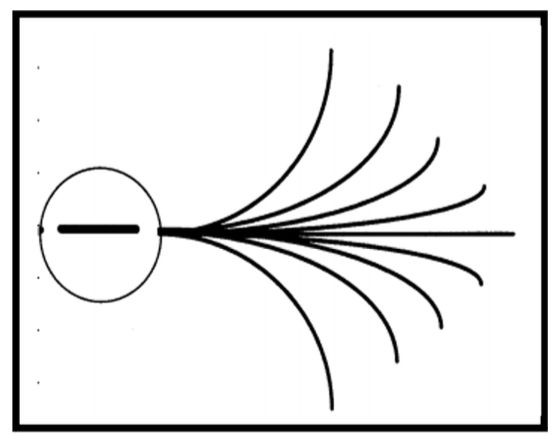}
         \caption{}
     \end{subfigure}
     \hfill
     
             \caption{ (A) Examples of the compatibilities around the central point of the image, derived from planar co-circularity. Brightness encodes compatibility values. Figure adapted from \cite{BZ04}. (B) Starting from the central initial oriented point, the solid line indicates a configuration between the patches where the association exists while the dashed line indicates a configuration where it does not. Figure adapted from \cite{FHH93}. (C) Association field of Field, Hayes and Hess. Figure adapted from \cite{FHH93}.}
        \label{assField2D}
\end{figure}

The visual signal propagates, in an anisotropic way, along cortical connectivity and connects more strongly cells with comparable orientations. 
This propagation establishes the connection between the geometry just developed and 2-dimensional contour integration. This is a formalization of the Gestalt law of good continuation \cite{ K63,K67}. It first arose in a simpler form, namely
co-circularity in the plane \cite{PZ89}, to describe the consistency and the compatibility of neighboring oriented points, in accordance with specific values of curvature. An example of these compatibilities can be found in Figure \ref{assField2D}, image (A). It is complemented by psychophysical experiments, e.g. \cite{IBR89,SV87,U83}. In particular, Field et al. in \cite{FHH93} describe the association rules  for 2-dimensional contour integration, introducing the concept of \textit{association fields}. A representation of these connections can be found in Figure \ref{assField2D}, images (B) and (C). Note that this is equivalent to the union (over curvature) in \cite{PZ89}.
Neurophysiological studies \cite{B92, BZSF97, HMD14, MAHG93, SGLS97} suggest that the cortical correlate of the association field is the long-range horizontal connectivity among cells of similar (but not necessarily identical)
orientation preference. 

Based on these findings, Citti and Sarti in \cite{CS06} modeled cortical propagation as propagation along integral curves of the vector fields $X_{1}$ and $X_{2}$, namely curves $\gamma: [0, T]\subset \R \longrightarrow \R^2\times\S^1$ described by the following differential equation:
\begin{equation}
\label{integralCurves2D}
\dot \gamma(t) = \vec{X}_{1, \gamma(t)}+k \vec{X}_{2, \gamma(t)}, \hspace{0.4cm} t\in [0,  T],
\end{equation}
obtained by varying the parameter $k \in \R$. ($k$ acts analogously as curvature.) An example of these curves is in  Figure \ref{simpleCells}(B). 
Their $2D$ projection is a close approximation of the association fields (Figure \ref{simpleCells}(B)). 

A related model has been proposed by Duits et al. \cite{DBRS13}. They study the geodesics of the sub-Riemannian structure to take into account all appropriate end-conditions of association fields. 

\subsubsection{Generalizing co-circularity for stereo}

The concept of co-circularity in $\R^2$ has been developed by observing that a bidimensional curve $\gamma$ can be locally approximated at $0$ via the
osculating circle. Zucker et al. in \cite{AZ00, LZ03, LZ06} generalize this concept with the Frenet differential geometry of a three dimensional curve. 

While in the two-dimensional case the approximation of the curve using the Frenet $2D$ basis causes the curvature to appear in the coefficient of the Taylor series development ($1st$ order), in the three-dimensional case the coefficients involve both the curvature and torsion. So, in \cite{AZ00} the authors propose heuristically to generalize the osculating circle for space curves with an osculating helix, with a preference for $r_3$-helices to improve stability in terms of camera calibration. In this way the orientation disparity is encoded in the behavior of the helix in the $3D$ space: there is no difference in orientation in the retinal planes if the helix is confined to be in the fronto-parallel plane (the helix becomes a circle), otherwise moving along the $3D$ curves the retinal projections have different orientations. 

\begin{figure}[tbh]
 \centering
 \begin{subfigure}[b]{0.45\textwidth}
         \centering
         \includegraphics[width=\textwidth]{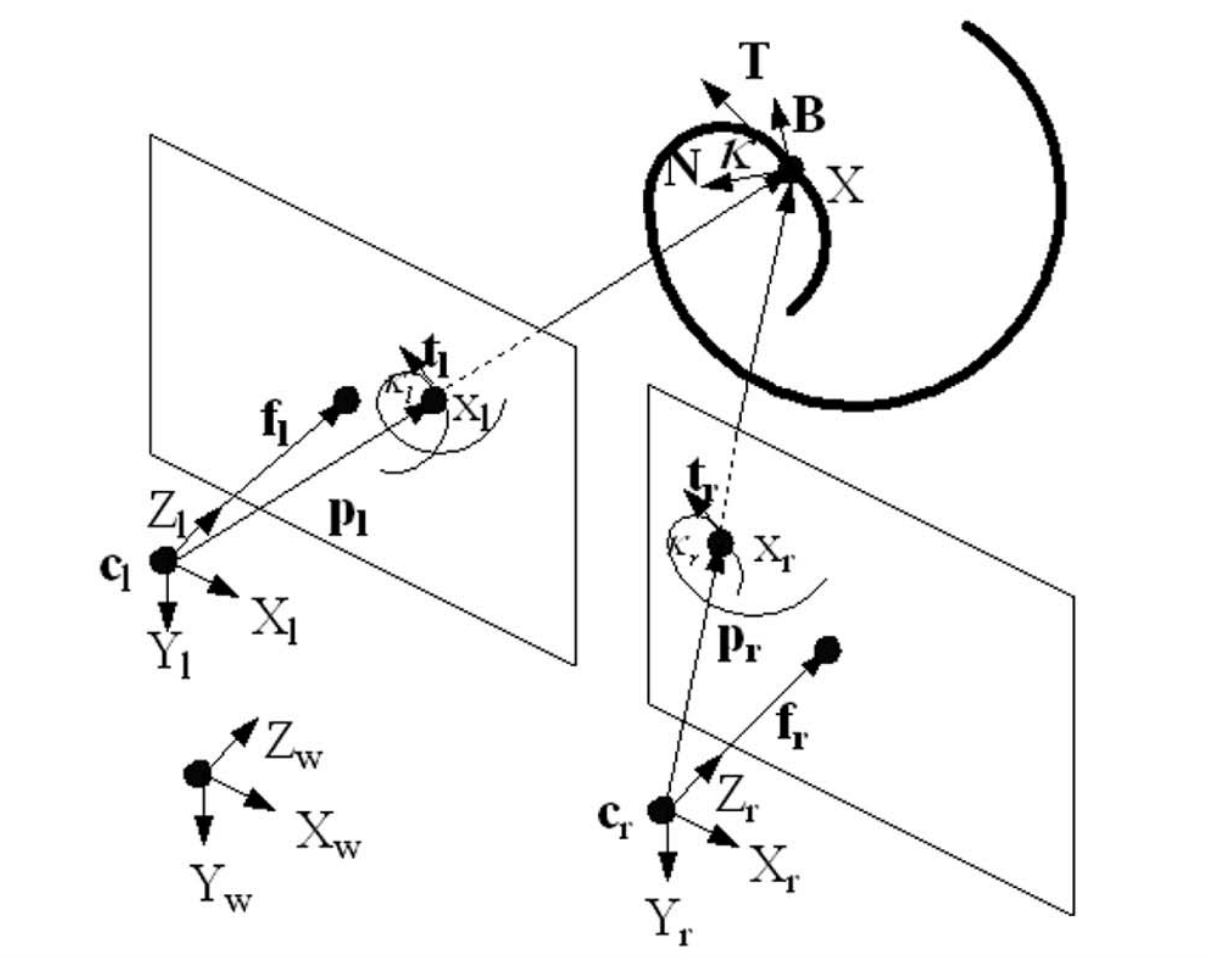}
         \caption{}
     \end{subfigure}
     \begin{subfigure}[b]{0.45\textwidth}
         \centering
         \includegraphics[width=\textwidth]{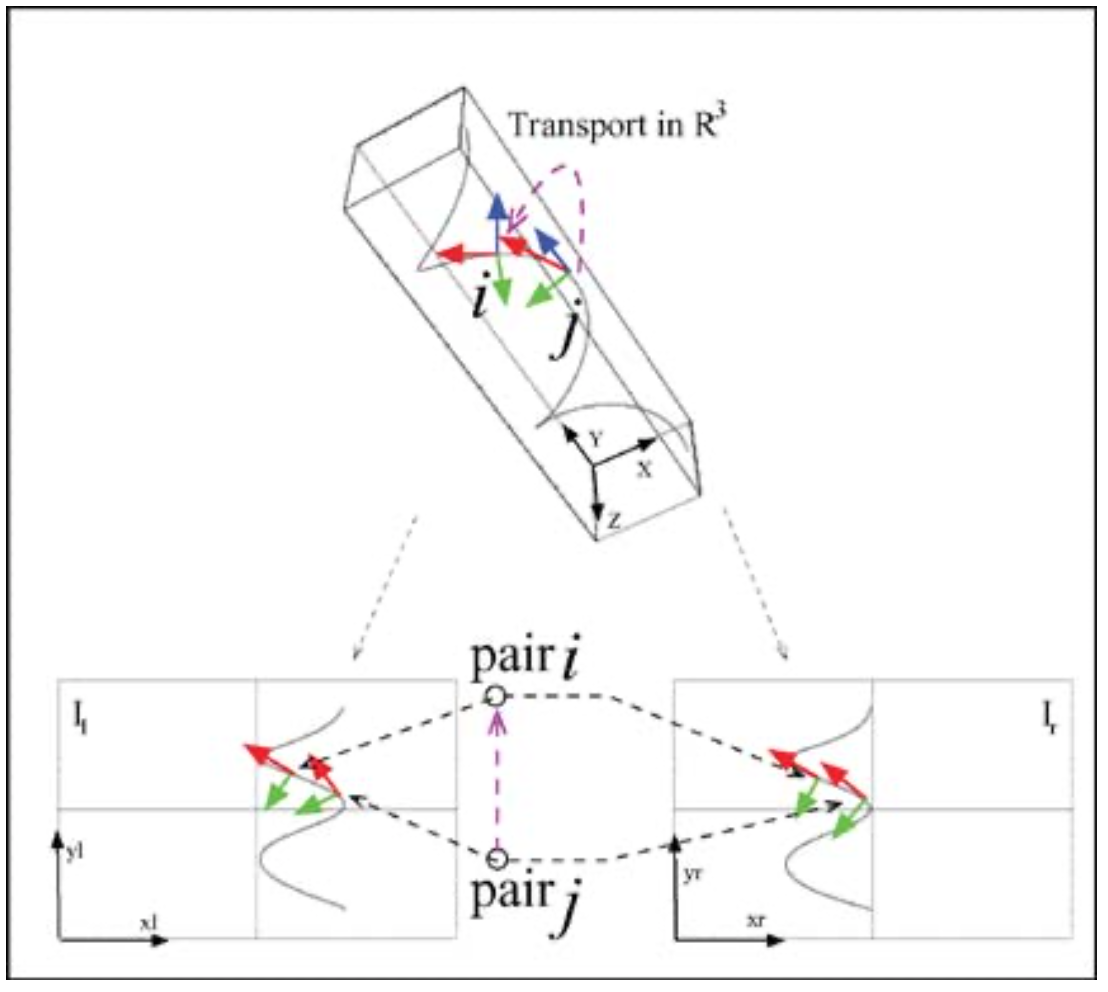}
         \caption{}
     \end{subfigure}
                 \caption{(A) Geometrical setup of Proposition \ref{geom3D}. The spiral curve $3D$ projects in the left and right retinal planes together with the Frenet structure. (B) Stereo correspondence between
pairs of (left-right) pairs of tangents. Both figures are taken from \cite{LZ06}.}
        \label{transportR3}
\end{figure}

In \cite{LZ03, LZ06} they observe that, by introducing the curvature variable as a feature in the two monocular structures, and assuming correspondence, it is possible to reconstruct the $3D$ Frenet geometry of the curve, starting from the two-dimensional Frenet geometry, up to the torsion parameter. In particular, they prove:

\begin{proposition}
\label{geom3D}
Given two perspective views of a $3D$ space curve with full calibration, the normal $N$ and curvature $k$ at a curve space point are uniquely determined from the positions, tangents, and curvatures of its projections in two images. Thus the Frenet frame $\{T, N, B\}$ and curvature $k$ at the space point can be uniquely determined. 
\end{proposition}

Hence, using the knowledge of the Frenet basis together with the fundamental addition of the curvature variable, Zucker et al. introduced the concept of \textit{transport}. This allowed moving the $3D$ Frenet frame in a consistent way with the corresponding $2D$ Frenet structures of the left and right retinal planes, to establish stereo correspondence between
pairs of (left and right) pairs of tangents. See Figure \ref{transportR3} image (B).

\begin{remark} The model that we propose in this paper is related to, but differs from, what has just been stated. In particular, to remain directly compatible with the previous neuro-geometric model, we will work only with the monocular variables of position and orientation. Rather than using curvature directly, we shall assume that these variables are encoded within the connections; mathematically they appear as parameters. A theoretical result of our model is that the heuristic assumption regarding the $r_3$-helix can now be established rigorously. 
\end{remark}

Let us also mention the paper \cite{ASFCSHR17}, where the curvature was considered as independent variable and helices have been obtained in the $2D$ space. 

\section{The neuromathematical model for stereo vision}
\label{sec:model_for_stereo}
\subsection{Binocular profiles}

\begin{figure}[tbh]
\centering
         \includegraphics[width=0.7\textwidth]{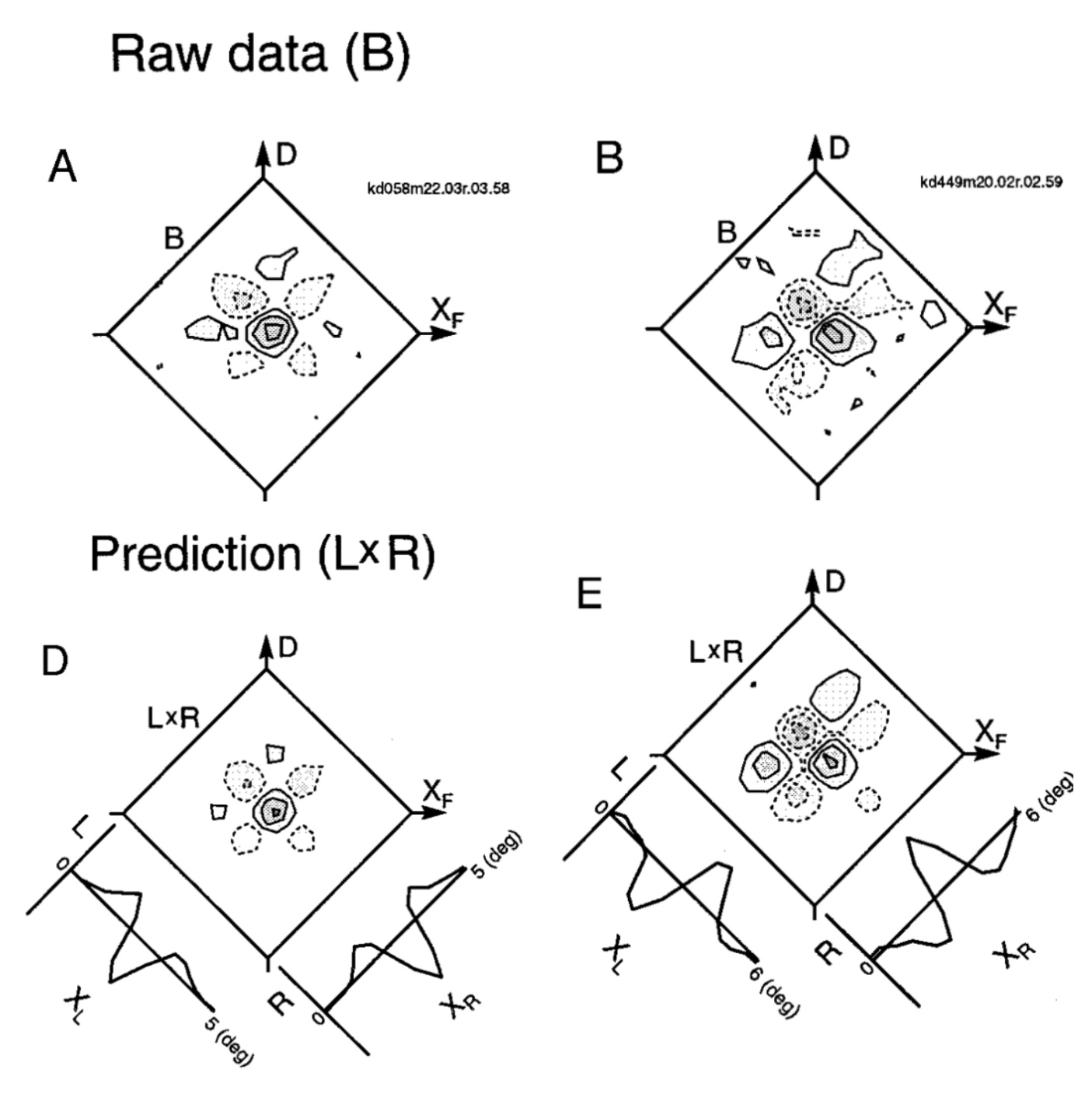}
\caption{ Comparisons between binocular interaction RPs and the product of left and right eye RPs, where left and right RPs are shown in Figure \ref{gabor}. Binocular interaction RPs (Raw data) of a cell is shown on the left. Contour plots for the product of left and right eye RPs (L$\times$R) are shown in the right along with 1-dimensional profiles of the left (L) and right (R) eye RPs.
Figure adapted from \cite{AOF99S}. }
\label{binProfiles}
\end{figure}

Binocular neurons receive inputs from both the left and right eyes. To facilitate calculations, we assume these inputs are first combined in simple cells in the primary visual cortex, a widely studied approach (\cite{AOF99S, CD01, KBK16, MF04ap}). It provides a first approximation in which binocular RPs are described as the product of monocular RPs; see Figure~\ref{binProfiles}, image (A). 
This is of course a simplification -- see \cite{sasaki2010complex}, for instance -- 
but it is compatible with existing neural findings.

This binocular model allows us to define disparity and frontoparallel coordinates as
\begin{equation}
 \label{changeVariables}
 \begin{cases}
  d = \frac{x_L-x_R}{2}\\
 x = \frac{x_R+x_L}{2},\\
 \end{cases}
 \end{equation} perfectly in accordance with the introduction of cyclopean coordinates in \eqref{dispDEF}. 
In this way $(x, y, d)$ correspond to the neural correlate of $(r_1, r_2, r_3)$, via the change of variables \eqref{changeVar3DNeur}. 

\subsection{The cortical fiber bundle of binocular cells}

The hypercolumnar structure of monocular simple cells (orientation selective) has been described as a jet fiber bundle in the works of Petitot and Tondut \cite{ PT99}, among many others. We concentrate on the fiber bundle $ \R^2 \times \S^1 $, with fiber $\S^1$; see e.g. \cite{BZ04} among many others.

In our setting, the binocular structure is based on monocular ones; recall the example illustrations from the Introduction. In particular, for each cell on the left eye there is an entire fiber of cells on the right, and vice versa, for each cell on the right there is an entire fiber of cells on the left. This implies that the binocular space is equipped with a symmetry that involves the left and right structures, allowing us to use the cyclopean coordinates $(x, y, d)$ defined in \eqref{changeVariables}.

Hence, we define the cyclopean retina $\mathcal{R}$, identified with $\R^2$, endowed with coordinates $(x,y)$. The structure of the fiber is $\mathcal{F} = \R \times \S^1 \times \S^1$, with coordinates $(d, \theta_L, \theta_R) \in \mathcal{F}$. 
The total space is defined in a trivial way, $\mathcal{E} = \mathcal{R}\times \mathcal{F} = \R^2 \times \R \times \S^1\times\S^1$, and the projection $\pi : \mathcal{E} \longrightarrow \mathcal{R}$ is the trivial projection $\pi(x, y, d, \theta_L, \theta_R) = (x, y)$.  The preimage of the projection $\mathcal{E}_{(x, y)} := \pi^{-1}(\{(x, y)\})$, for every $(x, y) \in \mathcal{R}$, is isomorphic to the fiber $\mathcal{F}$, and the local trivialization property is naturally satisfied.
 
\begin{figure}[tbh]
\begin{center}
\includegraphics[width = \textwidth]{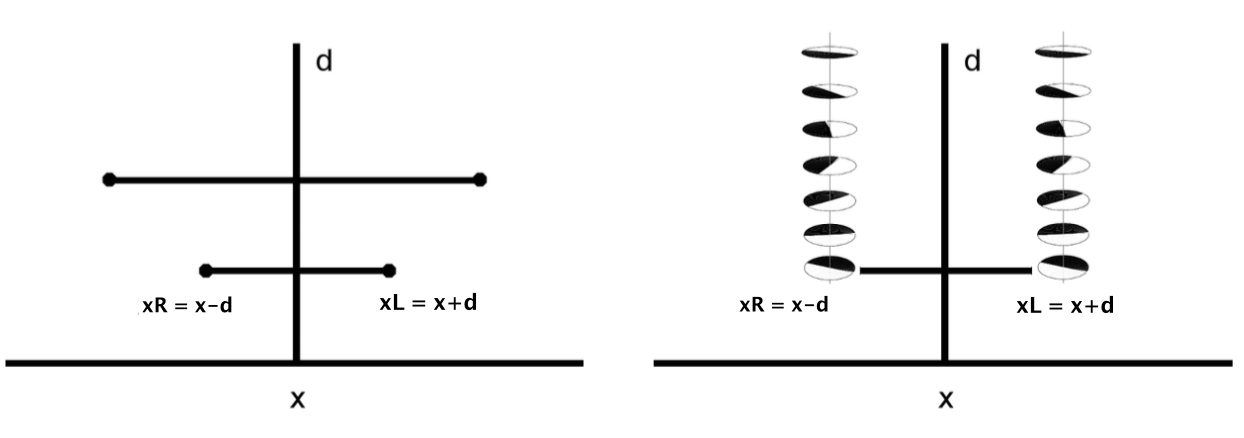}
\end{center}
\caption{Left: schematic representation of the fiber bundle in two dimension, with relationships between left and right retinal coordinates. Right: representation of the selection of a whole fiber of left and right simple cells, for every $x$ and for every $d$. 
}
\label{fiberBundleStr}
\end{figure}

A schematic representation can be found in Figure \ref{fiberBundleStr}. The base has been depicted as 1-dimensional, considering the restriction $\mathcal{R}_{|x}$ of the cyclopean retina $\mathcal{R}$ on the coordinate $x$. The left image displays only the disparity component of the fiber $\mathcal{F}$, encoding the relationships between left and right retinal coordinates. The right image shows the presence of the left and right monodimensional orientational fibers. 


\subsection{Binocular energy model}

To simplify calculations, as stated in the Introduction, we follow the classical binocular energy model \cite{AOF99S} for binocular RPs.  The basic idea is a binocular neuron  receives input from each eye; if the sum $O_L + O_R$ of the inputs from the left and right eye is positive, the firing rate of the binocular neuron is proportional to the square of the sum, and it vanishes, if the sum of the inputs is negative: 
\begin{equation}
\label{unitEnergyModel}
O_B=(Pos(O_L+O_R))^2,
\end{equation}with $Pos(x)=\max\{x,0\}$, $O_B$ the binocular output.

If $O_L+O_R >0$, then the output of the binocular simple cell can be explicitly written as $ O_B = O_L^2+O_R^2+2O_LO_R$. The first two terms represent responses due to monocular stimulation while the third term $2O_LO_R$ can be interpreted as the binocular interaction term. 

The activity of a cell is then measured from the output and will be strongest at points that have a higher probability of matching each other. The maximum value over $d$ of this quantity is the extracted disparity. 

It is worth noting that neurophysiological computations of binocular profiles displayed in Figure \ref{binProfiles} assume the mono-dimensionality of the monocular receptive profile, ignoring information about orientation of monocular simple cells. However, this information will be needed to encode different types of orientation disparity.

\begin{remark}[Orientation matters]
In 2001, the authors of \cite{BCP01} conducted investigations on the response of binocular neurons to orientation disparity, by extending the energy model of Anzai, Ohzawa and Freeman to incorporate binocular differences in receptive-field orientation.  More recently, the difference between orientations in the receptive fields of the eyes has been confirmed \cite{S10}.
\end{remark}

The binocular energy model is a type of minimal model. It serves as a starting point, allowing the combination of monocular inputs.  But is not sufficient to solve the stereo-matching problem. 

\begin{remark}[Connections]
\label{connections}
It is argued in \cite{P16, SPTL13} that, in addition to the neural mechanisms that couple characteristics (such as signals, stimuli, or particular features) relating the left and right monocular structures, there must be a system of connections between binocular cells, which characterizes the processing mechanism of stereo vision; see also
Samonds et al. in \cite{SPTL13} in particular.
\end{remark}

\subsection{Differential interpretation of binocular RPs}

It is possible to write the interaction term $O_LO_R$ coming from \eqref{unitEnergyModel}, in terms of the left and right receptive profiles: 
 \begin{equation}
 \begin{aligned}
O_LO_R =&\int \varphi_{\theta_L, x_{L}, y}(\tilde x_L,\tilde y_L)I_L(\tilde x_L,\tilde y_L)d\tilde x_Ld\tilde y_L \int\varphi_{\theta_R, x_{R}, y}(\tilde x_R,\tilde y_R) I_R(\tilde x_R,\tilde y_R)d\tilde x_Rd\tilde y_R\\
=&\int \int  \varphi_{\theta_L, x_{L}, y}(\tilde x_L,\tilde y_L) \varphi_{\theta_R, x_{R}, y}(\tilde x_R,\tilde y_R)I_L(\tilde x_L,\tilde y_L)I_R(\tilde x_R,\tilde y_R) d\tilde x_Rd\tilde y_Rd\tilde x_L d\tilde y_L .\\
 \end{aligned}
 \end{equation}
If we fix $(\tilde x_R, \tilde y_R, \tilde x_L, \tilde y_L)$, we derive the expression of the binocular profiles $\varphi_{L,R}= \varphi_{\theta_R, x_{R}, y}\varphi_{\theta_L, x_{L}, y}$ as the product of monocular left and right profiles. This is in accordance with the measured profiles of Figure \ref{binProfiles}. 


\begin{proposition}
\label{prop:crossProd}The binocular interaction term can be associated with the cross product of the left and right directions defined through \eqref{X3vf}, namely $\omega_L^\star$ and $\omega_R^\star$ of monocular simple cells:
\begin{equation}
O_LO_R = \omega_L^\star \times \omega_R^\star.
\end{equation}
\end{proposition}

\proof 
The idea is that the binocular output is the combined result of the left and right actions of monocular cells, thus identifying a direction in the space of cyclopean coordinates.
The detailed proof of this proposition can be found in Appendix \ref{app:proofProp}. 
\endproof

%
%

To better understand the geometrical idea behind Proposition \ref{prop:crossProd},  we recall that the retinal coordinates can be expressed in terms of cyclopean coordinates  \eqref{dispDEF} as $x_R = x-d$ and $x_L = x+d$, and so we can write $\omega_L^\star$ and $\omega_R^\star$ in the $3D$ space of coordinates $(x, y, d)$ as: 
\begin{equation}
\label{1FormVCyclopean}
\begin{aligned}
\omega_R^\star =& ( -\sin\theta_R, \cos\theta_R , \sin\theta_R)^T \hspace{1cm}
\omega_L^\star =& (-\sin\theta_L, \cos\theta_L ,-\sin\theta_L )^T.\\
\end{aligned}
\end{equation}
We define $\omega_{bin}: = \omega_L^\star\times \omega_R^\star$ as the natural direction characterizing the binocular structure: 
\begin{equation}
\label{omegaBin}
\omega_{bin}= \begin{pmatrix}\sin(\theta_R+\theta_L) \\ 2\sin\theta_R \sin\theta_L\\ \sin(\theta_R-\theta_L) \end{pmatrix}.
\end{equation}

\begin{figure}[tbh]
\begin{center}
\includegraphics[width =0.7\textwidth]{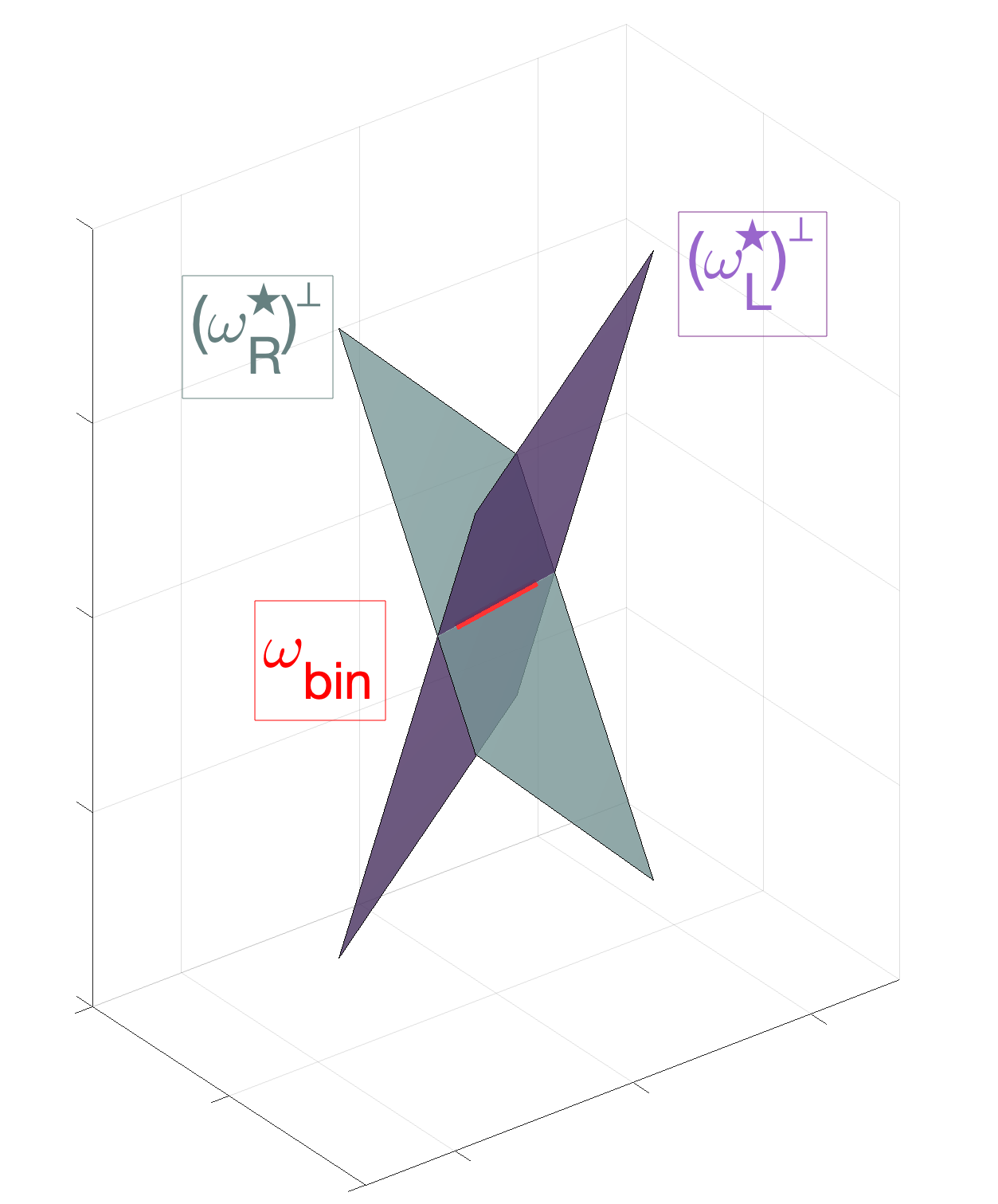}
\end{center}
\caption{Direction detected by $\omega_{bin}$ through the intersection of left and right planes generated by $ (\omega_R^\star)^\perp$ and $(\omega_L^\star)^\perp$. Red vector corresponds to the associated $2$-form $\omega_{bin}$.}
\label{planeInt}
\end{figure}

\begin{remark}
The vector $\omega_{bin}$ of equation \eqref{omegaBin} can be interpreted as the intersection of the orthogonal spaces defined with respect to  $\omega_R^\star$ and  $ \omega_L^\star$ when expressed in cyclopean coordinates $(x,y,d)$. More precisely, if
\begin{equation}
\begin{aligned}
&(\omega_L ^\star)^\perp= \Span\left\{\begin{pmatrix}\cos\theta_L\\ \sin\theta_L\\0\end{pmatrix},\begin{pmatrix}-1 \\0\\1 \end{pmatrix} \right\}, \hspace{1cm} ( \omega_R^\star)^\perp = \Span\left\{\begin{pmatrix}\cos\theta_R\\ \sin\theta_R\\0\end{pmatrix},\begin{pmatrix}1 \\0\\1 \end{pmatrix}\right\}\\
\end{aligned}
\end{equation}
then 
\begin{equation}
\omega_{bin} =( \omega_L ^\star)^\perp \cap (\omega_R^\star)^\perp.
\end{equation}
The result of the intersection of these monocular structures identifies a direction, as shown in Figure \ref{planeInt}.
\end{remark}

We earlier showed that the result of the action of a monocular odd simple cell is to select directions for the propagation of infomation. We now combine these, for the two eyes, to show that in the three-dimensional case the binocular neural mechanisms also lead to a direction. We will see in the next sections that this direction is the direction of the tangent vector to the $3D$ stimulus, provided points are corresponding.

\subsection{Compatibility with stereo geometry}
\label{sec:comp_SR_geom}
We consider the direction characterizing the binocular structure $\omega_{bin}$ defined in \eqref{omegaBin} and we show that it can be associated with the  $3D$ tangent vector to the $3D$ curve.
The idea is that this tangent vector is orthogonal both to $\omega_R^\star$ and to $\omega_L^\star$, and therefore it has the direction of the vector product $\omega_L^\star\times\omega_R^\star$. 

Precisely, we consider the normalized tangent vector $t_L$ and $t_R$ on retinal planes
\begin{equation}
t_R = (\cos\theta_R, \sin\theta_R)^T \hspace{0.3cm} t_L = (\cos\theta_L, \sin\theta_L)^T,
\end{equation} to the points $(x_R, y)$ and $(x_L, y)$ respectively. Taking into account that $f$ is the focal coordinate of the retinal planes in $\R^3$, then we associate to these points the correspondents in $\R^3$, namely $\tilde m_L= (x_L-c, y, f)^T$ , $\tilde m_R = (x_R+c, y, f)^T$.
Applying  equation \eqref{FaugerasTang}, it is possible to derive the tangent vector of the three dimensional contour: 
\begin{equation}
\label{UtlAndUtr}
\begin{aligned}
U_{t_L}&= P_L^{-1}\tilde m_L \times P_L^{-1}\tilde t_L &= \begin{pmatrix}x_L   \\ y_L \\ f  \end{pmatrix}\times \begin{pmatrix}\cos\theta_L \\ \sin\theta_L \\ 0 \end{pmatrix} &= \begin{pmatrix} - f \sin\theta_L \\ f \cos\theta_L \\ x_L\sin\theta_L -y_L\cos\theta_L  \end{pmatrix},\\
U_{t_R}&= P_R^{-1}\tilde m_R \times P_R^{-1}\tilde t_R &= \begin{pmatrix}x_R \\ y_R \\ f  \end{pmatrix}\times \begin{pmatrix}\cos\theta_R \\ \sin\theta_R \\ 0 \end{pmatrix} &= \begin{pmatrix} - f \sin\theta_R \\ f \cos\theta_R\\ x_R\sin\theta_R -y_R\cos\theta_R  \end{pmatrix},\\
\end{aligned}
\end{equation}
and the tangent direction is recovered by 
\begin{equation}
\label{utlXutr}
U_{t_L}\times U _{t_R}=
f\begin{pmatrix} \frac{x_L+x_R}{2}\sin({\theta_R-\theta_L}) - \frac{x_R-x_L}{2}\sin(\theta_L+\theta_R)\\ 
 y\sin(\theta_R-\theta_L) - (x_R-x_L)(\cos(\theta_R-\theta_L)-\cos(\theta_L+\theta_R))\\
f \sin(\theta_R-\theta_L)\\
\end{pmatrix}
\end{equation}

\begin{figure}[tbh]
\label{3Dreconstruction}
\begin{center}
\includegraphics[width = \textwidth]{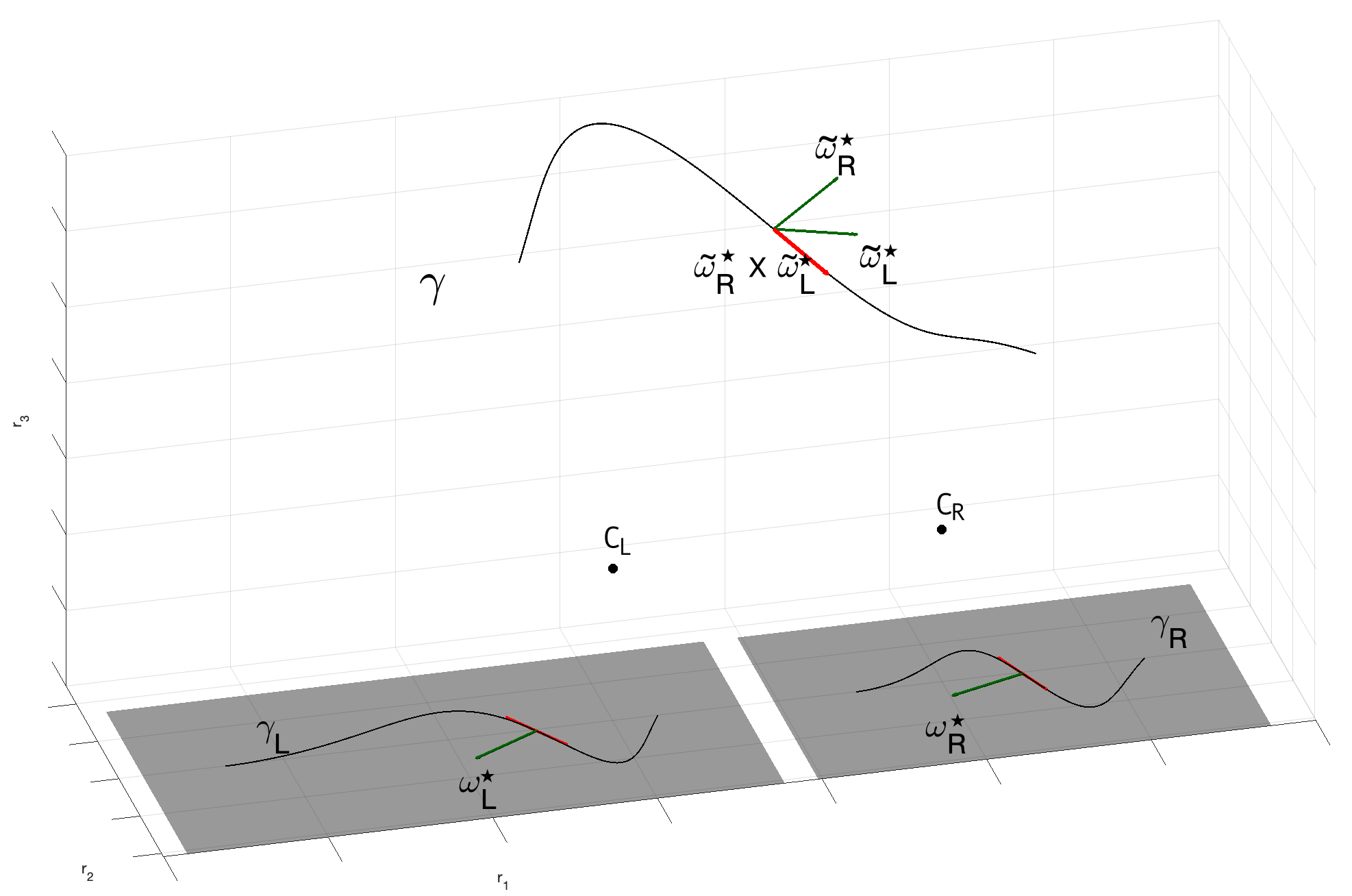}
\end{center}
\caption{Three dimensional reconstruction of the space from retinal planes. The $1$- forms $\omega_L^\star$ and $\omega_R^\star$ are identified with the normal to the curves $\gamma_L$ and $\gamma_R$. Their three dimensional counterpart $\tilde \omega_L^\star$ and $\tilde \omega_R^\star$ identify the tangent vector to the curve $\gamma: \R\rightarrow \R^3$ by the cross product $\tilde \omega_L^\star \times \tilde \omega_R^\star$. }
\label{projectedTangent}
\end{figure}
If we define
\begin{equation}
\tilde\omega_L^\star := \frac{d}{fc} U_{t_L}, \hspace{0.5cm} \tilde\omega_R^\star := \frac{d}{fc} U_{t_R}
\label{omegaTilde}
\end{equation} 
and  the corresponding 2 form $\omega_{\R^3} := \tilde\omega_L^\star\times\tilde\omega_R^\star$,
using the change of variables (\ref{changeVariables}) we observe that: 
\begin{equation}
\begin{aligned}
\tilde\omega_L^\star = \omega_L^\star, \hspace{0.3cm}\tilde\omega_R^\star = \omega_R^\star, \hspace{0.3cm} \omega_{\R^3}=\omega_{bin},
\end{aligned}
\end{equation}
up to a scalar factor. See Appendix \ref{compApp} for explicit computation.

In this way, the disparity binocular cells couple in a natural way positions, identified with points in $\R^3$, and orientations in $\S^2$, identified with three-dimensional unitary tangent vectors.  As already observed in Remark \ref{connections}, the geometry of the stereo vision is not solved only with these punctual and directional arguments, but there is the need to take into accounts suitable type of connections. In \cite{AZ00, LZ03, LZ06}, Zucker et al. proposed a model that considered the curvature of monocular structures as an additional variable. Instead, we propose to consider simple monocular cells selective for orientation, and to insert the notion of curvature directly into the definition of connection. It is therefore natural to introduce the perceptual space via the manifold $ \PO $, and look for appropriate curves.
 
\subsection{A perceptual model in the space of 3D position-orientation}

We now derive the objects in Fig.~\ref{fig:sphere}. 
We have clarified (end of section \ref{sec:comp_SR_geom}) that binocular cells are parametrized by  points in $\R^3$, and orientations in $\S^2$. 
An element $\xi$ of the space $\PO$ it is defined by a point $p = (p1, p2, p3)$ in $\R^3$ and an unitary vector $n \in \S^2$.  Since the topological dimension of this geometric object is $2$, we introduce the classical spherical coordinates $(\theta, \varphi)$ such that $n = (n_1, n_2, n_3)^T \in \S^2$ can be parameterized as:
\begin{equation}
\label{sphericalCoord}
\begin{aligned}
n_1 &= \cos\theta\sin\varphi\\
n_2 &= \sin\theta\sin\varphi\\
n_3 &= \cos\varphi\\
\end{aligned}
\end{equation}
with $\theta \in [0, 2\pi]$ and $\varphi \in (0, \pi)$. The ambiguity that arises using local coordinate chart is overcome by the introduction of a second chart, covering the singular points.

Translations and rotations are expressed using the group law of the three-dimensional special Euclidean group $SE(3)$, defining the group action 
\begin{equation}
    \sigma: \PO \times SE(3) \longrightarrow \PO \text{ s.t. } \sigma((p, n), (q, R)) = (Rp+q, Rn),
\end{equation}
with $(p, n) \in \PO$ , $(q, R)\in SE(3)$, namely $R \in SO(3)$ tridimensional rotation, and $q \in \R^3$.
%

\subsubsection{Stereo sub-Riemannian geometry}

The emergence of a privileged direction in $ \R^3$ (associated with the tangent vector to the stimulus) is the reason why we endow $ \PO $ with a sub-Riemannian structure that favors the direction in $3D$ identified by $\omega_{bin}$.
 
Formally, we consider admissible movements in  $\PO$ described by vector fields:  
%
\begin{equation}
\label{leftInvVec}
\begin{aligned}
Y_{\R^3, \xi} &= \sin\varphi\cos\theta\partial_1  + \sin\varphi\sin\theta\partial_2 + \cos\varphi \partial_3 \\
Y_{\theta, \xi} &=  -\frac{1}{\sin\varphi} \partial_\theta \\
Y_{\varphi,\xi} &= \partial_\varphi \\
\end{aligned}
\end{equation}
with $\xi \in \PO$ for $\varphi \neq 0, \varphi \neq \pi$. The admissible tangent space\footnote{see Appendix \ref{app:intro_SR} for the definition of admissible tangent space.} at a point $\xi$
\begin{equation}
\mathcal{A}_\xi := \Span \{Y_{\R^3, \xi}, Y_{\theta, \xi}, Y_{\varphi, \xi}\}
\end{equation}
encodes the coupling between position and orientations, as remarked by Duits and Franken in \cite{DF11}. 
In particular, the vector field $Y_{\R^3}$ identifies the privileged direction in $\R^3$, while $Y_\theta$ and $Y_\varphi$ allow changing this direction, involving just orientation variables of $\S^2$.
The vector fields $\{ Y_{\R^3}, Y_\theta, Y_\varphi\}$ and their commutators generate the tangent space of $\PO$ in a point, allowing to connect every point of the manifold using privileged directions (\textit{H\"ormander condition}). Furthermore, it is possible to define a sub-Riemannian structure by choosing a scalar product on the admissible tangent bundle $\mathcal{A}$: the simplest choice is to declare the vector fields $\{ Y_{\R^3},  Y_\theta, Y_\varphi\}$ orthonormal, considering on $\S^2$ the distance inherited from the immersion in $\R^3$ with the Euclidean metric.

\subsubsection{Change of variables} 

We have already expressed the change of variable in the variables $(x, y, d)$  to $(r_1, r_2, r_3)$ in equations \eqref{changeVar3DNeur}. However, the cortical coordinates also contain the angular variables $\theta_R$ and $\theta_L$ which involve the introduction of the spherical coordinates $\theta, \varphi$. 

To identify a change of variable among these variables, we first
 introduce the function \\
 $(r_1, r_2, r_3, \theta, \varphi) \xrightarrow{F} (x, y, d, \theta_L, \theta_R)$ : 
 
\begin{equation}
\begin{aligned}
F : \hspace{0.5cm}&\PO & \longrightarrow & \hspace{2cm} \PO\\
     & \begin{pmatrix} r_1\\ r_2 \\ r_3 \\ \theta \\ \varphi \\ \end{pmatrix}& \mapsto 
            & \begin{pmatrix}\frac{fr_1}{r_3}\\\frac{fr_2}{r_3}\\\frac{cf}{r_3}\\ 
            \tan^{-1}(\frac{r_3\sin\theta\cos\varphi-r_2\cos\varphi}{r_3\cos\theta\sin\varphi-(c+r_1)\cos\varphi})\\
            tan^{-1}(\frac{r_3\sin\theta\cos\varphi-r_2\cos\varphi}{r_3\cos\theta\sin\varphi-(c-r_1)\cos\varphi}\\\end{pmatrix}\\
\end{aligned},
\end{equation} where the retinal right angle 
$\theta_R = \tan^{-1}(\frac{r_3\sin\theta\cos\varphi-r_2\cos\varphi}{r_3\cos\theta\sin\varphi-(c+r_1)\cos\varphi})$ 
and the left retinal angle $\theta_L=  \tan^{-1}(\frac{r_3\sin\theta\cos\varphi-r_2\cos\varphi}{r_3\cos\theta\sin\varphi-(c-r_1)\cos\varphi})$
 are obtained considering equation \eqref{projDeriv}.
 
Analogously, it is possible to define the change of variable $(x, y, d, \theta_L, \theta_R) \xrightarrow{G} (r_1, r_2, r_3, \theta, \varphi)$: 

\begin{equation}
\begin{aligned}
G:\hspace{0.5cm} &\PO & \longrightarrow & \hspace{2cm}\PO \\
     & \begin{pmatrix} x\\ y \\ d \\ \theta_R \\ \theta_L \\ \end{pmatrix}& \mapsto 
            & \begin{pmatrix}\frac{cx}{d}\\\frac{cy}{d}\\\frac{cf}{d}\\ 
            \tan^{-1} ( \frac{2\sin\theta_R\sin\theta_L}{\sin(\theta_R+\theta_L)})\\
            \tan^{-1}(\frac{\sqrt{\sin^2(\theta_R+\theta_L)+4\sin^2\theta_R\sin^2\theta_L}}{\sin(\theta_R-\theta_L)})\\\end{pmatrix}\\
\end{aligned},
\end{equation} where the angles 
$\theta = \tan^{-1} ( \frac{2\sin\theta_R\sin\theta_L}{\sin(\theta_R+\theta_L)}) $ 
and $\varphi= \tan^{-1}(\frac{\sqrt{\sin^2(\theta_R+\theta_L) + 4\sin^2\theta_R\sin^2\theta_L}}{\sin(\theta_R-\theta_L)})$
 are obtained considering that $\tan\theta = \frac{(\vec{Y}_{\R^3})_2}{(\vec{Y}_{\R^3})_1}$ and $\tan\varphi=\frac{\sqrt{(\vec{Y}_{\R^3})_1^2+(\vec{Y}_{\R^3})_2^2}}{(\vec{Y}_{\R^3})_3}$.
 
\subsubsection{Integral curves}

The connectivity of the space is described by admissible curves of the vector fields spanning $\mathcal{A}$. In particular, a curve  $\Gamma : [0, T] \longrightarrow \PO$ is said to be \textit{admissible}\footnote{sometimes the term \textit{horizontal} is preferred. } if:
\begin{equation}
\label{horizontalCurves}
\dot\Gamma(t) \in \mathcal{A}_{\Gamma(t)}, \leftrightarrow \dot\Gamma(t) = a(t)\vec{Y}_{\R^3,\Gamma(t)}+ b(t)\vec{Y}_{\theta,\Gamma(t)} + c(t)\vec{Y}_{\varphi,\Gamma(t)}, 
\end{equation}
where $a, b, c$ are sufficiently smooth function on $[0, T]$.
We will consider a particular case of these admissible curves, namely constant coefficient integral curves with $a(t)= 1$, since the vector field $Y_{\R^3}$ represents the tangent direction of the $3D$ stimulus (and so it never vanishes): 
\begin{equation}
\label{integralCurves}
\dot\Gamma(t) = \vec{Y}_{\R^3,\Gamma(t)} + c_1\vec{Y}_{\theta,\Gamma(t)} + c_2\vec{Y}_{\varphi,\Gamma(t)}, 
\end{equation}
with $c_1$ and $c_2$ varying in $\R$. 

These curves can be thought of in terms of trajectories in $\R^3$ describing a movement in the $\vec{Y}_{\R^3}$ direction, which can eventually change according to $\vec{Y}_\theta$ and $\vec{Y}_\varphi$. An example of the fan of integral curves was shown in the Introduction in Figure \ref{fan}.

%
%


It is worth noting that in the case described by coefficients $c_1$ and $c_2$ equal to zero, the $3D$ trajectories would be straight lines in $\R^3$; by varying the coefficients $c_1$ and $c_2$ in $\R$, we allow the integral curves to follow curved trajectories, twisting and bending in all space directions.

Formally, the amount of "twisting and bending" in space is measured by introducing the notions of curvature and torsion. We then investigate how these measurements are encoded in the parameters of the family of integral curves, and what constraints have to be imposed to obtain different typologies of curves.

\begin{remark}\label{curvatureTorsionRMK}
The $3D$ projection of the integral curves \eqref{integralCurves} will be denoted $\gamma$ and satisfy  $\dot \gamma(t) = (\cos\theta(t)\sin\varphi(t),\sin\theta(t)\sin\varphi(t), \cos\varphi(t))^T$. Classical instruments of differential geometry let us compute the curvature and the torsion of the curve $\gamma(t)$:  
\begin{equation}
\begin{aligned}
k &= \sqrt{(\dot\varphi)^2+\sin^2\theta (\dot\theta)^2}, \\
\tau &= \frac{1}{k^2}(-\cos\varphi\sin^2\varphi(\dot\theta)^3- \sin\varphi \dot\varphi \ddot\theta + \dot\theta(-2\cos\varphi(\dot\varphi)^2+ \sin\varphi \ddot\varphi)).\\
\end{aligned}
\end{equation}
Using the explicit expression of the vector fields $Y_\theta$ and $Y_\varphi$ in equation \eqref{integralCurves}, we get
\begin{equation}
\dot \theta = -\frac{c_1}{\sin\varphi}, \hspace{1cm} \dot\varphi = c_2,
\end{equation}
from which it follows that: 
\begin{equation}
\label{ktau}
\begin{aligned}
k = & \sqrt{c_1^2+c_2^2}\\
\tau = & \frac{c_1^2-c_2^2}{k^2}c_1\cotan\varphi.\\
\end{aligned}
\end{equation}
\end{remark}

\begin{proposition}
\label{ktauProp}
By varying the parameters $c_1$ and $c_2$ in \eqref{ktau} where we explicit find solutions of \eqref{integralCurves},
we have:
\begin{itemize}
\item[1.] If $\varphi = \frac{\psi}{2}$ then $k = \sqrt{c_1^2}, \tau = 0$, and so the family of curves \eqref{integralCurves} are  circles of radius $1/c_1^2$ on the fronto-parallel plane $r_3 = \text{cost}$.  
\item[2.] If $\varphi = \varphi_0$, with $\varphi_0 \neq \pi/2$, then $k = \sqrt{c_1^2}$ and $\tau = c_1 \cotan\varphi_0$, and so the family of curves \eqref{integralCurves} are $r_3$-helices. 
\item[3.] If $\theta = \theta_0$ then $k = \sqrt{c_2^2}$, $\tau = 0$, and so the family of curves \eqref{integralCurves} are circles of radius $1/c_2^2$ in the osculating planes.
\item[4.] If $c_1 = \pm c_2$ then $\tau = 0$, and so the family of curves \eqref{integralCurves} are circles of radius $1/c_2^2$ in the osculating planes.
\end{itemize}
\end{proposition}

\proof The computation follows immediately from the computed curvature and torsion of \eqref{ktau} and classical results of differential geometry. 
\endproof

\begin{figure}[tbh]
     \centering

     \begin{subfigure}[b]{0.3\textwidth}
         \centering
         \includegraphics[width=\textwidth]{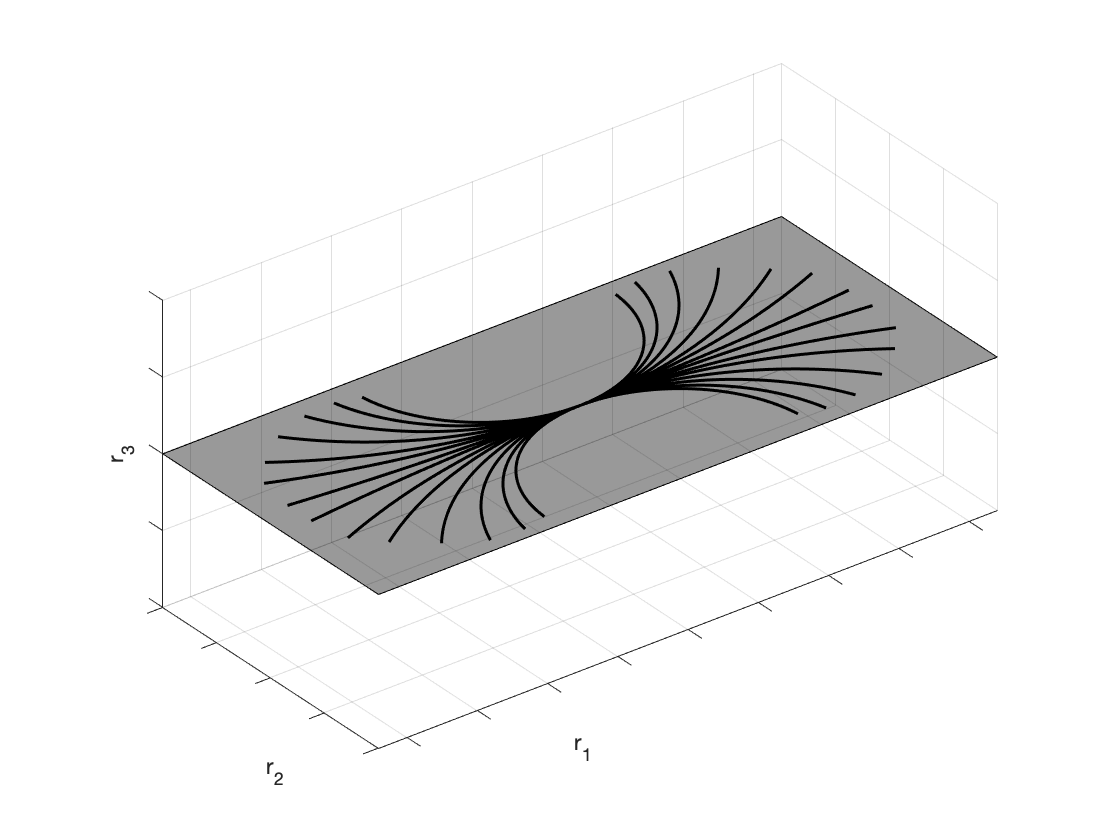}
         \caption{}
     \end{subfigure}
       \begin{subfigure}[b]{0.3\textwidth}
         \centering
         \includegraphics[width=\textwidth]{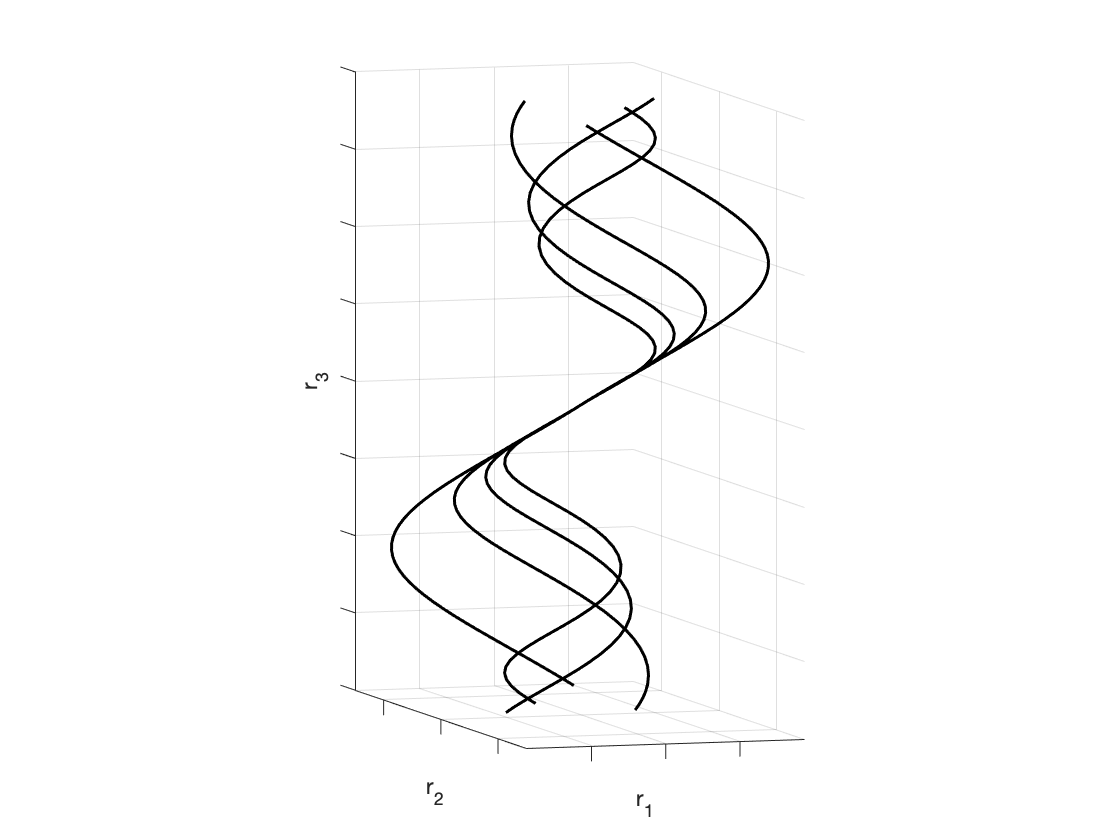}
         \caption{}
     \end{subfigure}
     \begin{subfigure}[b]{0.3\textwidth}
         \centering
         \includegraphics[width=\textwidth]{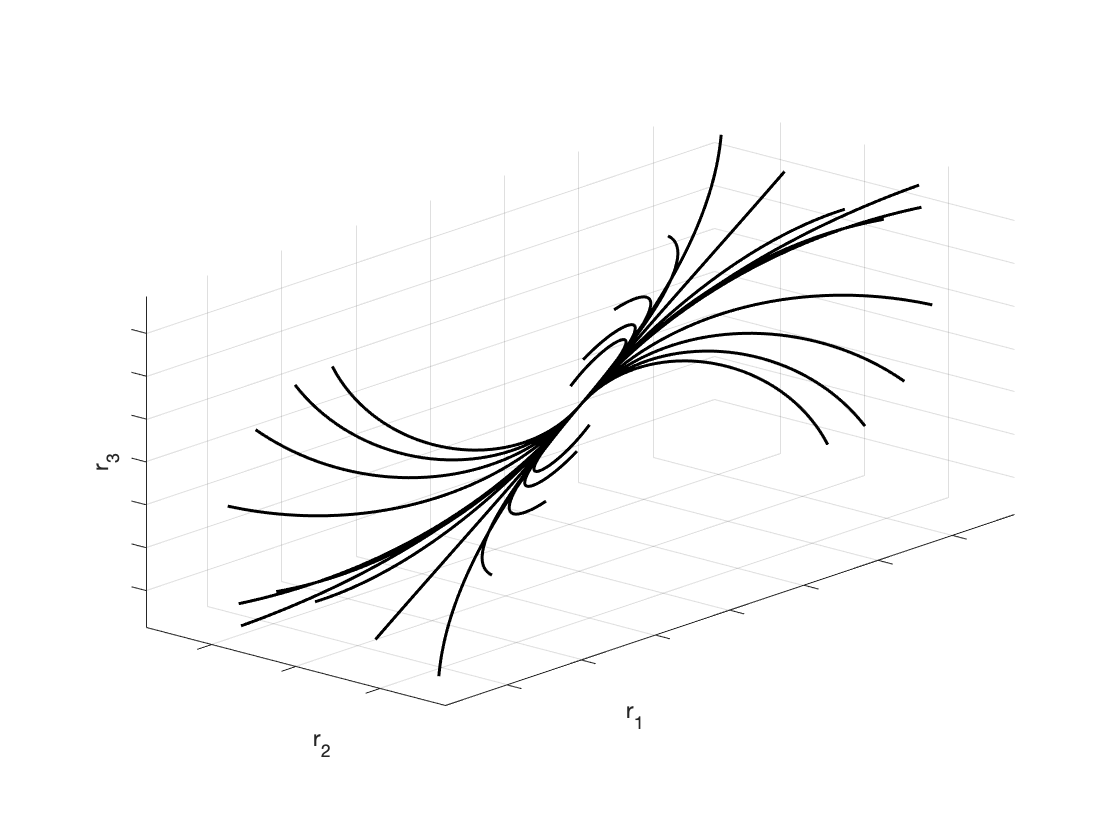}
         \caption{}
     \end{subfigure}
                \caption{Examples of integral curves obtained varying parameters $c_1$ and $c_2$. (A) Arc of circles for $ \varphi = \pi/2$. (B) $r_3$-helices for $\varphi = \pi/3$. (C) Family of curves with constant curvature $k$ and varying torsion parameter. }
        \label{integralCurvesFig}
\end{figure}

\begin{remark}
\label{knowK}If we know the value of the curvature $ k $, and we have one free parameter, $ c_2 $, in the definition of the integral curves \eqref {integralCurves}, then we are in the setting of Proposition \ref{geom3D}. In fact, the coefficient $c_1 $ is obtained by imposing $ c_1 = \pm \sqrt {k ^ 2-c_2 ^ 2} $, and in particular the component that remains to be determined is the torsion.\end{remark}

Examples of particular cases of the integral curves \eqref{integralCurves} according to Proposition \ref{ktauProp} and Remark \ref{knowK} are visualized in Figure \ref{integralCurvesFig}.

\section{Comparison with experimental data}
\label{sec:validation}

In this section we present results of compatibility between the proposed sub-Riemannian model and biological and psychophysical phenomena present in literature.

\subsection{Biological Connections}

The foundation for building our sub-Riemannian model of stereo was a sub-Riemannian model of curve continuation. 
This was motivated by the orientation column at each position, and the connections between cells in nearby columns. These connections were, in turn, a direct model of the long-range horizontal connections in visual cortex, for which there is beautiful biological data (e.g. \cite{BZSF97}). We further illustrated, in the Introduction, aspects of the cortical architecture that support binocular processing. Although the inputs from each eye are organized into ocular dominance bands, there is no direct evidence for "stereo columns" analogous to the monocular orientation columns. But, as we shall now show, there is evidence of long-range connections between binocular cells, and our model informs, concretely, what information should be carried by these long range connections. Thus, an organization for stereo is suggested, but it is implicit in the architecture. Nevertheless, there is evidence in support of it.

\begin{figure}[tbh]
\centering
\begin{subfigure}[b]{0.38\textwidth}
\centering
\includegraphics[width=\textwidth]{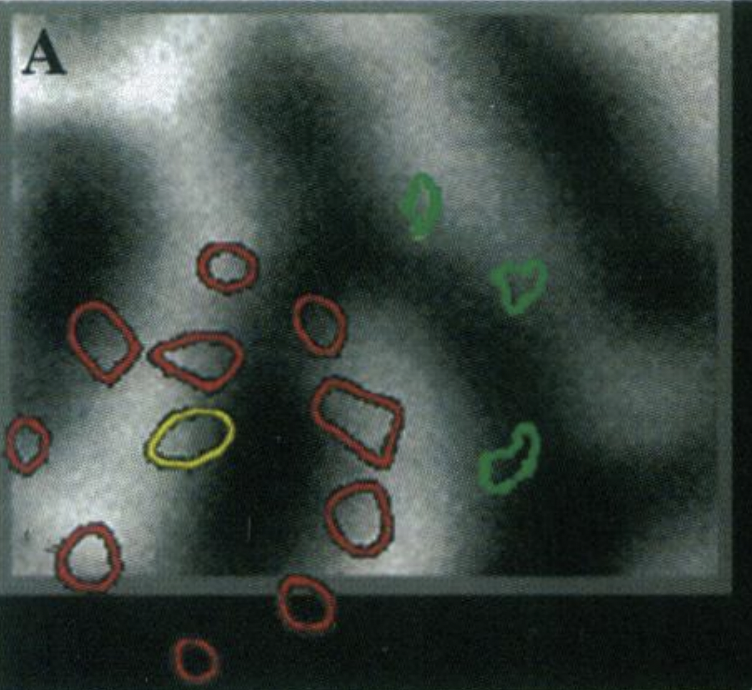} 
\caption{}
\end{subfigure}
\begin{subfigure}[b]{0.58\textwidth}
\centering
\includegraphics[width=.7\textwidth, height=.25\textheight]{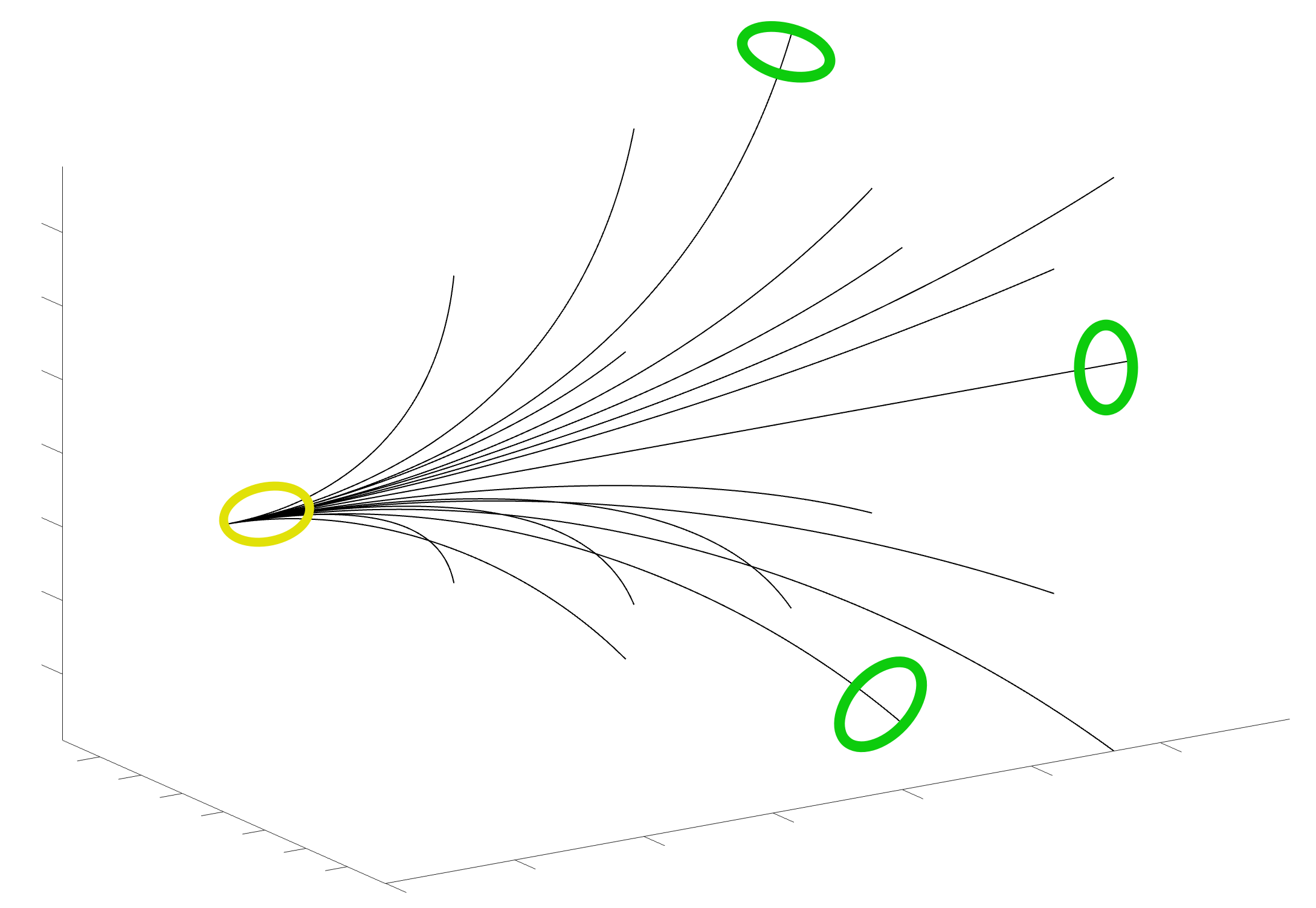} \caption{}
\end{subfigure}
\caption{ (A) A biocytin injection superimposed on a map of ocular dominance columns, image result from the work in \cite{MAHG93}. Binocular zones are in the middle of monocular zones (coded in black and white). Starting from the injection site (yellow circle in the center of a binocular zone) the patches' propagation (red corresponds to dense while green to sparsely labeled) tends to avoid highly monocular sites, bypassing the centers of ocular dominance columns, and are located in binocular zones.(B) $3D$ interpretation of the physiological image (A).}
\label{bin_connections}
\end{figure}

Just as information propagates to enforce monocular curve continuation, 
the binocular signal propagates to form a coherent binocular representations. The Grinwald group established this for stereo \cite{MAHG93} (see also Figure \ref{bin_connections}(A)), 
using biocytin injections, that propagate directly along neuronal processes and are deposited at excitatory synapses. Thus, this technique demonstrates the presence of long-range connections between binocular cells. These results were refined, more recently, by the Fitzpatrick group \cite{scholl2022binocular}, using in vivo calcium imaging. As shown in Fig.~\ref{fig:bands}(right) the authors demonstrated both the monocular and the binocular inputs for stereo, and (not shown) the dependence on orientations. 

More precisely, \cite{MAHG93} showed selective {\em anisotripic} connectivity among binocular regions: the biocytin tracer does not spread uniformly, but rather is highly directional with distance from the injection point. (This was the case with monocular biocytin injections as well.) Putting this together with \cite{scholl2022binocular}, we interpret the anisotropy as being related to (binocular) orientation (\cite{scholl2022binocular}), which is exactly the behavior of the integral curves of our vector fields. Our $3D$ association fields are strongly directional, and information propagates preferentially in the 
direction of (the starting point of) the curve. An example can be seen in Figure \ref{bin_connections}, image (B), where the fan of integral curves \eqref{integralCurves} is represented, superimposed with colored patches, following the experiment proposed in \cite{MAHG93}.


\subsection{Psychophysics and association fields}
In this section, we show that the connections described by the integral curves in our model can be related to the geometric relationships  from psychophysical experiments on perceptual organization of oriented elements in $\R^3$. The goal is to establish that our connections serve as a generalization of the concept of an association field in $3D$.

\subsubsection{Towards a notion of \textit{association field} for 3D contours}

The perception of continuity between two elements of position-orientation in $\R^3$  has been studied experimentally. To start, Kellman, Garrigan, and Shipley (\cite{KGS05, KGSYM05}) introduce \textit{3D relatability}, as a way to extend to $3D$ the experiments of Field, Heyes and Hess (\cite{FHH93}) in $2D$.

\begin{figure}[tbh]
              \centering
         \includegraphics[width=\textwidth]{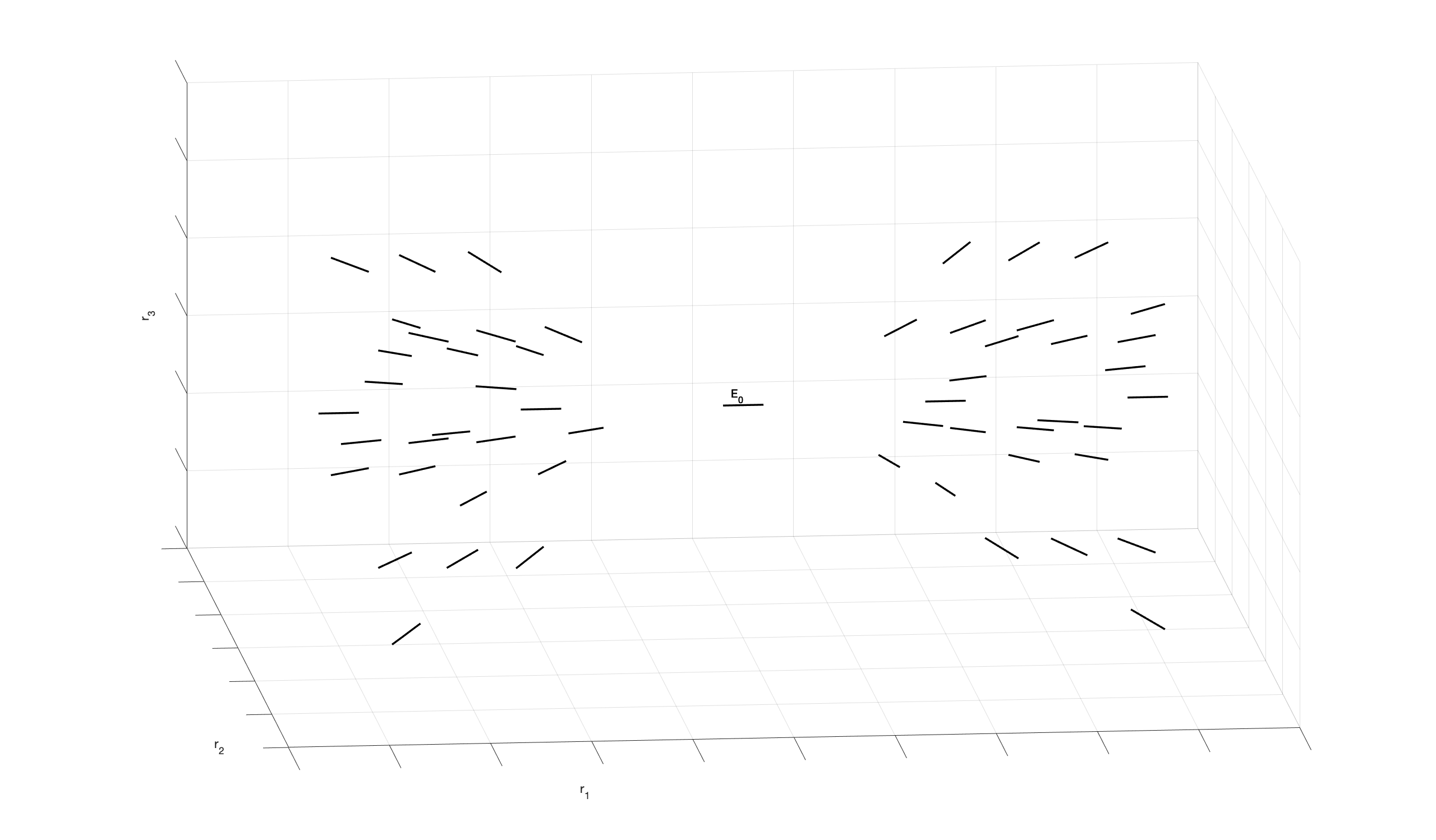}
       \caption{ Example of the fan of the $3D$ relatable edges with initial point $E_0$}
        \label{fanRelEdges}
\end{figure}

Particularly, 
in a system of $3D$ Cartesian coordinates, it is possible to introduce oriented edges $E$ at the application point  $(r_1, r_2, r_3)^T$ and with an orientation identified with the angles $\theta$ and $\varphi$. This orientation can be read, in our case, through the direction expressed by $(\cos\theta\sin\varphi, \sin\theta\sin\varphi, \cos\varphi)^T$. For an initial edge $E_0$, with application point on the origin of the coordinate system $(0, 0, 0)^T$ and orientation lying on the $r_1$-axis, described by $\theta = 0, \varphi = \pi/2$, the range of possible orientations $(\theta, \varphi)$ \footnote{The angle $\varphi$ here has been modified to be compatible with our set of coordinates. The relationship between the angle $\tilde \varphi$ in works \cite{KGS05, KGSYM05} can be expressed as : $\tilde \varphi = \acos(\sin \varphi)+\pi$.} for $3D$-relatable edges with $E_0$ is given by:

\begin{equation}
\tan^{-1} \left(\frac{r_2}{r_1}\right) \leq \theta \leq \frac{\pi}{2} \hspace{0.5cm} \text{and} \hspace{0.5cm}
\frac{\pi}{2} \leq \frac{3\pi}{2}-\varphi \leq \tan^{-1} \left(\frac{r_3}{r_1}\right).
\label{relEdges}
\end{equation}
The bound on these equations identified with the quantity $\frac{\pi}{2}$ incorporates the $90$ degree constraint in three dimensions, while the bounds defined by the inverse of the tangent express the absolute orientation difference between the reference edge $E_0$ and an edge positioned at the arbitrary oriented point $E_{(r_1, r_2, r_3)}$ so that its linear extension intersects $E_0$; see \cite{KGS05, KGSYM05} for further details.

Numerical simulations allow us to visually represent an example of the $3D$ positions and orientations that meet the $3D$ relatability criteria. Starting from an initial edge $E_0$ with endpoints in $(p_{01}, p_{02}, p_{03})^T$ and orientation on the $e_1$- axis, we represent for an arbitrary point $(p_1, p_2, p_3)^T$ the limit of the relatable orientation $(\theta, \varphi)$. Results are shown in Figure \ref{fanRelEdges}. 

\begin{figure}[tbh]
     \centering
     \begin{subfigure}[b]{0.5\textwidth}
         \centering
         \includegraphics[width=\textwidth]{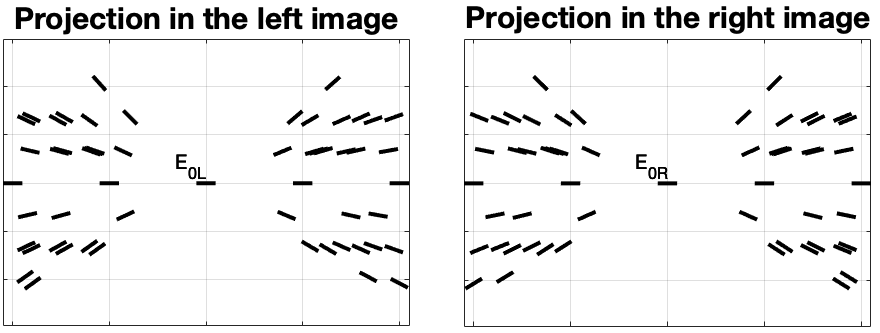} 
         \caption{}
     \end{subfigure}
     \begin{subfigure}[b]{0.41\textwidth}
         \centering
         \includegraphics[width=\textwidth]{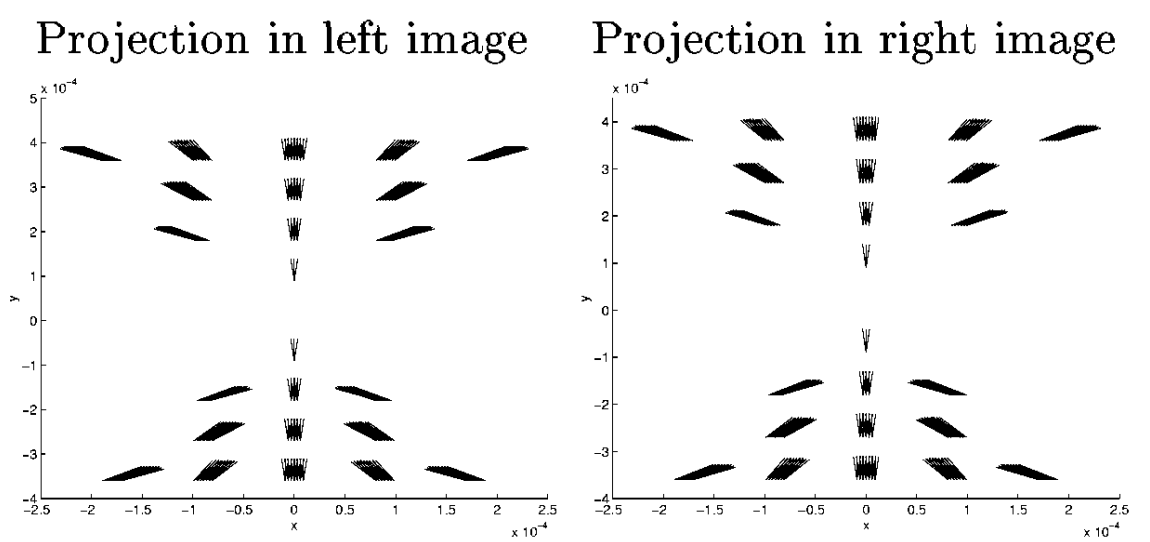}
         \caption{}
     \end{subfigure}
             \caption{ (A) Example of $3D$ association field in the two left and right retinal planes, generated with the geometry of $3D$ relatability. (B) Example of $3D$ compatibility field of \cite{AZ00}.  }
        \label{proj3DassField}
\end{figure}

\begin{remark}By projecting on the retinal planes of the $3D$ fan of relatable points, it is possible to notice that these projections are in accordance with the notion of $3D$ compatibility field of in \cite{AZ00}. See Figure \ref{proj3DassField}.
\end{remark}

Psychophysical studies, see \cite{DW15, HF95, HHK97}, have investigated the properties of the curves that are suitable for connecting these relatable points. These curves are well described by being smooth and monotonic. 
In particular, using non-oriented contour elements for contours,  Hess et al. in \cite{HHK97} indicate that contour elements can be effectively grouped based primarily on the good continuation of contour elements in depth. This statement is confirmed by the more recent work of Deas and Wilcox (\cite{DW15}), who in addition observe that detection of contours defined by regular depth continuity is faster than detection of discontinuous contours. All these results support the existence of depth grouping operations, arguing for the extension of Gestalt principles of continuity and smoothness in  three dimensional space. Finally, on the relationship of the three-dimensional curves to 2-dimensional association fields, see \cite{KGSYM05, KHK16}. These authors have assumed that the strength of the relatable edges in the co-planar planes of $E_0$ must meet the relations of the bi-dimensional association fields of \cite{FHH93}.

\subsubsection{Compatibility with the sub-Riemannian model}

\begin{figure}[tbh]
     \centering
     \begin{subfigure}[b]{0.42\textwidth}
         \centering
         \includegraphics[width=\textwidth]{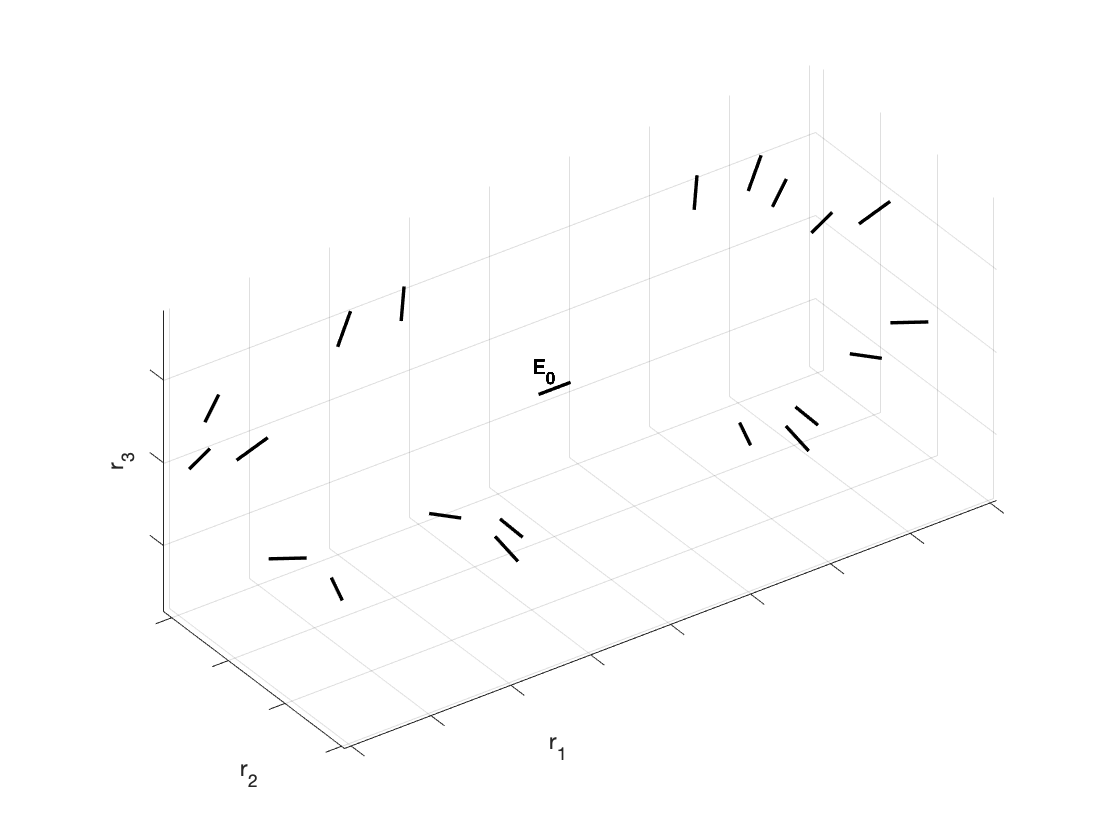}
         \caption{}
     \end{subfigure}
             \centering
     \begin{subfigure}[b]{0.45\textwidth}
         \centering
         \includegraphics[width=\textwidth]{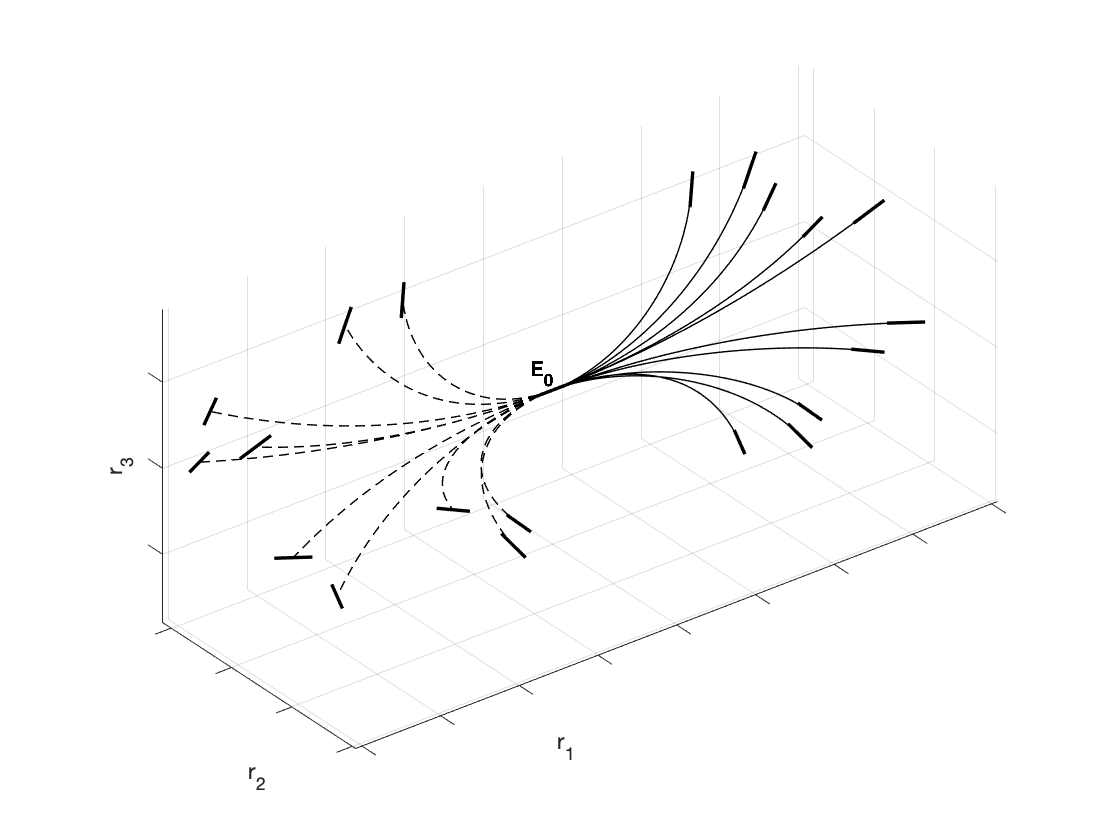}
         \caption{}
     \end{subfigure}
           \caption{(A)$3D$ relatable edges displayed on the right of the initial edge $E_0$. Unrelatable $3D$ edges displayed on the left. (B) Horizontal integral curves with filled lines connect $3D$ relatable edges with initial point $E_0$. Horizontal integral curves with dotted lines do not connect $3D$ unrelatable edges. }

        \label{3Dfit}
\end{figure}

To model associations underlying the $3D$ perceptual organization of the previous paragraph, we consider again the constant coefficient family of integral curves studied in \eqref{integralCurves}:
\begin{equation}
 \dot\Gamma(t) = \vec{Y}_{\R^3,\Gamma(t)}+ c_1\vec{Y}_{\theta,\Gamma(t)} + c_2\vec{Y}_{\varphi,\Gamma(t)}, \text{ with } c_1, c_2 \in \R.
\end{equation}
\noindent
Importantly, these curves locally connect the association fan generated by the geometry of $3D$ relatability. 
In particular, Figure \ref{3Dfit}, image(B) shows the family of the horizontal curves connecting the initial point $E_0$ with $3D$ relatable edges. These curves are computed using Matlab solver function \texttt{ode45}. 
In analogy with the experiment of Field , Hayes and Hess in \cite{FHH93}, we choose to represent non-relatable edges to the left of the starting point $E_0$, while on the right  are $3D$ relatable edges. So,  filled lines of the integral curves indicate the correlation between the central horizontal element $E_0$ and the ones on its right, while dotted lines connect the starting point $E_0$ with elements not correlated with it, as represented on the left part of the image.  

\begin{figure}[tbh]
             \centering
     \begin{subfigure}[b]{0.25\textwidth}
         \centering
         \includegraphics[width=\textwidth]{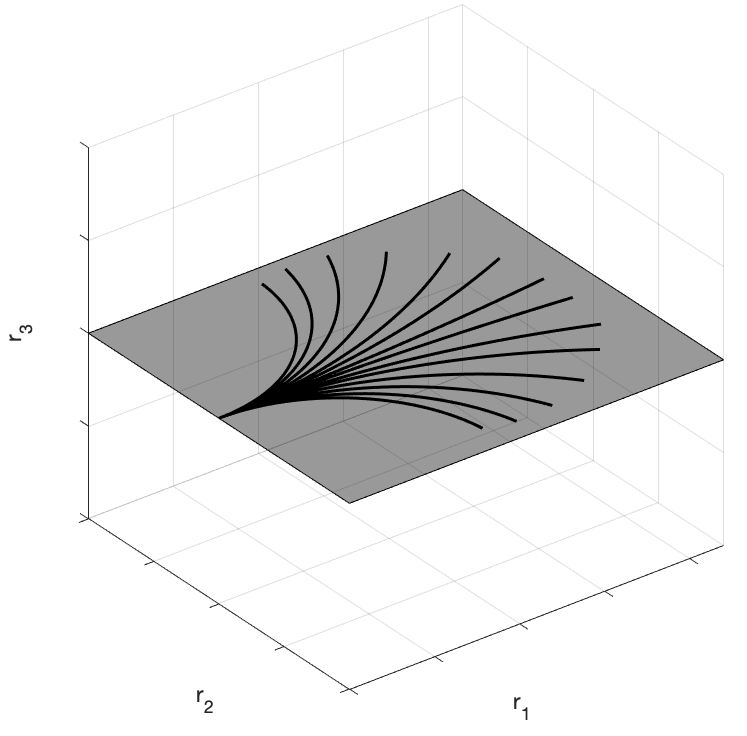}
         \caption{}
     \end{subfigure}
     \begin{subfigure}[b]{0.25\textwidth}
         \centering
         \includegraphics[width=\textwidth]{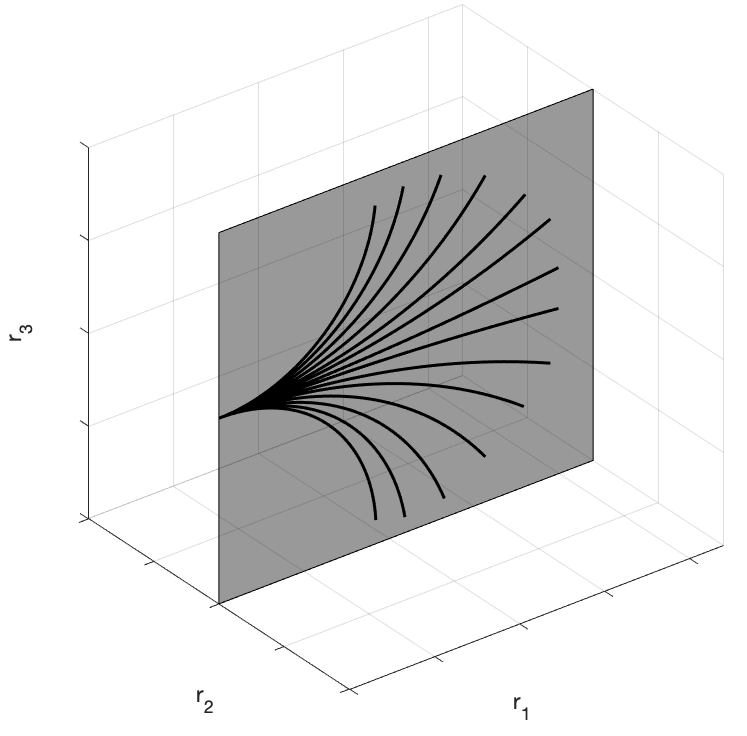}
         \caption{}
     \end{subfigure}
       \centering
     \begin{subfigure}[b]{0.35\textwidth}
         \centering
         \includegraphics[width=\textwidth]{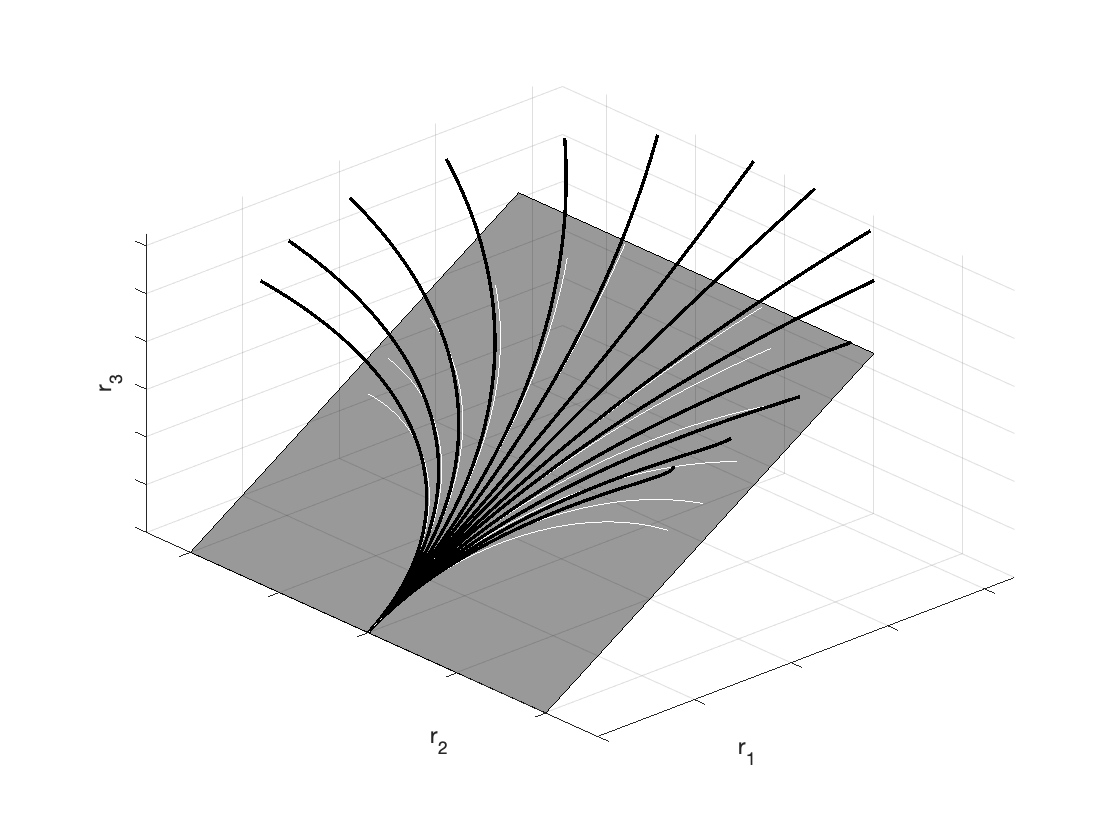}
         \caption{}
     \end{subfigure}
      \caption{(A) Restriction of the fan of the integral curves on the $e_1$-$e_2$ plane. (B)  Restriction of the fan of the integral curves on the $e_1$-$e_3$ plane. (C) Restriction of the fan at $\varphi = \varphi_0$. These curves (black lines) are not planar curves but helices. However, their projection (white lines) on the coplanar plane with initial edge satisfies the bidimensional constraints.}
\label{coPlanarFit}
\end{figure}

Restricting the curves on the neighborhood of co-planar planes with an arbitrary edge $E$, we have different cases. First, on the $r_1$-$r_2$ plane (fronto-parallel) and  the $r_1$-$r_3$ plane we have arcs of circle, as proved with Proposition \ref{ktauProp}. Furthermore, for an arbitrary plane in $\R^3$ containing an edge $E$, we observe that the curves generating with fixed angle $\varphi$ are helices, and locally they satisfy the bidimensional constraint in the plane.   Examples can be found in Figure \ref{coPlanarFit}.  In particular, the curves displayed in images (A) and (B) of Figure \ref{coPlanarFit} are well in accordance with the curves of the Citti-Sarti model, depicted in  Figure \ref{assField2D}. 

\subsection{Integration of contours and stereo correspondence problem}

Although the goal of this paper is not to solve the stereo correspondence problem, we can show how our geometry is helpful in understanding how to match left and right points and features. These ideas are developed more fully in \cite{phDthesis}.

\begin{figure}[tbh]
         \centering
         \includegraphics[width=0.9\textwidth]{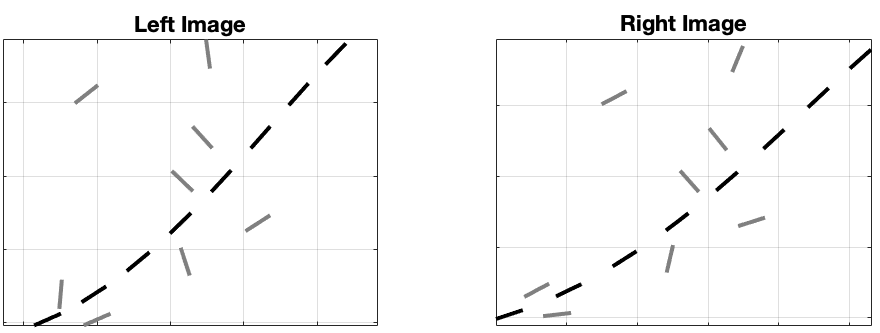}
         \caption{Left and right retinal images of the set $\Omega$. Black points are the projection of the point of the curve $\gamma$, while gray points are background random noise. }
     \label{retinalImages}
     \end{figure}

\begin{figure}[tbh]
     \centering
     \begin{subfigure}[b]{0.25\textwidth}
         \centering
         \includegraphics[width=\textwidth, height = 4.5cm]{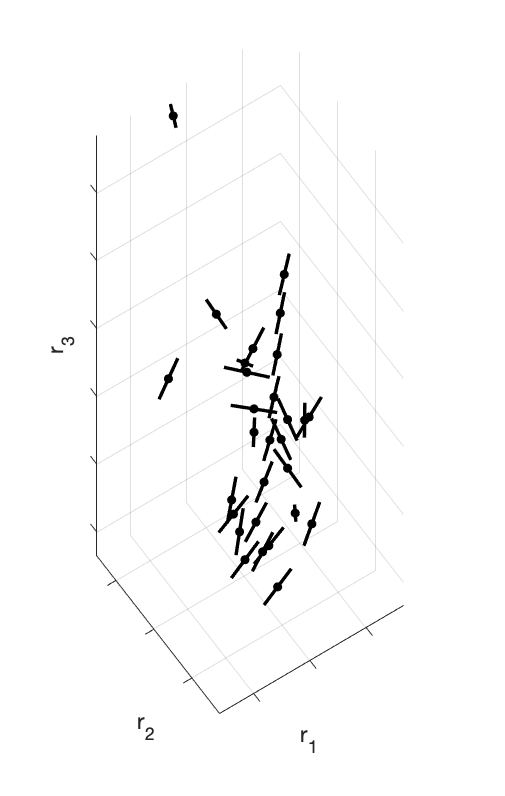}
         \caption{}
     \end{subfigure}
\centering
     \begin{subfigure}[b]{0.4\textwidth}
         \centering
         \includegraphics[width=\textwidth, height = 4.5cm]{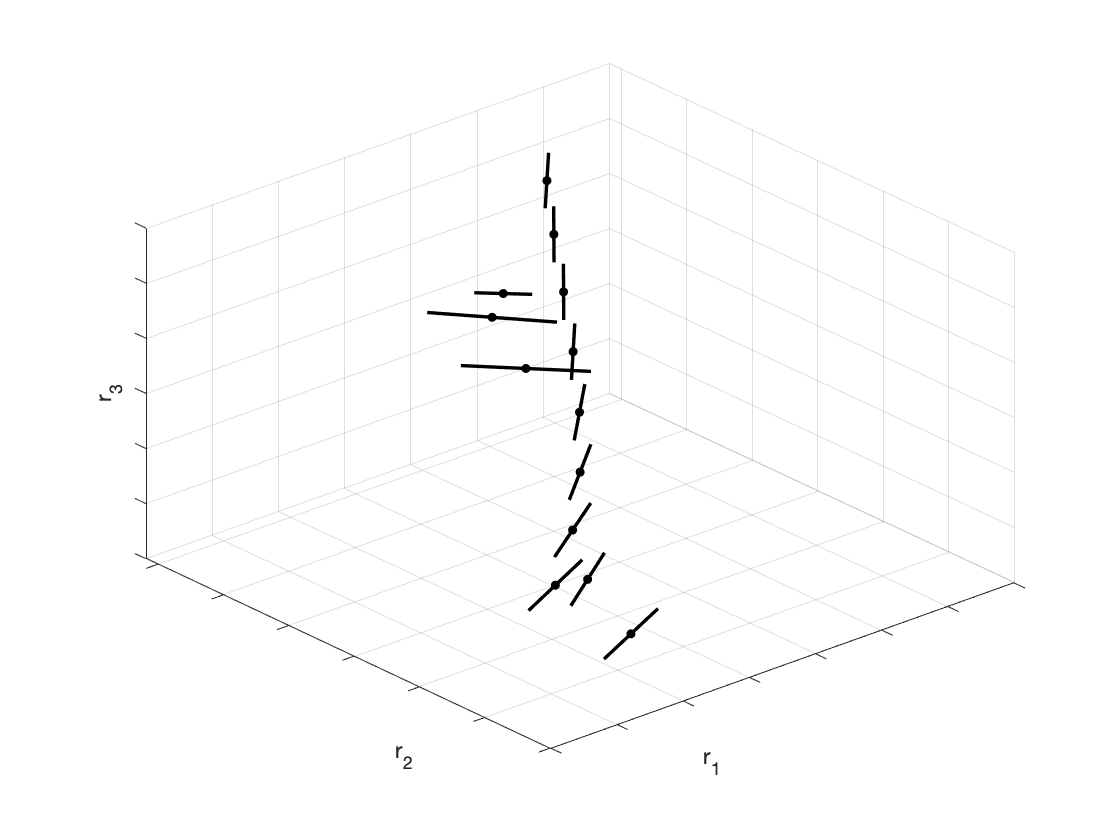}
         \caption{}
     \end{subfigure}
     \begin{subfigure}[b]{0.3\textwidth}
         \centering
         \includegraphics[width=\textwidth, height = 4.5cm]{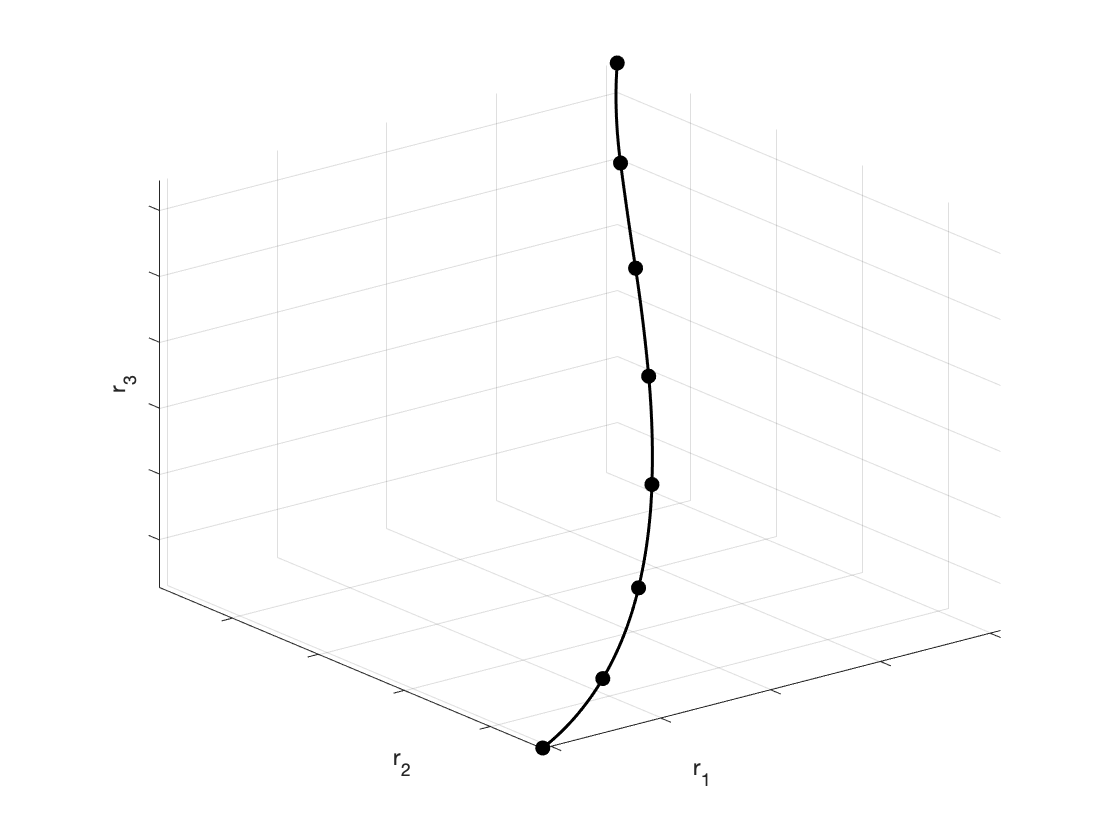}
         \caption{}
     \end{subfigure}
      \caption{ (A) Lifting of the two left and right retinal images of Figure \ref{retinalImages} in the space of position and orientation $\R^3\times\S^2$. (B) Selection of lifted points according to the binocular output. (C) Points of the stimulus $\gamma$ connected by integral curves \eqref{integralCurves}. }
      \label{integration}
\end{figure}

Inspired by the experiment of Hess and Field in \cite{HF95}, we consider a path stimulus $\gamma$ interpreted as a contour, embedded in a background of randomly oriented elements: left and right retinal visual stimuli are depicted in Figure \ref{retinalImages}.
We perform a first simplified lift of the retinal images to a set $\Omega$ subset of $\PO$. This set contains all the possible corresponding points, obtained by coupling left and right points which share the same $y$ retinal coordinate, see Figure \ref{integration}, image(A). The set $\Omega$ contains false matches, namely points that do not belong to the original stimulus. It is the task of correspondence to eliminate these false matches. 
     
We compute for every lifted point the binocular output $O_B$ of equation \eqref{unitEnergyModel}. This output can be seen as a probability measure that gives information on the correspondence of the couple of left and right points. We can simply evaluate which are the points with the highest probability of being in correspondence, applying a process of suppression of the non-maximal pairs over the fiber of disparity. In this way, noise points are removed (Figure \ref{integration}, image (B)). 

We now directly exploit good continuation in depth. The remaining noise elements are orthogonal to the directions of the elements of the curve that we would like to reconstruct. Calculating numerically the coefficients $ c_1 $ and $ c_2 $ of integral curves \eqref{integralCurves} that connect all the remaining pairs of points, we can obtain for every pair the value of curvature and torsion using \eqref{ktau}. 
\begin{figure}[tbh]
     \centering
     \begin{subfigure}[b]{0.4\textwidth}
         \centering
         \includegraphics[width=\textwidth]{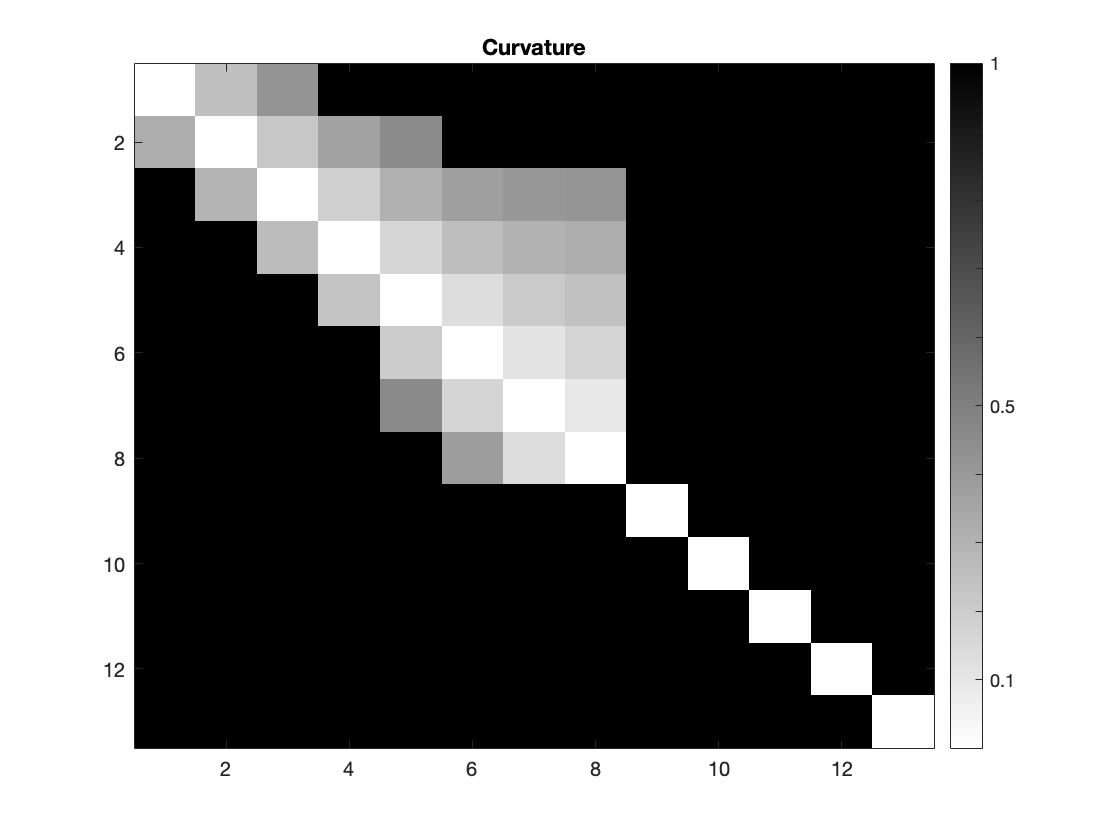}
         \caption{}
     \end{subfigure}
\centering
     \begin{subfigure}[b]{0.4\textwidth}
         \centering
         \includegraphics[width=\textwidth]{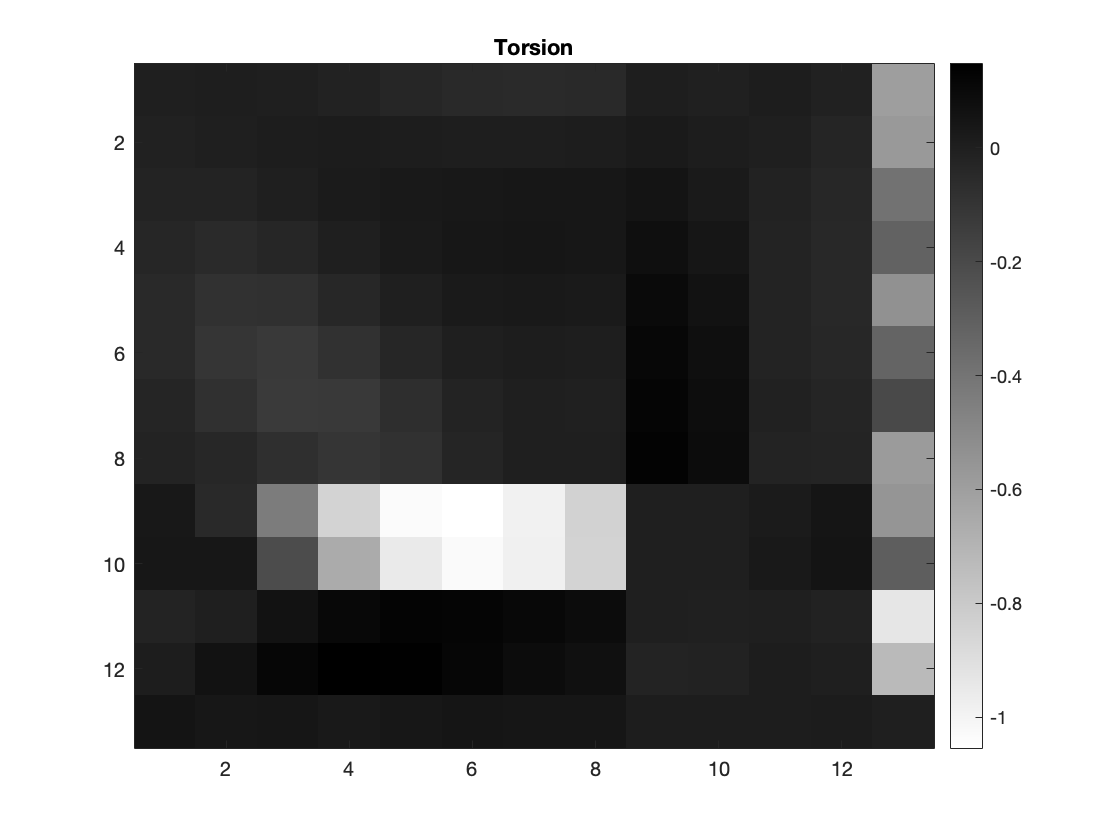}
         \caption{}
     \end{subfigure}
      \caption{Matrices $M$ which element $M_{ij}$ represents the value of curvature/ torsion for every couple of points $\xi_i, \xi_j$. The first eight points correspond to points of the curve $\gamma$ while the others are random noise.  (A) Curvature matrix. (B) Torsion matrix. }
      \label{KTmatrices}
\end{figure}
Figure \ref{KTmatrices} read it in terms of matrices $M$ representing the values of curvature or torsion for every couple of points $\xi_i, \xi_j$ in the element $M_{ij}$. In particular, we observe that random points are characterized by a very high curvature and in general also the torsion deviates from minimum magnitudes. So, by discarding these high values, we select only the three-dimensional points of the curve $\gamma$, which are well connected by the integral curves, as shown in image (C) of Figure \ref{integration}. This is in accordance with the idea developed in \cite {AZ00, LZ03, LZ06}, where curvature and torsion provide  constraints for reconstruction in 3D. 

\section*{Summary and Conclusions}

Understanding good continuation in depth, like good continuation for planar contours, can benefit from basic physiological constraints; from psychophysical performance measures, and from mathematical modeling. In particular, good continuation in the plane is supported by orientation selectivity and cortical architecture (orientation columns), by association field grouping performance, and by geometric modeling. We maintain that the same should be true for good continuation in depth. However, while the psychophysical data may be comparable, the physiological data are weaker and the geometry of continuation is not well understood.  In this paper, we introduced the neuro-geometry of stereo vision to fill this gap. It is strongly motivated by an analogical extension to $3D$ of $2D$ geometry, subject to respecting the psychophysics. In the end, it allowed us to be precise about the type of geometry that is relevant for understanding stereo abstractly, and concretely was highly informative toward the physiology. Although a "stereo columnar architecture" is not obvious from the anatomy, it is well-formed computationally.

The neuro-geometry of binocular cells are described through binocular RPs which are the product of left and right monocular RPs. Starting from binocular receptive profiles it is possible to reconstruct the three dimensional space using just the position and orientation of the visual stimulus recovered in the retinal planes (assuming one has  corresponding points). 

\vspace{0.2cm}

Technically, we proposed a sub-Riemannian model on the space of position and orientation $\PO$ for the description of the perceptual space of the neural cells involved. This geometrical structure favors the tangent direction of a $3D$ curve stimulus. The integral curves of the sub-Riemannian structure encode the notions of curvature and torsion within their coefficients, and are introduced to describe the connections between elements. This model can be seen as an extension in the three-dimensional scene of the 2-dimensional association field. In particular, the integral curves of the sub-Riemannian structure of the $3D$ space of position-orientation are exactly those that locally correspond to psychophysical association fields.

\vspace{0.2cm}

Although the goal of this paper is not to solve the stereo correspondence problem, we have seen how the geometry we propose is a good starting point to understand how to match left and right points and features. 
A future development of the model will consist in defining the probability of the co-occurrence between two elements, to
individuate percepts in $3D$ space. Individuation of percepts through harmonic analysis on the sub-Riemannian structure has been proposed in the past, both for $2D$ spatial stimuli \cite{SC15}  and in $2D$ + time spatio-temporal stimuli \cite{BCCS14}. It would be interesting to develop a similar analysis and extend it to stereo vision.

\section*{Acknowledgements}

MVB, GC, and AS were supported by
EU Project, GHAIA, Geometric and Harmonic Analysis with Interdisciplinary Applications, H2020-MSCA-RISE-2017 

SWZ was supported in part by US NIH EY031059 and by US NSF CRCNS 1822598. 

\begin{appendices}

\section{A gentle introduction to sub-Riemannian geometry}
\label{app:intro_SR}

In this paper we exploit techniques from differential geometry, and in particular sub - Riemannian geometry. In this appendix we provide an invitation to these ideas with a rather informal discussion. For the reader interested in a formal introduction on basic instruments of differential geometry (arguments of sections A.1 and A.2) please refer to \cite{T11}.
For a complete and formal mathematical (comprehensive) introduction to sub-Riemannian geometry we refer to \cite{ABB00}, while for a more informational point of view please consult \cite[Ch.\ 4.2]{sarti2019differential} and \cite{citti2015harmonic}. 

\subsection{Tangent bundle}

To start, imagine that you are standing at a point on a smooth surface in the world, far from any boundaries. Now, you can "walk away" from this point in any ($2D$) compass direction; for example, you could walk north or south or any direction in-between. If your steps were very very short, then the (flat) compass actually characterizes the $2D$ space of possible steps you might take. These same ideas are expressed more formally in differential geometry, as follows. One can attach to every point $p$ of a differentiable manifold $M$ (a generalized surface) a tangent space $T_pM$ (the compass plus some algebra describing vector operations). That is, the tangent space is a real vector space that contains the possible \textit{directions} in which one can tangentially pass through $p \in M$. If the manifold is connected, then the tangent space has, at every point, the same dimension as the manifold. So, if the manifold is a $2D$ surface, the tangent space at a point is a plane. In general, this tangent plane "approximates" the surface only locally.

The elements $\vec{X}_p$ of the tangent space $T_p M$ at $p$ are called \textit{tangent vectors} at $p$. Attached to a point on the surface, as above, these tangent vectors define the directions in which one could walk away from the point. But modern differential geometry provides another interpretation: it is possible to think of the elements of the tangent space in terms of directional derivatives.  Technically, for every smooth function $f$,   $Xf(p) = \vec{X}_p \cdot \nabla f(p)$ will denote the directional derivative of $f$ in the direction of the vector $\vec{X_p}$, with $\nabla$ denoting the gradient vector (expressed in an appropriate coordinate system) and $\cdot$ scalar product between these vectors. We will also denote $X_p  = \vec{X}_p \cdot \nabla_p $, omitting the function $f$.

We now consider pairs of directional derivatives 
 $X$ and $Y$. If $X$ and $Y$ are partial derivatives,  
for every regular function $f$ one has $XYf=YXf.$ If $X$ and $Y$ are  directional derivatives, in general $XYf\not=YXf.$ Explicit computation tell us that  at every point $p$
\begin{equation}
[X, Y]f(p) = (XY-YX)f(p) = (J_{\vec{Y}_p} \vec{X}_p- J_{\vec{X}_p}\vec{Y}_p)\cdot \nabla f(p),
\end{equation}
with $J_{\vec{X}_p}$ and $J_{\vec{Y}_p}$ Jacobian matrices of $\vec{X}_p$ and $\vec{Y}_p$. The quantity $[X, Y]f$ is called  \textit{commutator} since it expresses the fact that the two derivatives do not commute. The same notion can be expressed in terms of increments: one might visualize an increment from a point $p$ as the head of a vector $\vec{X}_p$ applied at the point $p$. 
Then the expression $XY-YX$ will be geometrical obtained as follows:
place $X$ down at a point, then the other $Y$ at its head, then the the first one backward finally the second one backward. The issue is whether the quadrilateral is closed. Formally this is captured by the \textit{commutator} of two elements $X$ and $Y$ at the point $p$. 

 In order to compute the second derivative $XY f$, we need to know $Yf$  at every point near $p$. This lead to the more general notion of  \textit{vector fields}, which are abstractions of the velocity field of points moving in the manifold. A vector field $X$ attaches to every point $p$ of the manifold $M$ a vector $\vec{X}_p $ from the tangent space at that point, in a smooth manner. There are no abrupt jumps between points.

\begin{figure}[tbh]
     \begin{subfigure}[b]{0.3\textwidth}
         \centering
         \includegraphics[width=\textwidth]{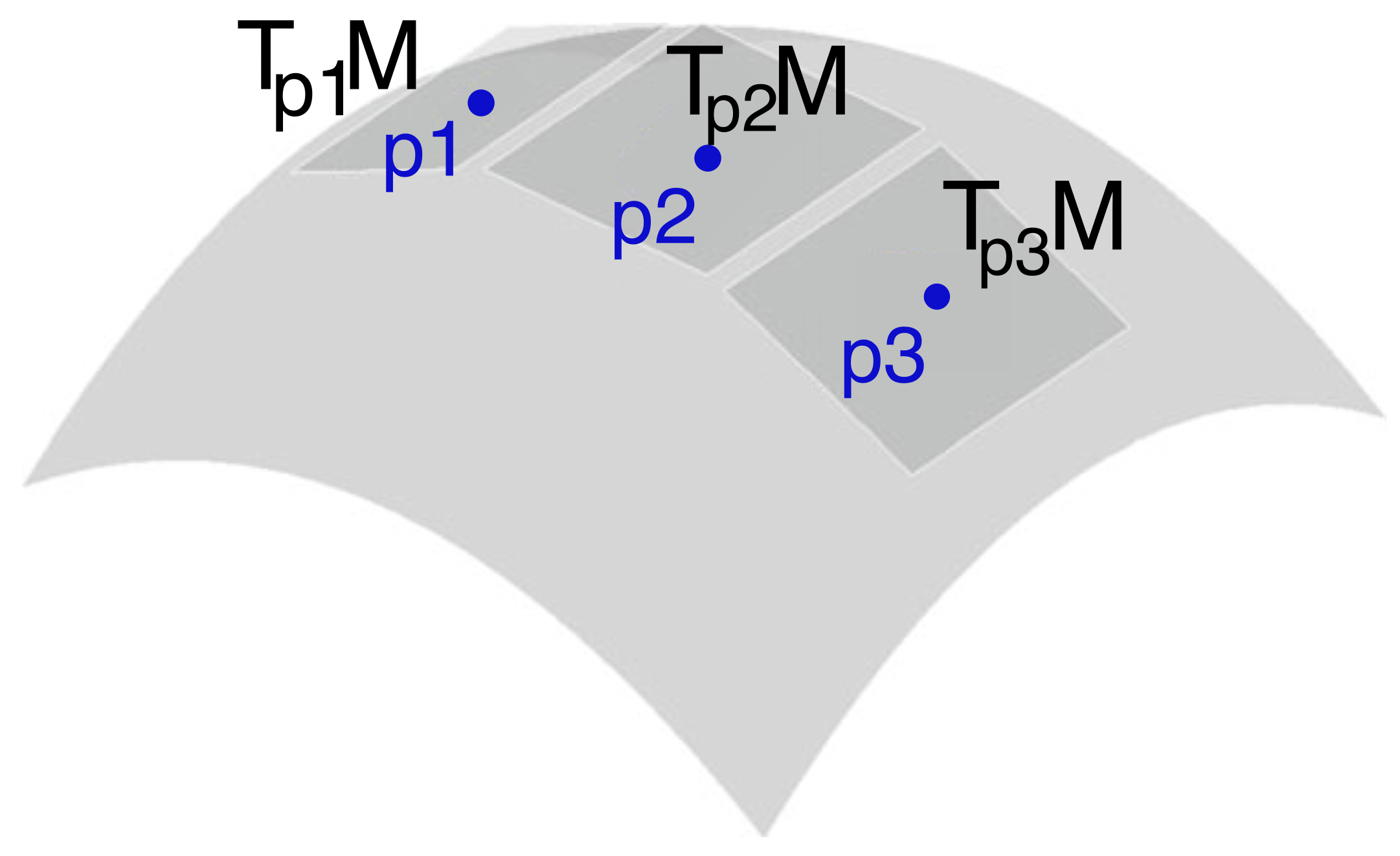}
         \caption{}
     \end{subfigure}
 \centering
     \begin{subfigure}[b]{0.3\textwidth}
         \centering
\includegraphics[width=\textwidth]{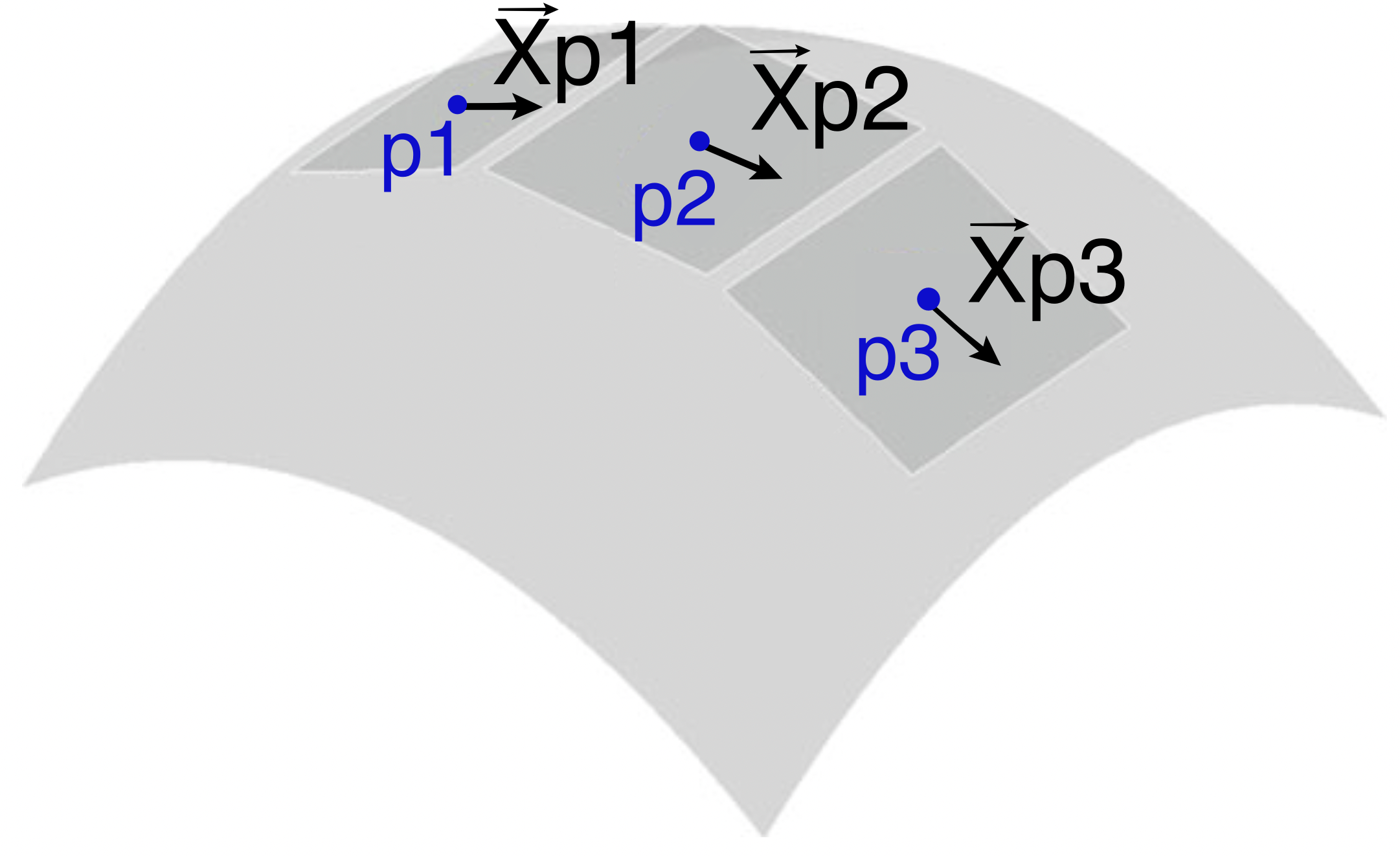}
         \caption{}
     \end{subfigure}
 \centering
     \begin{subfigure}[b]{0.3\textwidth}
         \centering
         \includegraphics[width=\textwidth]{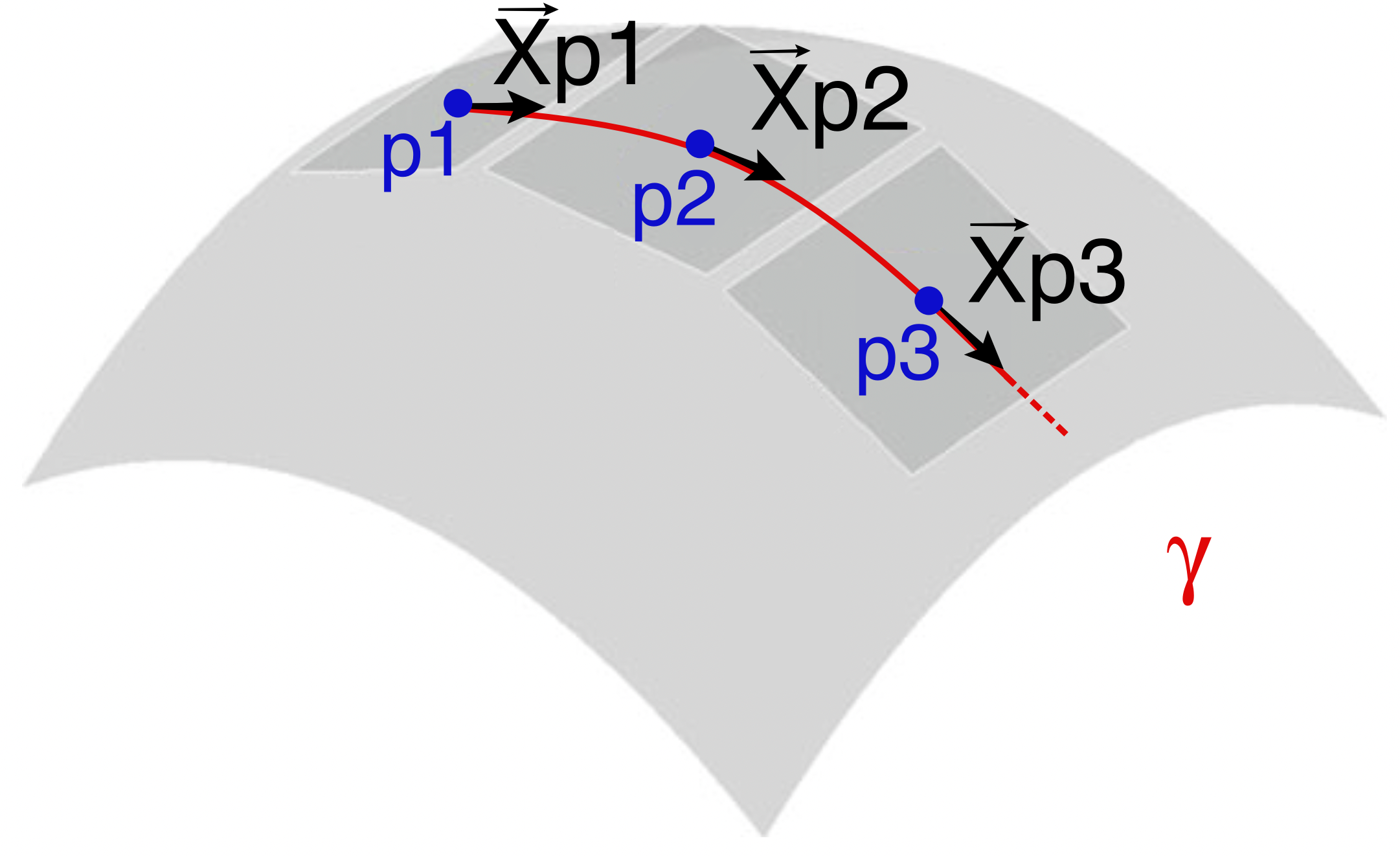}
         \caption{}
     \end{subfigure}
      \caption{(A) Tangent planes $T_{p_i}M$ (darker planes) at points $p_i, i = 1, 2, 3$ in the manifold $M$. (B) Vector field $X$ defined on $M$: to every point $p_i, i = 1, 2, 3$ of the manifold $M$ we have a vector $\vec{X}_{p_i}$ of the tangent space at that point. (C) Integral curve $\gamma$ associated with the vector field $X$ starting from $p_1 \in M$. }
      \label{tgSpaces}
\end{figure}

Since we related each tangent vector with a derivation above, we can now go further; see Fig.~\ref{tgSpaces}. Each vector field can be associated with an ordinary differential equation, whose solutions are called \textit{integral curves} of the vector field: they are parametric curves that represent specific solutions to the ordinary
differential equation depicted by the vector field. Think of it as follows: imagine you are starting at a point, and take an infinitesmimal step in the direction of a tangent vector at that point; you will now be at a neighboring point. So, again, you can take a step from this neighboring point in (possibly) another tangent direction. Continuing this process for a while, you geometrically trace out integral curves $\gamma: [t_1, t_2]\subseteq \R \longrightarrow M$. Importantly, the given vector field $X$ at the point $\gamma(t)$ is the tangent vector to the curve at that point. Importantly, this holds true everywhere along the curve, so that the integral curve satisfies a differential equation:
\begin{equation}
\dot \gamma (t) = \vec{X}_{\gamma(t)}.
\end{equation}

All the tangent spaces of a manifold may be "glued together" to form a new differentiable manifold with twice the dimension of the original manifold, called the \textit{tangent bundle} of the manifold. As a set, it is given by the disjoint union of the tangent spaces of $M$, that is:
\begin{equation}
\begin{aligned}TM&=\bigsqcup _{p\in M}T_{p}M =
\left\{(p, X_p)\mid p\in M,\, X_p \in T_{p}M\right\}.\end{aligned}
\end{equation}
In particular, an element of $TM$ can be thought of as a pair $(p,X_p)$, where $p$ is a point in $M$ and $X_p$ is a tangent vector to $M$ at $p$.
There exists a natural projection
 $\pi : TM \rightarrow M $
defined by $\pi (p,X_p)=p$. which maps each element of the tangent space $T_{p}M$ to the single point $p$.
 
\subsection{Group action on a manifold}
The operation of adding (real) numbers has an important algebraic structure, called a group. It requires, for example, that the sum of any two numbers is again a number; that there is an inverse operation "-"; and that there is an identity operation "0" that is, adding to any number yields the same number. 

When a group $G$ acts on a manifold (e.g. the real numbers, above), it means that each of its elements performs a certain operation on all the elements of the manifold in a way that is compatible with the manifold itself. More precisely, this action is described by a map $\sigma :G\times M\to M,(g,x)\mapsto g\cdot x$ which is the (left) group action of a group $G$ on a smooth manifold $M$, if the map $\sigma$  is differentiable.

\begin{figure}[tbh]
     \centering

         \includegraphics[width=0.6\textwidth]{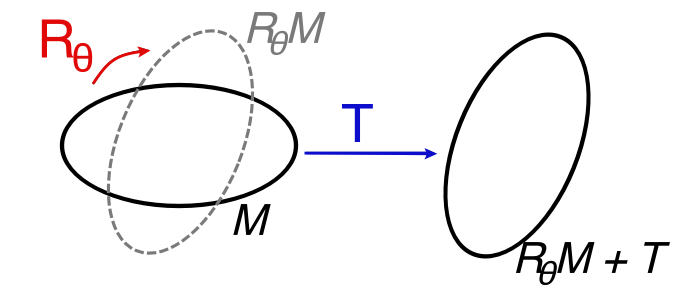}

      \caption{Group action of the roto-translation group $SE(2)$ on the manifold $M$ (black ellipse): first, the manifold is rotated through a rotation of angle $\theta$ obtaining $R_\theta M$, and then a translation is applied, moving the rotated manifold in space realizing $R_\theta M+ T$. }
      \label{figGroupAction}
\end{figure}

For example, we can take the bidimensional roto-translation group $SE(2)=\R^2\times \S^1$ and define its action on a smooth manifold $M \subseteq \R^2$ following the group law: first we apply a rotation and then a translation of the manifold itself. This is formalized through the map $\sigma: SE(2) \times M \longrightarrow M$, $\sigma(g, p)= (Rp+q)$, with $g = (q, R) \in SE(2)$, namely a point $q \in \R^2$ and $R$ bidimensional rotation of angle $\theta \in \S^1$.  A graphical example is shown in Figure \ref{figGroupAction}.

We are now ready to generalize these familiar ideas to cortical space, with its special position $\times$ orientation structure, or to stereo space.
 
\subsection{Sub-Riemannian geometry}
\label{srDef_appendix}
A point constraint to move on a manifold, illustrated above, dictates that one can move only along directions tangent to the manifold, since moving in the normal direction would leave the manifold. 
This means that, for every point $p$, the set of admissible directions of  displacement coincides with the tangent plane $T_pM$. In the presence of further constraints, some tangent directions could be forbidden. This leads to introducing, at every point $p$, the admissible tangent space $\mathcal{A}_p$, which is the subspace of $T_p M$ of admissible directions of movement.
 If the tangent space $T_p M$ has dimension $n$, the admissible tangent space $\mathcal{A}_p$ will have dimension $m \leq n$. Repeating the same construction for every point of the manifold, we call the \textit{admissible tangent bundle} the union of admissible tangent spaces at every point:
$\mathcal{A} =\bigsqcup _{p\in M}\mathcal{A}_p$.
If we introduce a scalar product on $\mathcal{A}_p$, then we are able to define a norm on vectors with the aim to measure the length of such vectors and the distance between points. The manifold with these properties is usually called sub-Riemannian manifold, while manifolds where movements are allowed in any direction are called Riemannian manifolds. 

\begin{figure}[tbh]
 \centering
     \begin{subfigure}[b]{0.4\textwidth}
         \centering
         \includegraphics[width=\textwidth]{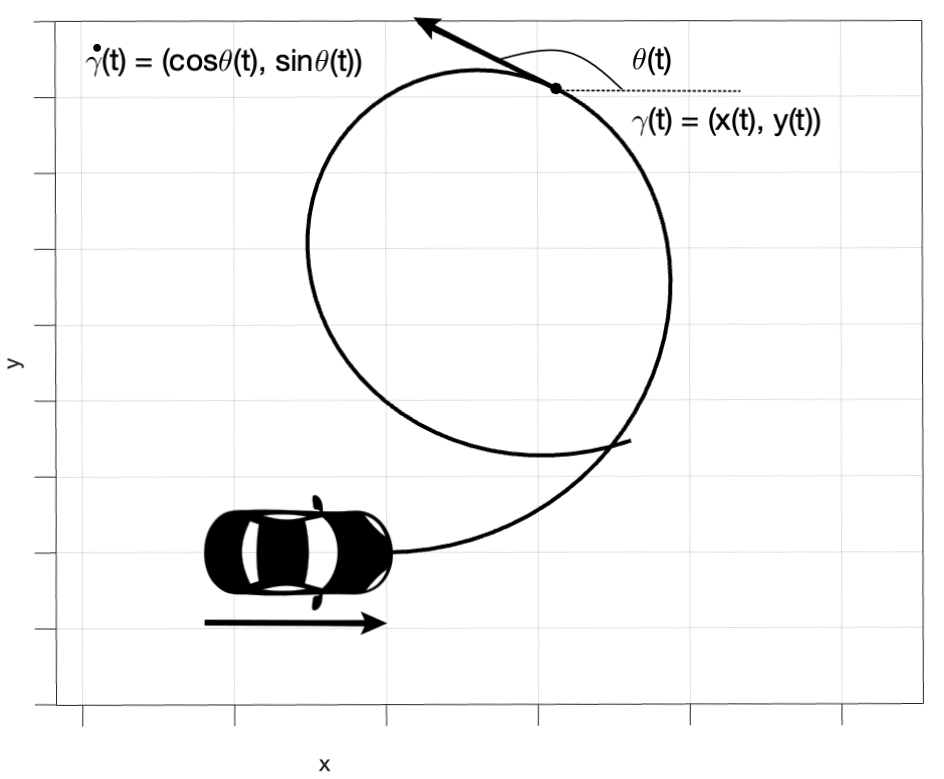}
         \caption{}
     \end{subfigure}
      \centering
     \begin{subfigure}[b]{0.55\textwidth}
         \centering
         \includegraphics[width=\textwidth]{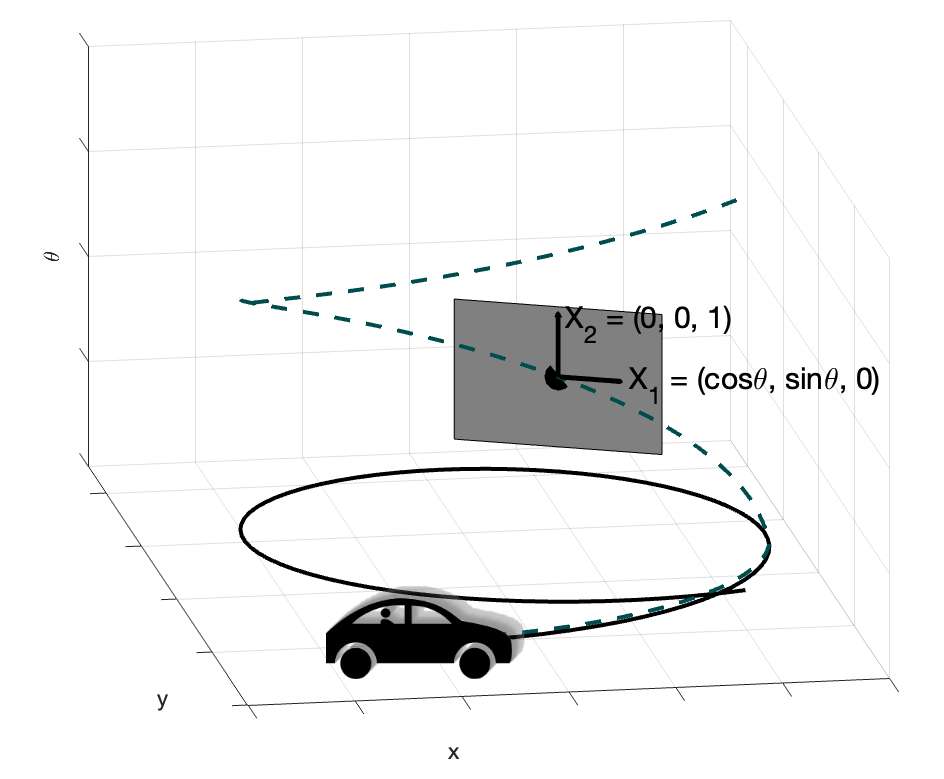}
         \caption{}
     \end{subfigure}
\caption{(A) Geometric set-up of the motion of a car moving on a plane.
(B)Sub-Riemannian formalization in $SE(2)$. Tangent vector of the path is constrained to be in the gray plane, span of $\vec{X}_{1, p}$ and $\vec{X}_{2, p}$, admissible directions of movement. }
\label{car}
\end{figure}

Let us explicitly note that while Riemannian geometry arises in presence of a physical constraints, sub-Riemannian geometry arises in presence of differential constraints, as for example in the description of the motion of vehicles.
A car moves on a bidimensional plane, but it can only move in its current direction or it can change its current orientation by rotating the steering wheel. 
These are the admissible directions. Moreover, the car cannot move "sideways" (forbidden direction): this prevents one from directly reaching any other direction while remaining in the initial position, restricting the allowable motions to a simultaneous combination of the two admissible movements. The trajectory described by the vehicle will therefore be a curve, whose tangent is constrained to follow the two admissible directions. The formalization of this sub-Riemannian problem takes place in $SE(2)$, considering for every $p \in SE(2)$ as admissible tangent space $A_pSE(2)$ the subspace generated by the current direction $\vec{X}_{1, p} = (\cos\theta, \sin\theta, 0)^T$ and the direction of rotation $\vec{X}_{2, p} = (0, 0, 1)^T$. See Figure \ref{car}.

Similarly, we can move from a retinotopic $(x,y)$ position to another retinotopic position, $(x',y')$, moving "up" or "down" through orientation columns from $\theta$ to $\theta'$, but we cannot reach $\theta'$ from $\theta$ maintaining the same initial position (running through the same orientation column): in order to reach the "forbidden direction" we have to walk simultaneously through positions and orientations.
This restriction of movement is what distinguishes a Euclidean (or Riemannian) geometry from a sub-Riemannian geometry.

\section{Proof of Proposition \ref{prop:crossProd}}
\label{app:proofProp}
\markboth{}{}
In this appendix, we show ho to prove Proposition 
\ref{prop:crossProd} using tools of differential geometry, and in particular the concept of differential $k-$form. 

\subsection{Differential forms} 
A differential $k$-form on an $n$-dimensional smooth manifold $M$ is any multilinear function $\omega :TM^k\longrightarrow \R$ which takes as input $k$ smooth vector fields and outputs a scalar element, satisfying the antisymmetry property:
$$\omega(X_1,\ldots,X_i,\ldots,X_j,\ldots, X_k) =-\omega(X_1,\ldots,X_j,\ldots,X_i,\ldots, X_k),$$ with $k\leq n$ and $k, n \in \N$. 

In the special case where $\omega$ is a $1$-form, it is worth noting that this is an element of the dual space to $TM$ (\textit{cotangent space}): $\omega \in TM^* \iff \omega: TM\longrightarrow \R$. If we have coordinates $(x_1, \ldots, x_n )$ on $M$, we can express the $1$-forms using the dual basis $\{\dd x_1, \ldots, \dd x_n \}$ of $TM^*$: 

$$\omega_ {p}= f_1(\bar x_1, \ldots, \bar x_n)\dd x_1 + \ldots + f_n(\bar x_1, \ldots, \bar x_n)\dd x_n, \text{ with } p = (\bar x_1, \ldots, \bar x_n),$$
with $f_i$ scalar smooth functions.

Furthermore, it is possible to multiply via the wedge product $\wedge$ a differential $k$-form, $\omega$,  with a differential $l$- form, $\eta$, obtaining a differential $k+l$-form $\omega \wedge \eta$. More precisely, we are interested in the wedge product of $1$-forms $\omega$ and $\eta$, where the wedge product can be computed as: $\omega\wedge \eta (X, Y) = \omega(X)\eta(Y ) - \omega(Y)\eta(X)$, with $X$ and $Y$ vector fields on $M$.

\subsection{Development of the proof}

\begin{proposition}
The binocular interaction term $O_LO_R$ can be associated with the cross product of the left and right directions defined through \eqref{X3vf}, namely $\omega_{p_L}^\star$ and $\omega_{p_R}^\star$ of monocular simple cells:
\begin{equation}
O_LO_R = \omega_{p_L}^\star \times \omega_{p_R}^\star.
\end{equation}
\end{proposition}

\proof
As noted in subsubsection \ref{CSmodel},
the output of simple cells \eqref{RP} in $SE(2)$ can then be locally approximated as
$ O(x, y, \theta) = - X_{3,p}(I_\sigma)(x,y)$ where $I_\sigma$ is a smoothed version of $I$, obtained by convolving it with a Gaussian kernel, the vector field 
\begin{equation}
 X_{3, p} = -\sin\theta\partial_x + \cos\theta\partial_y, 
 \end{equation}
with $p = (x, y, \theta) \in SE(2)$. Switching to the dual space, the action of simple cells induces a choice of a $1$-form separately on each cell:
\begin{equation}
\label{1form}
\omega_p = -\sin\theta \dd x + \cos\theta \dd y.
\end{equation}
Accordingly, it is possible to  re-write the binocular interaction term as:
 \begin{equation}
 \label{binRP}
 \begin{aligned}
O_LO_R =  \ X_{3,p_R}(I_{\sigma R})(x_{R},y)X_{3, p_L}(I_{\sigma L})(x_{L},y). \hspace{0.5cm}\\
 \end{aligned}
 \end{equation}
 In the following, we will see that this binocular action can be described by a $2$-form defined in terms of the two $1$-forms of monocular simple cells.
 
We will denote with the subscript $R$ the quantities corresponding to the right monocular structure, and we will use the subscript $L$ for the left one. So, we define $v_R := (J_{I_{\sigma_R}} \vec{X}_{3,p_R})X_{3,p_R}$ using the Jacobian (differential) of the smoothed version of the image $I$, in such a way that we have $\omega_{p_R}(v_R)= X_{3, p_R}(I_{\sigma_R}) = (J_{I_{\sigma_R}} \vec{X}_{3,p_R})$ since $\omega_{p_R}(X_{3, p_R})= 1$ and $J_{I_{\sigma_R}} \vec{X}_{3,p_R} \in \R$; the same reasoning holds for the left structure. It is then possible to recast \eqref{binRP} in the retinal coordinates as: 
 \begin{equation}
 \label{omegaBin1}
 \begin{aligned}
  O_LO_R = &\omega_{p_L}(v_L)\omega_{p_R}(v_R)\\
                    = & \omega_{p_L}\wedge\omega_{p_R}(v_L,v_R) + \underbrace{\omega_{p_R}(v_L)\omega_{p_L}(v_R)}_{=0}, \\
                    = & \omega_{p_L}\wedge\omega_{p_R}(v_L,v_R),
\end{aligned}
\end{equation}
exploiting the properties of the wedge product and the left and right retinal coordinates.  

The retinal coordinates can be expressed in terms of cyclopean coordinates  \eqref{dispDEF} as $x_R = x-d$ and $x_L = x+d$; then, the extended left and right $1$-form can be written as: 
\begin{equation}
\label{1FormCyclopean}
\begin{aligned}
\omega_{p_R} =& -\sin\theta_R\dd x+\cos\theta_R \dd y +\sin\theta_R \dd d\\
\omega_{p_L} =& -\sin\theta_L\dd x+\cos\theta_L \dd y -\sin\theta_L \dd d.\\
\end{aligned}
\end{equation}
Taking advantage of the isomorphism provided by the Hodge star between vectors and $2$-forms in $\R^3$, we relate the exterior and the cross product, using notations \eqref{1FormVCyclopean}
\footnote{Using the notation $\omega^\star$ we identify the vector whose components are the coefficients of the $1$-form $\omega$ with respect to the dual basis}, in the following way:
\begin{equation}
\star(\omega_{p_L}\wedge\omega_{p_R})= \omega_{p_L}^\star \times \omega_{p_R}^\star,
\end{equation}
from which it follows the thesis. 
\endproof

Throughout the paper, to lighten the notation, we will call $\omega_L = \omega_{p_L}$ and $\omega_R = \omega_{p_R}$.


\subsubsection{Meaning of the mathematical objects}
\label{app:proofProp_sec}

We conclude this section with a consideration on the mathematical tools introduced and used in this setting, to understand how the mathematical models proposed by Citti and Sarti, starting with \cite{CS06}, assign these different mathematical objects to the physical cell, to its action , and to the result of its action.

\begin{remark}
\label{rmk:diffModel}
    It is well known that an odd simple cell (selective for orientation) is activated as a result of the presence of a stimulus to select its direction (tangent vector to the perceptual curve). In this setting, the mathematical intuition behind the model proposed in \cite{CS06} is to identify each cell with a $1$-differential form, which is an element of the cotangent space. Roughly speaking, this differential form is able to grasp a vector that corresponds to the direction of the stimulus: this is the result of the action of the cell. Formally, this vector will be an element of the tangent space, and more precisely it will lay in the kernel of the $1$-form. This vector space is then associated with the action of the cell.
\end{remark}

The same reasoning is applied to different families of cells in a series of papers (\cite{SCP07, BCCS14, ASFCSHR17, BSC20}) even if these are characterized by distinct sub-Riemannian structures in various manifolds. The interested reader could refer to \cite{CS14} for a review. Similarly, we have found the same geometrical organization in the family of binocular cells.

\begin{remark}
     In this paper, we have dealt with binocular cells which are a combination of monocular simple cells. To these coupled simple cells (one for the left and one for the right eye) we formally associate a $2$-differential form, the wedge product of the two monocular left and right $1$-forms. This $2$-form can grasp again a vector, lying in the kernel of this mathematical object, identifying the three-dimensional stimulus direction. Thus, the same reasoning of Remark \ref{rmk:diffModel} also applies here to the binocular family of cells.
\end{remark}

Translating the results of Remark \ref{rmk:diffModel} into different spaces, with different dimensions, it is then possible to use the same mathematical objects to explain the behavior of families of different cells, identifying geometrically the mathematical objects at the basis of the functionality of the family of studied cells. 

\section{Change of variables}
\label{compApp}
\markboth{}{}
Let us recover the expression of the $1$-forms $\tilde \omega_L:= U_{t_L}$ and  $\tilde \omega_R:= U_{t_R} $.
Recall here the change of variable \eqref{changeVar3DNeur}: 
\begin{equation}
\begin{cases}
r_1 = \frac{xc}{d}\\
r_2 = \frac{yc}{d}\\
r_3 = \frac{fc}{d}\\
\end{cases},
\end{equation} 
and its differential: 
\begin{equation}
\begin{cases}
\dd r_1 = \frac{c}{d}\dd x -\frac{cx}{d^2}\dd d\\
\dd r_2 = \frac{c}{d}\dd y- \frac{cy}{d^2} \dd d\\
\dd r_3 = -\frac{fc}{d^2} \dd d\\
\end{cases}.
\end{equation}
Writing the quantity $U_{t_L}$, defined in \eqref{UtlAndUtr}, in term of a $1$-form in the variables $(r_1, r_2, r_3)$ we have: 
\begin{equation}
\begin{aligned}
\tilde \omega_L = & -f\sin\theta_L \dd r_1+ f\cos\theta_L \dd r_2 + (x_L\sin\theta_L-y\cos\theta_L)\dd r_3.
\end{aligned}
\end{equation}
Changing coordinates:
\begin{equation}
\begin{aligned}
\tilde \omega_L = &  -f\sin\theta_L \left (  \frac{c}{d}\dd x -\frac{cx}{d^2}\dd d \right)  + f\cos\theta_L \left (\frac{c}{d}\dd y- \frac{cy}{d^2} \dd d \right)\\
 & + (x_L\sin\theta_L-y\cos\theta_L)\left ( -\frac{fc}{d^2} \dd d \right)\\
 = & \frac{fc}{d} \left(  -\sin\theta_L\dd x+\cos\theta_L \dd y -\sin\theta_L \dd d\right) \\
 =& \frac{fc}{d} \omega_L.\\
\end{aligned}
\end{equation} 
So, up to a scalar factor, we have that $\tilde \omega_L = \omega_L$ in the variables $(x, y, d)$. The same reasoning holds for the right structure.

\end{appendices}

\bibliography{stereoBib}
\bibliographystyle{plain}

\end{document}